%
%
\input harvmac
\newcount\yearltd\yearltd=\year\advance\yearltd by 0

\def\journal#1&#2(#3){\unskip, \sl #1\ \bf #2 \rm(19#3) }
\def\andjournal#1&#2(#3){\sl #1~\bf #2 \rm (19#3) }

\def\ie{{\it i.e.}}
\def\eg{{\it e.g.}}

\def\frac#1#2{{#1\over#2}}

\def\half{\frac12}

\def\vev#1{\langle#1\rangle}

\def\IZ{\relax\ifmmode\mathchoice  
{\hbox{\cmss Z\kern-.4em Z}}{\hbox{\cmss Z\kern-.4em Z}}  
{\lower.9pt\hbox{\cmsss Z\kern-.4em Z}}  
{\lower1.2pt\hbox{\cmsss Z\kern-.4em Z}}\else{\cmss Z\kern-.4em  
Z}\fi}

\def\inbar{\,\vrule height1.5ex width.4pt depth0pt}
\def\IC{\relax\hbox{$\inbar\kern-.3em{\rm C}$}}
\def\IR{\relax{\rm I\kern-.18em R}}
\def\IP{\relax{\rm I\kern-.18em P}}
\def\Z{{\bf Z}}

%
%

%
\catcode`\@=11
\def\slash#1{\mathord{\mathpalette\c@ncel{#1}}}
\overfullrule=0pt

\def\BB{{\cal B}}

\def\FF{{\cal F}}

\def\MM{{\cal M}}
\def\NN{{\cal N}}
\def\OO{{\cal O}}

\def\WW{{\cal W}}

\def\underrel#1\over#2{\mathrel{\mathop{\kern\z@#1}\limits_{#2}}}

\catcode`\@=12


%

\def\vev#1{\left\langle #1 \right\rangle}

\def\tr{{\rm tr}}

\def\exp{{\rm exp}}


\def\ttr{{\widetilde {\rm tr}}}
\def\tj{{\tilde j}}

\def\bBB{{\bar \BB}}

\lref\AharonyUB{
O.~Aharony, M.~Berkooz, D.~Kutasov and N.~Seiberg,
``Linear dilatons, NS5-branes and holography,''
JHEP {\bf 9810}, 004 (1998)
[arXiv:hep-th/9808149].}

\lref\GiveonPX{
A.~Giveon and D.~Kutasov,
``Little string theory in a double scaling limit,''
JHEP {\bf 9910}, 034 (1999)
[arXiv:hep-th/9909110].
}

\lref\GiveonTQ{
A.~Giveon and D.~Kutasov,
``Comments on double scaled little string theory,''
JHEP {\bf 0001}, 023 (2000)
[arXiv:hep-th/9911039].
}

\lref\AharonyKS{
O.~Aharony,
``A brief review of 'little string theories',''
Class.\ Quant.\ Grav.\  {\bf 17}, 929 (2000)
[arXiv:hep-th/9911147].
}

\lref\KutasovUF{
D.~Kutasov,
``Introduction to little string theory,''
{\it Prepared for ICTP Spring School on Superstrings and Related
Matters, Trieste, Italy, 2-10 Apr 2001.}
}

\lref\AharonyVK{
O.~Aharony, B.~Fiol, D.~Kutasov and D.~A.~Sahakyan,
``Little string theory and heterotic/type II duality,''
Nucl.\ Phys.\ B {\bf 679}, 3 (2004)
[arXiv:hep-th/0310197].
}

\lref\GiveonZM{
A.~Giveon, D.~Kutasov and O.~Pelc,
``Holography for non-critical superstrings,''
JHEP {\bf 9910}, 035 (1999)
[arXiv:hep-th/9907178].
}

\lref\CallanAT{
C.~G.~Callan, J.~A.~Harvey and A.~Strominger,
``Supersymmetric string solitons,''
arXiv:hep-th/9112030.
}

\lref\DiFrancescoUD{
P.~Di Francesco and D.~Kutasov,
 ``World sheet and space-time physics in two-dimensional (Super)string
theory,''
Nucl.\ Phys.\ B {\bf 375}, 119 (1992)
[arXiv:hep-th/9109005].
}

\lref\McGreevyCW{
J.~McGreevy, L.~Susskind and N.~Toumbas,
``Invasion of the giant gravitons from anti-de Sitter space,''
JHEP {\bf 0006}, 008 (2000)
[arXiv:hep-th/0003075].
}
\lref\BalasubramanianNH{
V.~Balasubramanian, M.~Berkooz, A.~Naqvi and M.~J.~Strassler,
``Giant gravitons in conformal field theory,''
JHEP {\bf 0204}, 034 (2002)
[arXiv:hep-th/0107119].
}
\lref\CorleyZK{
S.~Corley, A.~Jevicki and S.~Ramgoolam,
``Exact correlators of giant gravitons from dual $N = 4$ SYM theory,''
Adv.\ Theor.\ Math.\ Phys.\  {\bf 5}, 809 (2002)
[arXiv:hep-th/0111222];
S.~Corley and S.~Ramgoolam,
``Finite factorization equations and sum rules for BPS correlators in  
$N = 4$ SYM theory,''
Nucl.\ Phys.\ B {\bf 641}, 131 (2002)
[arXiv:hep-th/0205221].
}
\lref\AharonyND{
O.~Aharony, Y.~E.~Antebi, M.~Berkooz and R.~Fishman,
```Holey sheets': Pfaffians and subdeterminants as D-brane operators in  
large $N$ gauge theories,''
JHEP {\bf 0212}, 069 (2002)
[arXiv:hep-th/0211152].
}
\lref\BerensteinAH{
D.~Berenstein,
``Shape and holography: Studies of dual operators to giant gravitons,''
Nucl.\ Phys.\ B {\bf 675}, 179 (2003)
[arXiv:hep-th/0306090].
}
\lref\thooft{
G.~'t Hooft,
``A planar diagram theory for strong
interactions,'' Nucl.\ Phys.\ B {\bf 72}, 461 (1974).}
\lref\ZamolodchikovBD{
A.~B.~Zamolodchikov and V.~A.~Fateev,
``Operator algebra and correlation functions in the two-dimensional 
Wess-Zumino $SU(2) \times SU(2)$ chiral model,''
Sov.\ J.\ Nucl.\ Phys.\  {\bf 43}, 657 (1986)
[Yad.\ Fiz.\  {\bf 43}, 1031 (1986)].
}

\lref\GiribetFT{
G.~Giribet and C.~Nunez,
``Correlators in AdS(3) string theory,''
JHEP {\bf 0106}, 010 (2001)
[arXiv:hep-th/0105200].
}

\lref\BershadskyIN{
M.~Bershadsky and D.~Kutasov,
``Comment on gauged WZW theory,''
Phys.\ Lett.\ B {\bf 266}, 345 (1991).
}
\lref\DijkgraafBA{
R.~Dijkgraaf, H.~Verlinde and E.~Verlinde,
``String propagation in a black hole geometry,''
Nucl.\ Phys.\ B {\bf 371}, 269 (1992).
}
\lref\GiveonUP{
A.~Giveon and D.~Kutasov,
``Notes on AdS(3),''
Nucl.\ Phys.\ B {\bf 621}, 303 (2002)
[arXiv:hep-th/0106004].
}

\lref\MaldacenaKM{
J.~M.~Maldacena and H.~Ooguri,
``Strings in AdS(3) and the SL(2,R) WZW model. III: Correlation  functions,''
Phys.\ Rev.\ D {\bf 65}, 106006 (2002)
[arXiv:hep-th/0111180].
}

\lref\MaldacenaHW{
J.~M.~Maldacena and H.~Ooguri,
``Strings in AdS(3) and SL(2,R) WZW model. I,''
J.\ Math.\ Phys.\  {\bf 42}, 2929 (2001)
[arXiv:hep-th/0001053].
}

\lref\ParnachevGW{
A.~Parnachev and D.~A.~Sahakyan,
``Some remarks on D-branes in AdS(3),''
JHEP {\bf 0110}, 022 (2001)
[arXiv:hep-th/0109150].
}

\lref\TeschnerFT{
J.~Teschner,
``On structure constants and fusion rules in the SL(2,C)/SU(2) WZNW  model,''
Nucl.\ Phys.\ B {\bf 546}, 390 (1999)
[arXiv:hep-th/9712256].
}

\lref\TeschnerUG{
J.~Teschner,
``Operator product expansion and factorization in the H-3+ WZNW model,''
Nucl.\ Phys.\ B {\bf 571}, 555 (2000)
[arXiv:hep-th/9906215].
}

\lref\KutasovXU{
D.~Kutasov and N.~Seiberg,
``More comments on string theory on AdS(3),''
JHEP {\bf 9904}, 008 (1999)
[arXiv:hep-th/9903219].
}

\lref\PeskinEV{
M.~E.~Peskin and D.~V.~Schroeder,
``An Introduction To Quantum Field Theory,''
Addison-Wesley, 1997,  section 7.2.
}

\lref\fzz{V. Fateev, A. B. Zamolodchikov and Al. B. Zamolodchikov, 
unpublished.}

\lref\KazakovPM{
V.~Kazakov, I.~K.~Kostov and D.~Kutasov,
``A matrix model for the two-dimensional black hole,''
Nucl.\ Phys.\ B {\bf 622}, 141 (2002)
[arXiv:hep-th/0101011].
}

\lref\AharonyTH{
O.~Aharony, M.~Berkooz, S.~Kachru, N.~Seiberg and E.~Silverstein,
``Matrix description of interacting theories in six dimensions,''
Adv.\ Theor.\ Math.\ Phys.\  {\bf 1}, 148 (1998)
[arXiv:hep-th/9707079].
}
\lref\WittenYU{
E.~Witten,
``On the conformal field theory of the Higgs branch,''
JHEP {\bf 9707}, 003 (1997)
[arXiv:hep-th/9707093].
}
\lref\AharonyDW{
O.~Aharony and M.~Berkooz,
``IR dynamics of $d = 2$, $\NN = (4,4)$ 
gauge theories and DLCQ of 'little  string theories',''
JHEP {\bf 9910}, 030 (1999)
[arXiv:hep-th/9909101].
}
\lref\SethiZZ{
S.~Sethi,
``The matrix formulation of type IIB five-branes,''
Nucl.\ Phys.\ B {\bf 523}, 158 (1998)
[arXiv:hep-th/9710005].
}
\lref\GanorJX{
O.~J.~Ganor and S.~Sethi,
``New perspectives on Yang-Mills theories with sixteen supersymmetries,''
JHEP {\bf 9801}, 007 (1998)
[arXiv:hep-th/9712071].
}

\lref\MinwallaPX{
S.~Minwalla, M.~Van Raamsdonk and N.~Seiberg,
``Noncommutative perturbative dynamics,''
JHEP {\bf 0002}, 020 (2000)
[arXiv:hep-th/9912072].
}
\lref\VanRaamsdonkRR{
M.~Van Raamsdonk and N.~Seiberg,
``Comments on noncommutative perturbative dynamics,''
JHEP {\bf 0003}, 035 (2000)
[arXiv:hep-th/0002186].
}

\lref\KutasovUA{
D.~Kutasov and N.~Seiberg,
``Noncritical Superstrings,''
Phys.\ Lett.\ B {\bf 251}, 67 (1990).
}
\lref\BerkoozCQ{
M.~Berkooz, M.~Rozali and N.~Seiberg,
``On transverse fivebranes in M(atrix) theory on T**5,''
Phys.\ Lett.\ B {\bf 408}, 105 (1997)
[arXiv:hep-th/9704089].
}
\lref\SeibergZK{
N.~Seiberg,
 ``New theories in six dimensions and matrix description of M-theory on  T**5
and T**5/Z(2),''
Phys.\ Lett.\ B {\bf 408}, 98 (1997)
[arXiv:hep-th/9705221].
}
\lref\KlebanovQA{
I.~R.~Klebanov,
``String theory in two-dimensions,''
arXiv:hep-th/9108019.
}

\lref\GinspargIS{
P.~Ginsparg and G.~W.~Moore,
``Lectures On 2-D Gravity And 2-D String Theory,''
arXiv:hep-th/9304011.
}

\lref\PolchinskiMB{
J.~Polchinski,
``What is string theory?,''
arXiv:hep-th/9411028.
}

\lref\TeschnerRV{
J.~Teschner,
``Liouville theory revisited,''
Class.\ Quant.\ Grav.\  {\bf 18}, R153 (2001)
[arXiv:hep-th/0104158].
}

\lref\DiFrancescoSS{
P.~Di Francesco and D.~Kutasov,
``Correlation functions in 2-D string theory,''
Phys.\ Lett.\ B {\bf 261}, 385 (1991).
}

\lref\TakayanagiSM{
T.~Takayanagi and N.~Toumbas,
``A matrix model dual of type 0B string theory in two dimensions,''
JHEP {\bf 0307}, 064 (2003)
[arXiv:hep-th/0307083].
}
\lref\KlebanovHB{
I.~R.~Klebanov and M.~J.~Strassler,
``Supergravity and a confining gauge theory: Duality cascades and
$\chi$SB-resolution of naked singularities,''
JHEP {\bf 0008}, 052 (2000)
[arXiv:hep-th/0007191].
}
\lref\DoreyPP{
N.~Dorey,
``S-duality, deconstruction and confinement for a marginal deformation of N =
4 SUSY Yang-Mills,''
arXiv:hep-th/0310117.
}

\lref\DouglasUP{
M.~R.~Douglas, I.~R.~Klebanov, D.~Kutasov, J.~Maldacena, E.~Martinec 
and N.~Seiberg,
``A new hat for the c = 1 matrix model,''
arXiv:hep-th/0307195.
}

\lref\GiveonWN{
A.~Giveon, A.~Konechny, A.~Pakman and A.~Sever,
``Type 0 strings in a 2-d black hole,''
JHEP {\bf 0310}, 025 (2003)
[arXiv:hep-th/0309056].
}

\lref\McGreevyDN{
J.~McGreevy, S.~Murthy and H.~Verlinde,
``Two-dimensional superstrings and the supersymmetric matrix model,''
arXiv:hep-th/0308105.
}

\lref\AtickSI{
J.~J.~Atick and E.~Witten,
``The Hagedorn transition and the number of degrees of freedom of string
theory,''
Nucl.\ Phys.\ B {\bf 310}, 291 (1988).
}
\lref\KutasovJP{
D.~Kutasov and D.~A.~Sahakyan,
``Comments on the thermodynamics of little string theory,''
JHEP {\bf 0102}, 021 (2001)
[arXiv:hep-th/0012258].
}
\lref\MaldacenaCG{
J.~M.~Maldacena and A.~Strominger,
``Semiclassical decay of near-extremal fivebranes,''
JHEP {\bf 9712}, 008 (1997)
[arXiv:hep-th/9710014].
}
\lref\AharonyTT{
O.~Aharony and T.~Banks,
``Note on the quantum mechanics of M theory,''
JHEP {\bf 9903}, 016 (1999)
[arXiv:hep-th/9812237].
}
\lref\BerkoozMZ{
M.~Berkooz and M.~Rozali,
``Near Hagedorn dynamics of NS fivebranes, or a new universality class  of
coiled strings,''
JHEP {\bf 0005}, 040 (2000)
[arXiv:hep-th/0005047].
}
\lref\HarmarkHW{
T.~Harmark and N.~A.~Obers,
``Hagedorn behaviour of little string theory from string corrections to
NS5-branes,''
Phys.\ Lett.\ B {\bf 485}, 285 (2000)
[arXiv:hep-th/0005021].
}
\lref\BuchelDG{
A.~Buchel,
``On the thermodynamic instability of LST,''
arXiv:hep-th/0107102.
}
\lref\NarayanDR{
K.~Narayan and M.~Rangamani,
``Hot little string correlators: A view from supergravity,''
JHEP {\bf 0108}, 054 (2001)
[arXiv:hep-th/0107111].
}
\lref\DeBoerDD{
P.~A.~DeBoer and M.~Rozali,
``Thermal correlators in little string theory,''
Phys.\ Rev.\ D {\bf 67}, 086009 (2003)
[arXiv:hep-th/0301059].
}
\lref\TeschnerFT{
J.~Teschner,
``On structure constants and fusion rules in the $SL(2,C)/SU(2)$ WZNW  model,''
Nucl.\ Phys.\ B {\bf 546}, 390 (1999)
[arXiv:hep-th/9712256].
}
\lref\TeschnerUG{
J.~Teschner,
``Operator product expansion and factorization in the $H_3^+$ WZNW model,''
Nucl.\ Phys.\ B {\bf 571}, 555 (2000)
[arXiv:hep-th/9906215].
}

\lref\WittenZW{
E.~Witten,
``Anti-de Sitter space, thermal phase transition, and confinement in  gauge
theories,''
Adv.\ Theor.\ Math.\ Phys.\  {\bf 2}, 505 (1998)
[arXiv:hep-th/9803131].
}

\lref\AntoniadisSW{
I.~Antoniadis, S.~Dimopoulos and A.~Giveon,
``Little string theory at a TeV,''
JHEP {\bf 0105}, 055 (2001)
[arXiv:hep-th/0103033].
}
\lref\ItzhakiDD{
N.~Itzhaki, J.~M.~Maldacena, J.~Sonnenschein and S.~Yankielowicz,
``Supergravity and the large N limit of theories with sixteen  supercharges,''
Phys.\ Rev.\ D {\bf 58}, 046004 (1998)
[arXiv:hep-th/9802042].
}
\lref\BerensteinJQ{
D.~Berenstein, J.~M.~Maldacena and H.~Nastase,
``Strings in flat space and pp waves from N = 4 super Yang Mills,''
JHEP {\bf 0204}, 013 (2002)
[arXiv:hep-th/0202021].
}
\lref\BerensteinKK{
D.~Berenstein,
``A toy model for the AdS/CFT correspondence,''
arXiv:hep-th/0403110.
}

\lref\deAlwisPR{
S.~P.~de Alwis, J.~Polchinski and R.~Schimmrigk,
``Heterotic Strings With Tree Level Cosmological Constant,''
Phys.\ Lett.\ B {\bf 218}, 449 (1989).
}

\lref\ArgyresJJ{
P.~C.~Argyres and M.~R.~Douglas,
``New phenomena in $SU(3)$ supersymmetric gauge theory,''
Nucl.\ Phys.\ B {\bf 448}, 93 (1995)
[arXiv:hep-th/9505062].
}

\lref\ArgyresXN{
P.~C.~Argyres, M.~Ronen Plesser, N.~Seiberg and E.~Witten,
``New $N=2$ Superconformal Field Theories in Four Dimensions,''
Nucl.\ Phys.\ B {\bf 461}, 71 (1996)
[arXiv:hep-th/9511154].
}

\lref\EguchiVU{
T.~Eguchi, K.~Hori, K.~Ito and S.~K.~Yang,
``Study of $N=2$ Superconformal Field Theories in $4$ Dimensions,''
Nucl.\ Phys.\ B {\bf 471}, 430 (1996)
[arXiv:hep-th/9603002].
}
\lref\BanksVH{
T.~Banks, W.~Fischler, S.~H.~Shenker and L.~Susskind,
``M theory as a matrix model: A conjecture,''
Phys.\ Rev.\ D {\bf 55}, 5112 (1997)
[arXiv:hep-th/9610043].
}
\lref\DijkgraafVV{
R.~Dijkgraaf, E.~Verlinde and H.~Verlinde,
``Matrix string theory,''
Nucl.\ Phys.\ B {\bf 500}, 43 (1997)
[arXiv:hep-th/9703030].
}
\lref\MotlTH{
L.~Motl,
``Proposals on nonperturbative superstring interactions,''
arXiv:hep-th/9701025.
}
\lref\ArkaniHamedIE{
N.~Arkani-Hamed, A.~G.~Cohen, D.~B.~Kaplan, A.~Karch and L.~Motl,
``Deconstructing $(2,0)$ and little string theories,''
JHEP {\bf 0301}, 083 (2003)
[arXiv:hep-th/0110146].
}

\lref\McGreevyKB{
J.~McGreevy and H.~Verlinde,
``Strings from tachyons: The c = 1 matrix reloaded,''
JHEP {\bf 0312}, 054 (2003)
[arXiv:hep-th/0304224].
}

\lref\KlebanovKM{
I.~R.~Klebanov, J.~Maldacena and N.~Seiberg,
``D-brane decay in two-dimensional string theory,''
JHEP {\bf 0307}, 045 (2003)
[arXiv:hep-th/0305159].
}

\lref\TakayanagiSM{
T.~Takayanagi and N.~Toumbas,
``A matrix model dual of type 0B string theory in two dimensions,''
JHEP {\bf 0307}, 064 (2003)
[arXiv:hep-th/0307083].
}

\lref\DouglasUP{
M.~R.~Douglas, I.~R.~Klebanov, D.~Kutasov, J.~Maldacena, E.~Martinec and N.~Seiberg,
``A new hat for the c = 1 matrix model,''
arXiv:hep-th/0307195.
}

\lref\RangamaniIR{
M.~Rangamani,
``Little string thermodynamics,''
JHEP {\bf 0106}, 042 (2001)
[arXiv:hep-th/0104125].
}


\rightline{WIS/09/04-APR-DPP, RI-03-04, EFI-04-11}
\Title{
\rightline{hep-th/0404016}}
{\vbox{\centerline{LSZ in LST}}}
\centerline{\it Ofer Aharony${}^{a}$, Amit Giveon${}^{b}$ {\rm and}
David Kutasov${}^{c}$}
\bigskip
\centerline{${}^a$Department of Particle Physics, Weizmann Institute of 
Science, Rehovot 76100, Israel}
\centerline{\tt Ofer.Aharony@weizmann.ac.il}
\smallskip
\centerline{${}^b$Racah Institute of Physics, The Hebrew University,
Jerusalem 91904, Israel}
\centerline{\tt giveon@vms.huji.ac.il}
\smallskip
\centerline{${}^c$EFI and Department of Physics, University of Chicago}
\centerline{5640 S. Ellis Av., Chicago, IL 60637, USA }
\centerline{\tt kutasov@theory.uchicago.edu}

\bigskip
\noindent
We discuss the analytic structure of off-shell correlation functions 
in Little String Theories (LSTs) using their description as 
asymptotically linear dilaton backgrounds of string theory. We focus 
on specific points in the LST moduli space where this description 
involves the spacetime $\IR^{d-1,1}\times SL(2)/U(1)$ times a compact 
CFT, though we expect our qualitative results to be much more general. 
We show that $n$-point functions of vertex operators $\OO(p_{\mu})$ 
have single poles as a function of the $d$-dimensional momentum 
$p_{\mu}$, which correspond to normalizable states localized near the 
tip of the $SL(2)/U(1)$ cigar. Additional poles arise due to the 
non-trivial dynamics in the bulk of the cigar, and these can lead to 
a type of UV/IR mixing. Our results explain some previously puzzling 
features of the low energy behavior of the Green functions. As another 
application, we compute the precise combinations of single-trace and 
multi-trace operators in the low-energy gauge theory which map to 
single string vertex operators in the $\NN=(1,1)$ supersymmetric $d=6$ 
LST. We also discuss the implications of our results for two dimensional 
string theories and for the (non-existence of a) Hagedorn phase 
transition in LSTs.

\vfill

\Date{}


\newsec{Introduction and Summary}

Little String Theory (for reviews see \refs{\AharonyKS,\KutasovUF}
and section 4 of \AharonyVK) describes the physics of
defects, such as $NS5$-branes and/or singularities, in string
theory. It can be isolated from the rest of the dynamics by 
taking the decoupling limit in which the string coupling far 
from the defect goes to zero \refs{\BerkoozCQ,\SeibergZK}. 
The effective coupling near the defect remains non-vanishing 
and grows as one approaches the defect.

A number of methods to study the dynamics of Little String Theory 
(LST) have been proposed. One uses discrete light-cone quantization 
\refs{\AharonyTH,\WittenYU,\AharonyDW}, in the spirit of Matrix
theory. Another, which we will focus on here, is provided by 
string propagation in the near-horizon geometry of the defect 
\AharonyUB, as in the AdS/CFT correspondence. LSTs are 
holographically equivalent to asymptotically linear dilaton backgrounds 
(which are sometimes called ``non-critical string theories''). 
In the near horizon geometry, the radial direction away from the 
defect is described by a non-compact scalar $\phi$. For large 
positive $\phi$, the dilaton $\Phi$ depends on $\phi$ as follows:
\eqn\ggss{g_s= e^\Phi\simeq e^{-{Q\over2}\phi},}
where the slope of the linear dilaton, $Q$, is real and positive. 
The worldsheet central charge of $\phi$ is given by
\eqn\cqq{c_\phi=1+3Q^2~.} 
As $\phi\to\infty$, the string coupling \ggss\ goes to zero and
interactions turn off. Thus, $\phi=\infty$ can be thought of as 
the boundary of the near-horizon geometry of the defect. As 
$\phi$ decreases, the string coupling \ggss\ grows, and 
there are two basic possibilities. If the string coupling remains 
small everywhere in the near-horizon spacetime, we can study the system 
using perturbative string techniques. If, on the other hand, the string 
coupling becomes of order one or larger anywhere, the system may not be 
perturbative, and one may have to resort to other means of studying it. 
In this paper, we will restrict the discussion to the weakly coupled case. 

The full geometry around a defect corresponding to a $d$-dimensional
LST has the typical form (near the boundary at $\phi=\infty$) 
\eqn\typform{\IR^{d-1,1}\times\IR_\phi\times\MM}
where $\IR^{d-1,1}$ labels the worldvolume of the defect, $\IR_{\phi}$
is the real line labeled by $\phi$, and $\MM$ labels the angular 
directions at fixed distance from the defect. For example, for $k$ 
parallel $NS5$-branes in type II string theory one has \CallanAT\ $d=6$ 
and $\MM=SU(2)_k$, the supersymmetric level $k$ $SU(2)$ WZW model. At 
finite $\phi$, the geometry \typform\ must be deformed to avoid the strong 
coupling singularity at $\phi=-\infty$.  

The spectrum of normalizable states in weakly coupled asymptotically linear
dilaton spacetimes falls into two classes. One consists of delta-function
normalizable states whose vertex operators behave at large $\phi$ as 
$\exp(-{Q\over2}+i\lambda)\phi$ with real $\lambda$. These scattering
states correspond to incoming and outgoing waves carrying momentum $\lambda$ 
in the (radial) $\phi$ direction. They are quite analogous to standard 
scattering states in critical string theory. From the point of view of 
the $d$ dimensional theory on $\IR^{d-1,1}$, they have a continuous mass 
spectrum, which generally starts above some mass gap (which depends on the 
theory and on the operator).   

In addition to the delta-function normalizable scattering states, the
theory typically includes normalizable states which live at finite
$\phi$. The spectrum of such states is discrete; they can be thought 
of as bound states associated with the defect. They are described by 
normalizable vertex operators, with wavefunctions that decay rapidly as 
$\phi\to\infty$. For example, in the $d=6$, $\NN=(1,1)$ LST corresponding
to parallel $NS5$-branes in type IIB string theory that we will study
in detail below, these states include the massless gauge bosons living 
on the fivebranes, and their superpartners.

In critical string theory, the physical 
observables are vertex operators corresponding to on-shell
states. Their correlation functions give the S-matrix elements 
of these states. Such observables exist in the asymptotically 
linear dilaton (non-critical) case \typform\ as well. Indeed, 
one can use the (delta-function) normalizable vertex operators 
corresponding to both types of states described above to compute 
their S-matrix. 

A very interesting feature of linear dilaton backgrounds is the existense 
of additional observables, corresponding to off-shell operators in the 
$d$-dimensional theory of the defect. These observables correspond to 
non-normalizable vertex operators which go like $\exp(\beta\phi)$ with 
$\beta>-Q/2$ as $\phi\to\infty$. Like in anti-de-Sitter space, correlation 
functions of these operators correspond  to off-shell Green functions in 
LST. The main purpose of this paper is to elucidate the analytic structure 
of these correlation functions. 

We will show that the off-shell Green functions of 
weakly coupled asymptotically linear dilaton string theory satisfy an
analog of LSZ reduction, which is familiar from local quantum field 
theory\foot{Since LST is not a local Quantum Field Theory (QFT), 
the usual arguments for the LSZ reduction (see \eg\ \PeskinEV) do not 
apply in this case.}. They exhibit poles (which we will refer to as LSZ 
poles) at the locations of the normalizable states. The residues of these 
poles are on-shell correlation functions which involve the normalizable 
vertex operators creating these states from the vacuum; they can be used 
to study the interactions of these states.

Unlike local QFT, where all poles can be interpreted as due to 
particles going on-shell, it is well known that in asymptotically 
linear dilaton spacetimes there is another type of poles, associated 
with the contribution of the (semi-)infinite region $\phi\to\infty$ in
\typform. Scattering processes that can occur uniformly at any value
of $\phi$ are enhanced by the volume of the $\phi$ coordinate; 
this leads to poles
in correlation functions. As we will see below, these poles play an
important role in understanding the analytic structure of the LST 
correlation functions. 

The off-shell Green functions of LST also exhibit other singularities,
such as branch cuts associated with creating the continuum of scattering
states mentioned above, and poles corresponding to intermediate states
going on-shell. These singularities are rather standard, and we will not
study them in detail in this paper. 

Most of our discussion below will focus on a particular class of weakly
coupled asymptotically linear dilaton spacetimes, which have the form
(up to discrete identifications)
\eqn\rrpphh{ \IR^{d-1,1}\times {SL(2)_k\over U(1)} \times {\tilde \MM}.}
As before \typform, $\IR^{d-1,1}$ labels the worldvolume of the defect; 
$SL(2)_k\over U(1)$ is the well studied ($\NN=2$ supersymmetric)
cigar CFT; ${\tilde \MM}$ is a compact CFT, 
closely related to $\MM$ in \typform.
The linear dilaton direction labeled by $\phi$ \typform\ is in this case
the direction along the cigar, and the boundary at $\phi=\infty$ corresponds
to the asymptotic region far from the tip of the cigar. The linear dilaton 
slope $Q$ in \ggss\ is related to the level of $SL(2)$, $k$, via the relation
\eqn\kkqq{Q^2={2\over k}.}
In the $SL(2)/U(1)$ theory the asymptotic form of the dilaton is given 
by \ggss, but the dilaton does not grow indefinitely; the string coupling
reaches some maximal value $g_s^{(tip)}$ at the tip of the cigar.

One motivation for studying the case \rrpphh\ is that it arises
naturally in the physics of $NS5$-branes and singularities of Calabi-Yau
manifolds, where the process of going from \typform\ to \rrpphh\ corresponds
to smoothing the singularity or separating the fivebranes. In particular, the
case $d=6$, ${\tilde \MM}={SU(2)_k\over U(1)}$ 
that will be of interest to us below, 
is obtained by studying a system of parallel $NS5$-branes in type II string 
theory, spread out at equal distances on a circle in the transverse $\IR^4$.  

The background \rrpphh\ describes a particularly symmetric deformation of 
the singularity, and thus is more tractable than the generic 
case. While some of the techniques we use are specific to this background, 
we expect most of the results we obtain to be much more general.   

The dynamics of LST in spacetimes of the form \rrpphh\ was discussed by 
\refs{\GiveonZM,\GiveonPX,\GiveonTQ,\AharonyVK} and others. In particular, 
in \refs{\GiveonPX,\GiveonTQ} a study of two and three-point functions was
undertaken in the six dimensional LST corresponding to type IIB fivebranes 
distributed on a circle. 
The expectation was that at low energies the non-normalizable 
vertex operators in the background \rrpphh\ should reduce to local, 
gauge-invariant operators in the low energy field theory on the branes, which 
in this case is an $SU(k)$ gauge theory with sixteen supercharges ($\NN=(1,1)$ 
supersymmetry in six dimensions), at a particular point in its Coulomb
branch. The string theory correlation functions should reduce to off-shell 
Green functions of these operators.    

Surprisingly, it was found that this is not the case. While the string 
theory correlation functions do exhibit poles which agree with gauge 
theory expectations, they also exhibit some additional poles. In particular,
the string theory two-point functions have some unexpected poles at
$p_\mu^2=0$. On general grounds, one expects such poles to signal 
the creation of massless single particle states from the vacuum. 
However, in the case of the additional poles found in \refs{\GiveonPX,\GiveonTQ}
there were no candidate states in the $U(1)^{k-1}$ gauge theory with the right
quantum numbers. Also, the residues of some of the poles were negative,
which in a particle interpretation would signal non-unitarity. This 
puzzling behavior motivated the work described in this paper.  

Our general analysis of the correlation functions of non-normalizable vertex
operators in backgrounds of the form \rrpphh\ leads to the following 
explanation of the analytic structure of the amplitudes studied in 
\refs{\GiveonPX,\GiveonTQ}. Some of the poles of the two and three-point
functions are of the LSZ type, and correspond to processes where the
non-normalizable vertex operators act on the vacuum and create  normalizable
states on the cigar, that belong to the principal discrete series of $SL(2)$. 
The massless states of this type are in one to one correspondence with single 
particle states in the low energy field theory. As in the LSZ reduction in
quantum field theory, the residue of the LSZ poles in a correlation function
of off-shell observables may be interpreted as an S-matrix of single-particle
states, or (equivalently) as a correlation function of normalizable 
observables.

All the poles found in \refs{\GiveonPX,\GiveonTQ} that 
are not expected from the low energy gauge theory analysis 
are of the bulk type. This seems surprising since such poles 
are associated with the infinite region far from the tip of 
the cigar, and one might expect this region to contribute only 
above the energy scale at which one can create states belonging 
to the continuum of scattering states that live there. Since
this continuum starts (in the six dimensional example)
above a gap of order $M_s/\sqrt k$ (for $k$ fivebranes),
one might expect that it should not give rise to interesting effects at low
energies. Nevertheless, as we will see, bulk effects lead to poles at 
$p_\mu^2=0$.
This is a sort of UV/IR mixing (reminiscent of the mixing observed 
in non-commutative field theory \refs{\MinwallaPX,\VanRaamsdonkRR}), 
where a massless pole is due to the contribution of massive states.

It is important to emphasize that if we restrict attention to the
S-matrix of the massless single particle states, \ie\ to correlation 
functions of the normalizable observables corresponding to the relevant
principal discrete series states, we find a much more conventional 
picture. In particular, as far as is known, one can match the S-matrix 
elements computed from string theory in the background \rrpphh\ with a 
standard low energy effective Lagrangian written in terms of the light 
fields. The bulk singularities do not arise in these correlation functions, 
which behave much like scattering amplitudes in other string theory 
backgrounds. The new element in asymptotically linear dilaton backgrounds 
is that off-shell observables make sense (unlike, say, in flat spacetime 
with a constant dilaton), and it is these observables that exhibit the
new behavior. 

Let us elaborate this point.
LST provides us with a list of off-shell observables and their
correlation functions. Taking all the momenta uniformly to zero
in these correlation functions should normally lead one to a
{\it conformal field theory}. On general grounds, we know that
in any unitary six dimensional CFT, a scalar field $\OO(x_\mu)$
with scaling dimension two, which satisfies
\eqn\ttwwoo{\langle \OO(x)\OO(0) \rangle \simeq {c\over x^4}}
is free and decoupled. Associated with it via the
state-operator correspondence is a massless particle.

Thus, it is natural to argue that two-point functions
which exhibit bulk poles at $p_\mu^2=0$ should be interpreted as
signalling the presence of extra massless particles in the theory.
As mentioned above, there are at least two problems with this
conclusion. One is that there are in fact no normalizable massless
states with the right quantum numbers, either in the low energy
gauge theory, or in the cigar description.
The second is that some of the relevant two-point functions \ttwwoo\
have $c<0$, in seeming contradiction with unitarity.

One possible response to this conundrum is that the limit taken
to decouple the fivebranes from the bulk is inconsistent, and
one cannot consistently consider the model in which the linear
dilaton behavior \ggss\ persists up to arbitrarily large $\phi$. 
Indeed, if one introduces a cutoff in the form of an upper bound 
on $\phi$, the offending bulk poles are smoothed out and replaced
by large but finite contributions to the low energy Green functions.

We do not expect this resolution to be correct, for several
reasons. First, models of the asymptotic form \typform\ 
seem to be well defined without such a cutoff. Second, we 
provide below a consistent interpretation of the analytic
structure without appealing to the existence of a region
outside the linear dilaton one. Third, general considerations
of open-closed string duality to which we will return in \S11.4 
suggest that one might be able to reproduce the same
analytic structure by studying D-branes localized in the strong
coupling region (\eg\ near the tip of the cigar \rrpphh). 

Instead, what seems to be going on here is a  ``violation of 
folklore'' (to quote \GinspargIS). The UV and IR do not decouple,
and one cannot describe the extreme IR behavior of LST
correlation functions in terms of a standard six dimensional CFT.
Clearly, it would be interesting to understand this phenomenon
better.

The plan of this paper is as follows.  We begin in section 2 with a
detailed analysis of the analytic structure of correlation functions
in spacetimes of the form \rrpphh, in the bosonic string.  This is a
useful warm up exercise for the superstring, which exhibits all the
non-trivial elements that are important for our purposes. We show that
in order to study the analytic structure of correlation functions on
the cigar as a function of the external momenta, it is useful to write
the operators in the coset CFT in terms of the natural observables in
the underlying CFT on $AdS_3$, $\Phi_j(x,\bar x)$. Here, $(x,\bar x)$
label positions on the boundary of $AdS_3$, and in order to study
observables on $SL(2,\IR)/U(1)$ one has to perform a transform
from $(x,\bar x)$ to the conjugate variables $(m,\bar m)$. Correlation
functions in the coset theory are naturally given by integrals over
$x_i$ of the corresponding correlation functions in CFT on
$AdS_3$. This integral representation is very useful for studying the
analytic structure of the amplitudes. In particular, we show that LSZ
poles corresponding to principal discrete series states come from
$x_i\to0,\infty$, while bulk poles are associated with regions in the
integrals over $x_i$ where some or all of them approach each other.

In section 3 we briefly comment on the extension of the general
results of section 2 to the superstring. Most of the analysis goes
through, with only minor changes.  In section 4 we discuss the six
dimensional LST corresponding to $k$ $NS5$-branes spread out on a
circle in type IIB string theory. We find that the spectrum of
massless principal discrete series states on the cigar is in one to
one correspondence with the spectrum of single particle states in the
low energy gauge theory.

In sections 5 and 6, we apply our results to the
analysis of several correlation functions in the 
${\cal N}=(1,1)$ supersymmetric $d=6$ LST mentioned 
above, and present the details of the picture
described earlier in this introduction. In particular,
we show that all the massless poles found in LST that 
do not correspond to states in the low energy gauge theory
are bulk poles. Our improved understanding of the relation 
between the string theory and the low-energy 
gauge theory also allows us to find the precise mapping 
between string theory operators and low-energy gauge theory 
operators in this example, including the precise combination 
of single-trace and multi-trace operators that corresponds to
single string vertex operators in the relevant geometry \rrpphh. 
As we vary the angular momentum of the operators in the compact 
space, we find that they change from 
single-trace operators, for very small angular momentum, to 
sub-determinant type operators when the angular momentum is close 
to its maximal value. This is in agreement with previous discussions 
of ``giant gravitons'' in the context of the AdS/CFT correspondence. 

In sections 7-9  we comment on some additional applications of our 
results. Section 7 contains a brief discussion of four dimensional 
LSTs corresponding to isolated singularities of Calabi-Yau manifolds 
or equivalently wrapped fivebranes. In section 8 we discuss two 
dimensional bosonic string theory from the point of view of our 
analysis. In section 9 we discuss the thermodynamics of LSTs, and 
we argue that the Hagedorn temperature is a maximal temperature for 
LSTs. Section 10 contains a further discussion of our results, and 
section 11 mentions some future directions. Two appendices contain 
some useful technical results.

\newsec{Bosonic string theory on $SL(2,\IR)/U(1)$}

In this section we study the analytic structure of correlation
functions in string theory on spacetimes of the form
\rrpphh. As in many other instances in string theory, it is
convenient to first consider the technically simpler 
bosonic case, and then generalize the discussion to
the superstring (which we will do in the next sections).

\subsec{Generalities}

To make the transition to the superstring smoother we will
denote the level of $SL(2,\IR)$, which the model possesses
before modding out by $U(1)$, by $k+2$. The central charge
of the $SL(2,\IR)/U(1)$ coset CFT is then given by
\eqn\cbossl{c_{sl}={3(k+2)\over k}-1=2+{6\over k}.}
With this definition, the slope of the linear dilaton $Q$ in
\ggss\ is given by \kkqq. The condition that \rrpphh\ is
a consistent background for the bosonic string takes the form
\eqn\critbos{d+c_{sl}+c_{\tilde\MM}=26.}
In studying perturbative string theory on the space \rrpphh,
one is interested in computing correlation functions of physical
vertex operators on this space. A large class of such operators, which
will be sufficient for our purposes, can be constructed as follows.

Let $\WW$ be a conformal primary on $\tilde\MM$, with
scaling dimension $(\Delta_L,\Delta_R)$.  We can form a
physical vertex operator by ``dressing'' $\WW$ by an
$SL(2,\IR)/U(1)\times \IR^{d-1,1}$ vertex operator,\foot{See 
appendix A for a description of vertex operators in $SL(2,\IR)$ 
and $SL(2,\IR)/U(1)$.} $V_{j;m,\bar m}e^{ip\cdot x}$,  to
construct a physical observable\foot{This is not the most general
fundamental string excitation on \rrpphh\ -- we have suppressed the
towers of transverse oscillators associated with
$\IR^{d-1,1}\times SL(2,\IR)/U(1)$.}
\eqn\physw{\CO_\WW(p)=\WW V_{j;m,\bar m}e^{ip\cdot x}.}
The $SL(2,\IR)/U(1)$ part of $\CO_\WW$, $V_{j;m,\bar m}$,
corresponds to a fundamental string carrying momentum and
winding around the cigar. Its worldsheet scaling dimension is
given by
\eqn\dimvjm{\eqalign{
\Delta_{j;m}=&-{j(j+1)\over k}+{m^2\over k+2},\cr
\bar\Delta_{j;\bar m}=&-{j(j+1)\over k}+{\bar m^2\over k+2}.\cr
}}
At large $\phi$, the Euclidean cigar $SL(2,\IR)/U(1)$ 
looks like a semi-infinite cylinder, and the operators 
$V_{j;m,{\bar m}}$ have well-defined momentum $n$ and 
winding number $w$ ($n,w\in \Z$) around the cylinder, 
given by
\eqn\momwin{\eqalign{
m=&\half[w(k+2)+n],\cr
\bar m=&\half[w(k+2)-n].\cr
}}
The physical state condition reads
\eqn\masssh{\half p_\mu^2+\Delta_{j;m}+\Delta_L=
\half p_\mu^2+\bar\Delta_{j;\bar m}+\Delta_R=1~,}
where $\alpha'=2$ and the signature convention is 
$p_{\mu}^2 =p_i^2-p_0^2 $. For given $\WW$, and 
fixed momentum and winding $(n,w)$ (or fixed 
$(m,\bar m)$ \momwin), the physical state condition 
can be used to determine $j$ as a function of $p_\mu^2$. 
For $p_\mu^2$ above a certain critical value (which 
depends on $\WW$, $m$, $\bar m$), $j$ is real and the 
operator $V_{j;m,\bar m}$ is non-normalizable. On the 
other hand, for $p_\mu^2$ smaller than that value, the 
solution of \masssh\ is $j=-\half+i\lambda$, with real 
$\lambda$. The resulting wave function is delta-function 
normalizable.

The dimension formula \dimvjm\ is invariant under $j\to -j-1$. In fact,
the operators $V_{j;m,\bar m}$ and $V_{-j-1;m,\bar m}$ are related
via the reflection property \refs{\TeschnerFT,\TeschnerUG}
\eqn\reflrel{\eqalign{
&V_{j;m,\bar m}=R(j,m,\bar m;k) V_{-j-1;m,\bar m},\cr
&R(j,m,\bar m;k)=\nu(k)^{2j+1}{\Gamma(1-{2j+1\over k})
\Gamma(j+m+1)\Gamma(j-\bar m+1)\Gamma(-2j-1)\over
\Gamma(1+{2j+1\over k})\Gamma(m-j)\Gamma(-j-\bar m)\Gamma(2j+1)}
,\cr}}
where\foot{As explained in \GiveonUP, $\nu(k)$ is actually
a free parameter, which depends on the couplings 
$\lambda$, $\mu$ introduced in \S2.4. The value of $\nu(k)$
given here corresponds to the choice made in 
\refs{\TeschnerFT,\TeschnerUG}.}
\eqn\nukk{\nu(k)={1\over\pi}{\Gamma(1+{1\over k})\over
\Gamma(1-{1\over k})}.}
For real $j$, one can use \reflrel\ to restrict to $j>-1/2$, and we will
usually do so below.

The  relation \reflrel\ degenerates when the reflection coefficient
$R(j,m,\bar m;k)$ develops singularities. Poles of $R$ have two
sources that will be discussed below. One (associated with the first
and last $\Gamma$ functions in the numerator of the expression for
$R$) has to do with operator mixing due to bulk interactions of
$V_{j;m,\bar m}$ with the background. The other (associated with
the factor $\Gamma(j+m+1)\Gamma(j-\bar m+1)$ in \reflrel) signals 
the presence of normalizable states in the theory. Zeroes of $R$, 
associated with the second and third $\Gamma$ functions in the 
denominator of \reflrel, are related to the existence of degenerate 
operators in the theory (see \eg\ \S4 in \GiveonUP).

Correlation functions of the observables \physw\ are given by $n$-point
functions
\eqn\npointf{\langle\CO_{\WW_1}(p_1)\CO_{\WW_2}(p_2)
\cdots \CO_{\WW_n}(p_n)\rangle,}
where $(n-3)$ vertex operators are integrated over the
worldsheet\foot{Although many of our considerations
below are more general, we will mainly discuss in this
paper tree-level string theory, \ie\ a worldsheet with
spherical topology.}, and three are placed, as usual, at
arbitrary points, say $0,1,\infty$. When all the vertex operators
in \npointf\ obey $j > -1/2$, these amplitudes
correspond in spacetime to off-shell Green functions in
a $d$-dimensional LST. Our main focus here is on their analytic
structure.

There are two sources of dependence on the momenta $(p_1,\cdots, p_n)$
in the correlation function \npointf. One is the standard dependence on the
Mandelstam invariants $p_i\cdot p_j$, which comes from the correlation
function of  the $\exp(ip_i\cdot x)$ factors in $\CO_{\WW_i}$ \physw.
This contribution leads to singularities of the integrated correlation function,
which occur when intermediate states go on mass-shell, and are well understood.

The second source of momentum dependence in \npointf\ comes from
the unintegrated $SL(2,\IR)/U(1)$ correlation function,
\eqn\sltwocon{
\langle V_{j_1;m_1,\bar m_1}\cdots V_{j_n;m_n,\bar m_n}\rangle~.
}
This $n$-point function has singularities as a function of the $j_i$, which
correspond via \masssh\ to singularities in momentum space. Our main
purpose in this paper is to better understand these singularities, and in
particular their spacetime interpretation.

Of special interest to us will be external leg poles, which occur
as a single $j$ (or $p_\mu^2$ \masssh) is tuned to a particular value.
Below, we will exhibit the origin of these poles, and show that
they are always {\it single} poles, which are associated with
asymptotic single-particle states.

These poles can be used to implement an analog of the LSZ reduction
in LST; thus we will refer to them below as LSZ poles.
As one approaches an LSZ pole, we will see that one has
\eqn\resnorm{\CO_\WW(p)\sim {1\over p^2+M^2}\CO^{\rm (norm)}_\WW(p),}
where $\CO^{\rm (norm)}_\WW$ is a normalizable vertex operator, which
creates from the vacuum a particle with mass $M$. In the bosonic string, $M^2$
can be negative. In spacetime supersymmetric theories it will always be non-negative.

Near a pole \resnorm\ as a function of one of the momenta in \npointf,
say $p_n$, the $n$-point function has the form
\eqn\nptfact{
\langle0|\CO_{\WW_1}(p_1)\CO_{\WW_2}(p_2)\cdots \CO_{\WW_n}(p_n)|0\rangle
\simeq {1\over p_n^2+M_n^2}
\langle 0|\CO_{\WW_1}(p_1)\cdots \CO_{\WW_{n-1}}(p_{n-1})|\WW_n,p_n\rangle}
where $|\WW_n,p_n\rangle=\CO^{\rm (norm)}_{\WW_n}(p_n)|0\rangle$ is the state
created by the normalizable operator \resnorm\ acting on the vacuum.

As we approach poles of the form \resnorm\ in all the external momenta $p_i$ in
\npointf, the amplitude behaves as
\eqn\smatr{\langle\CO_{\WW_1}(p_1)\cdots \CO_{\WW_n}(p_n)\rangle
\sim \left(\prod_i{1\over p_i^2+M_i^2}\right)
\langle0|\CO^{\rm (norm)}_{\WW_1}(p_1)\cdots
\CO^{\rm (norm)}_{\WW_n}(p_n)|0\rangle.}
As in QFT, the residue of the poles -- the $n$-point function of the 
normalizable operators
$\CO^{\rm (norm)}_{\WW_i}(p_i)$ -- is proportional to the S-matrix of the
particles with masses $M_i$ created by these operators from the vacuum\foot{To
obtain the S-matrix precisely one needs to suitably normalize the states
$|\WW_i,p_i\rangle$. Note that apriori one might have thought that the
poles \smatr\ themselves were just an artifact of choosing a wrong
normalization for the vertex operators $\CO_{\WW}(p)$, but our discussion
will make it clear that this is not the case.}.

In the next two subsections we will exhibit the origin of the poles
\resnorm\ in LST backgrounds of the form \rrpphh. Later, we will
also discuss other singularities of the amplitudes \sltwocon\ that have
a different origin and play an important role in some applications.

\subsec{LSZ in bosonic LST: (I) a semiclassical analysis}

In order to exhibit the poles \resnorm\ we have to understand better
the structure of the $SL(2,\IR)/U(1)$ operators $V_{j;m,\bar m}$ which
 give rise to them.  One way to construct these operators is to start
 with the corresponding operators $\Phi_{j;m,\bar m}$ in CFT on the
 $SL(2,\IR)$ group manifold, and remove from them the $U(1)$ part. In
 this subsection we use a semiclassical analysis to do this. In the
 next subsection we generalize it to the full quantum worldsheet
 theory.

The $SL(2,\IR)$ group manifold can be thought of as Minkowski
$AdS_3$. For our purposes, it is convenient to analytically continue
to Euclidean $AdS_3$, which we can parametrize by the Poincar\'e
coordinates $(\phi,\gamma,\bar\gamma)$, with the metric
\eqn\adsmet{ds^2=d\phi^2+e^{Q\phi}d\gamma d\bar\gamma~.}
A natural set of observables in CFT on Euclidean  $AdS_3$ is given by the
eigenfunctions of the Laplacian \refs{\TeschnerFT,\TeschnerUG}
\eqn\phij{\Phi_j(x,\bar x)={1\over\pi}\left(|\gamma-x|^2e^{Q\phi\over2}+
e^{-{Q\phi\over2}}\right)^{-2(j+1)},}
with $j > -{1\over 2}$.
The parameters $(x,\bar x)$, on which these observables depend,
label positions on the boundary of $AdS_3$. The operators \phij\
transform as primaries of dimension $h={\bar h}=j+1$ under
conformal transformations of this boundary. The coordinates $(x,\bar x)$
should not be confused with the string worldsheet coordinates $(z,\bar z)$,
which are mostly suppressed in this paper.

For the application to the coset theory it will be useful to expand $\Phi_j$
around $\phi=\infty$ \KutasovXU,
\eqn\largephi{\Phi_j(x,\bar x)\simeq {1\over 2j+1} e^{Qj\phi}\delta^2(\gamma-x)
+O(e^{Q(j-1)\phi})+{e^{-Q(j+1)\phi}\over\pi|\gamma-x|^{4(j+1)}}+
O(e^{-Q(j+2)\phi}).}
For generic $j$, the large $\phi$ expansion naturally splits into two
independent series. One includes the leading term,
$e^{Qj\phi}\delta^2(\gamma-x)$
and an infinite series of corrections of the form
$ e^{Q(j-n)\phi}\partial_x^n\bar\partial_{\bar x}^n\delta^2(\gamma-x)$.
This series is relevant only near $\gamma=x$.
The second series starts with the dominant term for generic $\gamma$,
$e^{-Q(j+1)\phi}|\gamma-x|^{-4(j+1)}$ and includes corrections
that are down by powers of $e^{-Q\phi}/|\gamma-x|^2$. This
series consists purely of terms which decay exponentially as
$\phi\to\infty$, and are thus normalizable there.

The $AdS_3$ ancestor of the $SL(2,\IR)/U(1)$ observable $V_{j;m,\bar m}$
discussed above  is the operator
\eqn\furtrans{\Phi_{j;m,\bar m}\equiv \int d^2x x^{j+m}\bar x^{j+\bar m}
\Phi_j(x,\bar x)~.}
Plugging the large $\phi$ expansion \largephi\ into \furtrans\ we find:
\eqn\largejmm{\eqalign{
\Phi_{j;m,\bar m}&=
{1\over 2j+1} e^{Qj\phi}\gamma^{j+m}\bar \gamma^{j+\bar m}
+O(e^{Q(j-1)\phi)})+\cr
&{\Gamma(j+m+1)\Gamma(j-\bar m+1)\Gamma(-2j-1)\over
\Gamma(m-j)\Gamma(-j-\bar m)\Gamma(2j+2)}
e^{-Q(j+1)\phi}\gamma^{m-j-1}\bar\gamma^{\bar m-j-1}+
O(e^{-Q(j+2)\phi}).\cr}}
In computing the second line of \largejmm\ we have
rescaled $x$ by $\gamma$ and used the result
\eqn\useint{\int d^2x x^{j+m}\bar x^{j+\bar m}|1-x|^{-4(j+1)}=
\pi{\Gamma(j+m+1)\Gamma(j-\bar m+1)\Gamma(-2j-1)\over
\Gamma(m-j)\Gamma(-j-\bar m)\Gamma(2j+2)}.}

\noindent
Using standard techniques (see \eg\ \BershadskyIN) one can write
the vertex operator \largejmm\ as a product of a plane wave living
in the $U(1)$ CFT and an operator in the coset, whose asymptotic form
far from the tip of the cigar is
\eqn\largecos{
V_{j;m,\bar m}={e^{i\sqrt{{2\over {k+2}}}(mY-\bar m\bar Y)}\over 2j+1}
\left[e^{Qj\phi}+
{\Gamma(j+m+1)\Gamma(j-\bar m+1)\Gamma(-2j-1)\over
\Gamma(m-j)\Gamma(-j-\bar m)\Gamma(2j+1)}
e^{-Q(j+1)\phi}+\cdots\right].}
Here $Y$ is the angular coordinate around the cigar.
It lives on a circle of radius $\sqrt{2(k+2)}$.

We see that \largecos\ has the structure anticipated 
in \resnorm. The leading non-normalizable\foot{Recall 
that $j>-1/2$.} contribution to the vertex operator is 
finite, while the leading normalizable term has poles 
at various values of $j$. Note that the relative coefficient
of the $e^{-Q(j+1)\phi}$ and $e^{Qj\phi}$ terms in \largecos\
coincides with the large $k$ limit of the 
reflection coefficient $R$ in \reflrel\ (the power of 
$\nu(k)$ in $R$ is unimportant since we can generate it
by shifting $\phi\to\phi+\phi_0$ in \largecos). At finite $k$,
the ratio of $\Gamma$ functions in \largecos\ should indeed
be replaced by the exact reflection coefficient \reflrel.

An important point is that not all of the poles in 
\largecos\ give rise to normalizable states as indicated 
in \resnorm. To see which ones do and which ones do not, 
it is convenient to go back to the integral representation 
\useint\ that gave rise to the second term in \largecos. 
The poles associated with the three $\Gamma$ functions in 
the numerator come from the vicinity of $x=0$, $x=\infty$ 
and $x=1$, respectively. Let us consider these singularities
in turn, starting with the region $x\to 0$. The contribution 
of this region to the integral \furtrans\ is given by
\eqn\intdi{\int_{|x|<\epsilon} d^2 x x^{j+m}\bar x^{j+\bar m}
\Phi_j(x,\bar x)}
where $\epsilon\ll 1$ is a cutoff, whose role is to isolate the singularity
arising from $x=0$. We will see shortly that the integral \intdi\ has poles at
\eqn\poles{j=M-1,M-2,\cdots>-{1\over 2}~,
\quad M=\min\{|m|,|\bar m|\}~, \quad m,\bar m<-{1\over 2}~.}
Note that this is the same as the set of poles of the second term in
\largecos\ associated with the first $\Gamma$ function in the
numerator, as advocated above. We will see that the residues of
the poles agree with \largejmm, \largecos\ as well.

To prove \poles, we assume (without loss of generality) that
the momentum on the cigar
\eqn\mmn{n \equiv m-\bar m}
is a non-negative integer. If $n<0$, one can interchange
the roles of $(x,m)\leftrightarrow (\bar x, \bar m)$ and repeat
the analysis below.

We may rewrite \intdi\ as
\eqn\intdiv{\int_{|x|<\epsilon}
d^2 x x^{j+m}\bar x^{j+\bar m}\Phi_j(x,\bar x)
=\int_{|x|<\epsilon}
d^2 x |x|^{2(j+m)}\bar x^{-n}\Phi_j(x,\bar x)~.}
The leading singularity of \intdiv\ is obtained as follows.
Using the fact that $\Phi_j(x,\bar x)$ \phij\ is analytic at $x=0$
and expanding it in powers of $x$, $\bar x$, we see that the
first term that survives the integral over the phase of $x$
is $\Phi_j(x,\bar x)=\cdots+{1\over n!}\bar\partial^n\Phi_j(0)
\bar x^n+\cdots$. Its contribution to \intdiv\ is
\eqn\expint{{1\over n!}\int_{|x|<\epsilon}d^2 x
|x|^{2(j+m)}\bar\partial^n \Phi_j(0)~.}
This integral diverges when $j+m\to -1$,
where it has a simple pole:
\eqn\leaddiv{\int_{|x| < \epsilon} d^2 x |x|^{2(j+m)}
\bar\partial^n\Phi_j(0)\simeq
{\pi\bar\partial^n\Phi_j(0)\over j+m+1}~.}
For $j+m<-1$ the integral \intdiv\ is also divergent and should
be treated via analytic continuation, as is familiar from studies of
Shapiro-Virasoro amplitudes in string theory.
To study the subleading singularities,
one needs to expand $\Phi_{j}(x,\bar x)$
to $l$'th order in $x$, and $(n+l)$'th order in $\bar x$,
and consider the resulting integral:
\eqn\subdiv{\int_{|x|<\epsilon} d^2 x |x|^{2(j+m+l)}
\partial^l\bar\partial^{n+l}\Phi_j(0)\simeq
{\pi\partial^l\bar\partial^{n+l}\Phi_j(0)\over j+m+l+1}~.}

We conclude that for $m\geq \bar m$, the singularities of
\intdiv\ occur at
\eqn\singxzero{
j+m+l+1=0;\;\;\; l=0,1,2,\cdots~.}
Since $j>-{1\over 2}$, we see that \singxzero\ has
solutions iff $m<-{1\over 2}$, and since $m\geq \bar m$
this also implies $\bar m<-{1\over 2}$. This concludes the
proof of \poles.

We see that as $j$ approaches one of the values
\poles, the vertex operator $V_{j;m,\bar m}$ 
develops a pole associated with the $\Gamma(j+m+1)$ 
factor in \largecos. The residue of this pole is a 
normalizable operator, as indicated in \resnorm. 
This operator creates from the vacuum a physical 
state in the spacetime theory. It can be obtained 
by starting with the operator $\partial^l\bar\partial^{n+l}
\Phi_j(0)$ in CFT on $AdS_3$, and removing the $U(1)$ part. 

The region $x\to\infty$ can be studied in a
very similar way. In this limit, the
vertex operator \phij\ behaves as
\eqn\phijinf{\Phi_j(x,\bar x)\propto |x|^{-4(j+1)}.}
Thus, the leading singularities from this region
are due to the behavior of the integral
\eqn\intinf{\int_{|x|>1/\epsilon}
d^2 x x^{j+m}\bar x^{j+\bar m} |x|^{-4(j+1)},}
while subleading singularities receive contributions 
from subleading terms in the large $x$ expansion of 
$\Phi_j$. 
A similar analysis to the previous case leads 
to poles at
\eqn\polesinf{j=M-1,M-2,\cdots>-{1\over 2}~,
\quad M=\min\{m,\bar m\}~, \quad m,\bar m>{1\over 2}~.}
This is the same set of poles as that associated with the
$\Gamma(j-\bar m+1)$ factor on the right-hand side of \largecos, in agreement
with our comments above. One can check that the residues agree
as well.

Note that the physical states \poles, \polesinf\ appear only when
the signs of $m$ and $\bar m$ are the same. Looking back at equation
\momwin\ we see that this means that these states are ``winding
dominated.'' As we will discuss later (in \S10.1), this is very
natural from the
spacetime perspective. Winding modes feel an attractive potential
towards the tip, while momentum modes feel a repulsive one
\DijkgraafBA. Thus, fundamental string modes that are dominated
by winding can bind to the tip of the cigar. The normalizable states
corresponding to the poles \poles, \polesinf\ are precisely such bound states.

So far we have discussed the first two $\Gamma$ functions
in the numerator of the second term in \largecos, attributed
them to the contributions from $x=0,\infty$ in the integral
\useint, and interpreted them as corresponding to normalizable
physical states created from the vacuum by the operators
$V_{j;m,\bar m}$, as in the discussion following equation \resnorm.

The third $\Gamma$ function in \largecos, $\Gamma(-2j-1)$,
leads to divergences at $2j+1=0,1,2,\cdots$. These can be seen
to arise from the region $x\to 1$ in the integral \useint, or before
rescaling $x$ by $\gamma$, from $x\to \gamma$. Interestingly,
these poles do not signal the appearance of additional physical
states. Instead, they signal the breakdown of the large $\phi$
expansion \largephi. Indeed, the expansion parameter for
$x\not=\gamma$ in \largephi, $e^{-Q\phi}/|\gamma-x|^2$,
becomes large when $x\simeq \gamma$. In order to evaluate the
contribution of this region to the integral \furtrans\ we have 
to sum the full expansion indicated in \largephi. A related 
complication is that the two series in the large $\phi$ expansion 
\largephi, both of which are important for $x\simeq\gamma$, mix 
when $j$ takes half-integer values. We will see later (in \S2.4)
that the behavior of the vertex operator \largecos\ near the poles 
at $2j+1\in \Z_+$ is in general dominated by non-trivial interactions 
that occur at large positive $\phi$. 

At any rate, to see the actual behavior as $x\to \gamma$, 
we substitute \phij\ in \furtrans\ and evaluate
\eqn\xeqgamma{{1\over\pi}\int_{|x-\gamma|<\epsilon} d^2x x^{j+m}
\bar x^{j+\bar m}\left(|\gamma-x|^2e^{Q\phi\over2}+
e^{-{Q\phi\over2}}\right)^{-2(j+1)}.}
This integral is actually finite for positive integer $2j+1$. 
We conclude that the contribution of the region $x\simeq \gamma$ 
to the integral \furtrans\ is finite. The apparent 
singularities of $\Gamma(-2j-1)$ in \largecos\ do not in 
fact lead to a divergence of the full vertex operator, and 
do not correspond to physical states via \resnorm.

To summarize, we see that at least semiclassically, the observables
\physw\ indeed exhibit the behavior \resnorm\ near poles determined
by the physical state conditions \dimvjm\ -- \masssh\
and \poles, \polesinf. The normalizable vertex operators
$\CO_\WW^{({\rm norm})}$ behave as $\phi\to\infty$ (far from
the tip of the cigar) like $\exp(-Q(j+1)\phi)$, with $j$ given by the
appropriate value \poles, \polesinf. A fact that will be useful below
is that in the representation of these operators in terms of observables
in $AdS_3$ integrated over the boundary variables $(x,\bar x)$, these
poles come from the regions $x\to 0,\infty$.

\subsec{LSZ in bosonic LST: (II) exact results}

In the previous subsection we studied the behavior of the vertex
operators $V_{j;m,\bar m}$ semiclassically. In particular, we treated
the $AdS_3$ operators $\Phi_j$ \phij\ as functions and discussed their
properties. This provides a good approximation in the limit $k\to\infty$
($k$ is related to the level of $SL(2,\IR)$, see equation \cbossl), in which the
$\sigma$-model on the cigar becomes weakly coupled and fluctuations are small.
At finite $k$, the operators $\Phi_j$ and $V_{j;m,\bar m}$ should be thought
of as fluctuating quantum operators, and cannot be treated as functions.
Nevertheless, as we will see in this subsection, the results of the previous
subsection can be extended to the quantum theory, 
with only minor modifications.

In the full quantum worldsheet theory, the $AdS_3$ observables
\phij, \furtrans\ correspond to local operators with worldsheet
scaling dimension
\eqn\dimphijmm{\Delta_j=\bar\Delta_j=-{j(j+1)\over k}~.}
Removing from them the $U(1)$ part, one finds the observables on
the cigar, $V_{j;m,\bar m}$, whose scaling dimensions are given by \dimvjm.

The data that we have access to in the full CFT on the cigar is the
set of worldsheet correlation functions \sltwocon. The symmetries of 
the problem imply that momentum on the cigar is conserved, \ie\ (see 
\momwin) $\sum_i n_i=\sum_i(m_i-\bar m_i)=0$. Winding, on the other 
hand, need not be conserved, since strings wound around the cigar can 
unwind at the tip. While winding violating amplitudes are not much more 
difficult to study, in this paper we will (for simplicity) focus on 
amplitudes that conserve winding. We will comment on the more general 
case in \S2.6 below.

Imposing both momentum and winding conservation leads to the constraint
\eqn\windcons{\sum_i m_i = \sum_i {\bar m}_i = 0.}
In this case, we can compute the amplitude \sltwocon\ directly
in the $SL(2,\IR)$ (or $AdS_3$) CFT, since
\eqn\liftads{
\langle V_{j_1;m_1,\bar m_1}\cdots V_{j_n;m_n,\bar m_n}\rangle
=C_{1\cdots n}\langle \Phi_{j_1;m_1,\bar m_1}\cdots
\Phi_{j_n;m_n,\bar m_n}\rangle}
where $C_{1\cdots n}$ is a known  function of the moduli (including the
positions of the operators on the worldsheet), coming from a correlation
function in a $U(1)$ CFT. For the purpose of our
discussion here, it gives an uninteresting overall constant.

Using \furtrans\ we see that the $SL(2,\IR)/U(1)$ correlator \liftads\ can be
written as
\eqn\densga{\langle V_{j_1;m_1,\bar m_1}\cdots V_{j_n;m_n,\bar m_n} \rangle=
C_{1\cdots n}\prod_{i=1}^n\int d^2 x_i x_i^{j_i+m_i}\bar x_i^{j_i+\bar m_i}
\vev{\Phi_{j_1}(x_1,\bar x_1)\cdots\Phi_{j_n}(x_n,\bar x_n)}.}
The semiclassical analysis of the previous subsection 
leads us to expect that the contributions of the regions
$x_i\to0,\infty$ in \densga\ give rise to poles at values 
of $j_i$ corresponding to \poles, \polesinf, respectively.

Consider, for example, the region $x_1\to 0$ in the $x_1$
integral in \densga. The correlator
$\vev{\Phi_{j_1}(x_1,\bar x_1)\cdots\Phi_{j_n}(x_n,\bar x_n)}$
in $SL(2,\IR)$ is analytic\foot{In our discussion here (and below) we
implicitly assume that the worldsheet locations $z_i$ of the operators
are generic.} near $x_1=0$ with fixed generic
$(x_2,\cdots, x_n)$. This can be seen by thinking about it
as an expectation value in an AdS/CFT-dual conformal field
theory in $x$ space --  singularities are expected to occur 
only when some of the $x_i$ coincide, and there is nothing special
about the point $x_1=0$. Therefore, the analysis 
following equation \intdi\ is directly applicable, and 
leads to the same conclusions as there: the region near 
$x_1=0$ in the integral \densga\ gives rise to singularities 
when $(j_1,m_1)$ are related as in \poles. For example, for 
$m_1=\bar m_1<-1/2$, the leading singularity is at $j_1=-m_1-1$, 
near which \densga\ behaves as (compare to \nptfact):
\eqn\polezero{\eqalign{
\langle V_{j_1;m_1,\bar m_1}\cdots &V_{j_n;m_n,\bar m_n} \rangle
\simeq {\pi C_{1\cdots n}\over j_1+m_1+1}\times\cr
&\prod_{i=2}^n\int d^2 x_i x_i^{j_i+m_i}\bar x_i^{j_i+\bar m_i}
\vev{\Phi_{j_1}(0)\Phi_{j_2}(x_2,\bar x_2)\cdots\Phi_{j_n}(x_n,\bar x_n)}.\cr
}}
Note that (as discussed above) 
the operator $\Phi_{j_1}(0)$ in \polezero\ is a normalizable
vertex operator which thus corresponds to an asymptotic state,
$\langle0| \Phi_{j_1}(0)$. This state exists already in $AdS_3$,
and focusing on the residue of the pole \polezero\ picks out its
contribution to the correlation function \densga.
Subleading singularities can be analyzed as in \subdiv; 
they correspond to the asymptotic states $\langle 0|
\partial^l\bar\partial^{\bar l}\Phi_j(0)$.

At the same time that $x_1\to 0$, we can also send another
of the $x_i$, say $x_n$, to infinity. The $SL(2,\IR)$
Ward identities imply that as $x_n\to\infty$, the correlation
functions of the $\Phi_{j_i}$ in \densga\ behave as\foot{On the 
right hand side of this formula, and below, we define
$\Phi_{j}(\infty)\equiv
\lim_{x\to\infty}|x|^{4(j+1)}\Phi_{j}(x)$; it is easy
to see that this limit is well-defined in any correlation function by using
the conformal transformation $x\to 1/x$.}
\eqn\inftbeh{\langle\Phi_{j_1}(x_1,\bar x_1)\cdots\Phi_{j_n}(x_n,\bar x_n)
\rangle\propto |x_n|^{-4(j_n+1)}\langle\Phi_{j_1}(x_1,\bar x_1)\cdots
\Phi_{j_n}(\infty)\rangle~.}
This can again be compared to the semiclassical result \phijinf,
and proceeding as there one concludes that the region
$x_n\to\infty$ gives rise to potential poles at \polesinf.
For example, for $m_n=\bar m_n>1/2$, the leading singularity is at
$j_n=m_n-1$.

The joint contribution of $x_1\to 0$ and $x_n\to\infty$
behaves near the leading LSZ poles as
\eqn\polezinf{\eqalign{
\langle V_{j_1;m_1,\bar m_1}&\cdots V_{j_n;m_n,\bar m_n} \rangle
\simeq {\pi^2 C_{1\cdots n}\over (m_1+j_1+1)(m_n-j_n-1)}\times\cr
&\prod_{i=2}^{n-1}\int d^2 x_i x_i^{j_i+m_i}\bar x_i^{j_i+\bar m_i}
\vev{0|\Phi_{j_1}(0)\Phi_{j_2}(x_2,\bar x_2)
\cdots\Phi_{j_{n-1}}(x_{n-1},\bar x_{n-1})
\Phi_{j_n}(\infty)|0}.\cr}}
The residue of the poles in \polezinf\ can be interpreted
as the expectation value of $n-2$ off-shell operators
between two normalizable states,  $\langle 0|\Phi_{j_1}(0)$
and $\Phi_{j_n}(\infty)|0\rangle$. 

Suppose we want to take more of the external legs
on-shell, \eg\ to compute the S-matrix, as in \smatr.
Naively, there is a problem, since after sending one
of the $x_i$ in \densga\ to zero, and one to infinity,
if we try to send additional $x_i$ to zero or
infinity, they will in particular approach $x_1$ or
$x_n$, which will spoil the above analysis. For
example, it is no longer true that the correlator
$\vev{\Phi_{j_1}(0)\Phi_{j_2}(x_2,\bar x_2)
\cdots\Phi_{j_{n-1}}(x_{n-1},\bar x_{n-1})
\Phi_{j_n}(\infty)}$ is regular as $x_2\to 0$,
due to the short distance singularities of
$\Phi_{j_2}(x_2,\bar x_2)$ and $\Phi_{j_1}(0)$.

The problem is evidently due to interactions between the
two asymptotic states $\langle 0|\Phi_{j_1}(0)$ and
$\langle 0|\Phi_{j_2}(0)$. In order to study this issue
we need to understand the behavior of
\eqn\twooneope{\langle 0|\Phi_{j_1}(0)\Phi_{j_2}(x_2,\bar x_2)}
in the limit $x_2\to 0$. This OPE\foot{Note that here we are
discussing the OPE on the boundary of $AdS_3$ parametrized
by the coordinates $(x,\bar x)$ and not on the string worldsheet,
which is labeled by $(z,\bar z)$.} was analyzed in
\refs{\TeschnerUG,\MaldacenaKM}, where it was
found that it receives two types of contributions. One is given
by an integral over the delta function normalizable states
corresponding to $\Phi_j$ with $j=-\half+i\lambda$. This
contribution has to do with interactions between the states
associated with $\Phi_{j_1}(0)$ and  $\Phi_{j_2}(0)$. It gives
a different kinematic structure from what we are looking for
here; we will return to contributions of this type in the next subsection.

In addition to the integral over continuous series states, the
OPE \twooneope\ contains a discrete
sum over states with real $j$ \refs{\TeschnerUG,\MaldacenaKM}.
This contribution is given by a power series in $x_2$, $\bar x_2$.
The leading term goes like $|x_2|^0$; its coefficient can be thought
of as the two particle state $\langle 0|\Phi_{j_1}\Phi_{j_2}(0)$
(where the operator $\Phi_{j_1}\Phi_{j_2}(0)$ is defined as the
operator appearing in the $|x_2|^0$ term in the OPE; it is a regularized
product of the two operators).
Higher order terms correspond to states of the form
$\langle 0|\Phi_{j_1}\partial^n\bar\partial^{\bar n}\Phi_{j_2}(0)$.
In our analysis, these contributions give rise to poles of
\densga\ coming from $x_2\to 0$, with the leading pole,
corresponding to the state $\langle 0|\Phi_{j_1}\Phi_{j_2}(0)$,
occurring at $j_2=-m_2-1$, etc.

We expect a similar analysis to apply when we send any number of
operators to $x=0$ and/or to $x=\infty$.  Thus, we see that the
correlation function \densga\ indeed has the structure expected from
the LSZ reduction. The regions $x_i\to 0,\infty$ give poles in the
$j_i$, at the locations \poles, \polesinf\ respectively\foot{There is
a slight subtlety associated with the last of the $n$ poles in the
$n$-point function, which we will briefly discuss in \S2.5.}. 
The residues of these poles are matrix elements involving the
normalizable vertex operators creating the relevant states from the
vacuum.

We finish this subsection with some comments:
\item{(1)} We see that a vertex operator of the form
\physw\ can create many different normalizable states 
when acting on the vacuum as in \resnorm, corresponding 
to different values of $j$ in \poles, \polesinf. Thus, 
in general the mapping between states and operators is not
one to one. This is standard in non-conformal theories,
and for example is expected to be generically true 
in large $N$ confining gauge theories. Generally, even if we
normalize the operator such that it creates a particular
one-particle state with canonical normalization, other
states it creates will not be canonically normalized.
These normalizations need to be taken into account when
extracting the S-matrix from the correlation functions using
\smatr.
\item{(2)} Like in QFT, the behavior \smatr\ appears 
only in $n$-point functions with $n\ge 3$. The two 
point function has single poles, rather than the 
double poles implied by \smatr. We will exhibit this
and discuss it further in some examples in \S5.
\item{(3)} An interesting feature of the discussion
of this section is the role played by the parameters
$x_i$ (see \eg\ \densga) in the analysis of the analytic
structure of the amplitudes. The $x_i$ can be thought
of as complexified Schwinger parameters in the spacetime
theory. In QFT, one way to introduce the Schwinger parameter 
(or proper time) is by replacing
\eqn\schwing{{1\over p^2+M^2}\to \int_0^\infty dt e^{-t(p^2+M^2)}.}
The divergence as $p^2\to -M^2$ is due to the region of 
large Schwinger parameter. In our discussion in the 
previous section, the role of the Schwinger parameter 
is played by $\pm\log |x|$ (see \eg\ \leaddiv, \intinf).
\item{(4)} A new effect that needs to be taken into account 
at finite $k$ is the upper bound on $j$, $j<(k-1)/2$ 
\refs{\GiveonPX,\MaldacenaKM}. In string theory on 
$AdS_3$, operators $\Phi_j$ with $j>(k-1)/2$ are 
expected not to exist, and the coset should presumably
inherit this bound. The role of this bound  in string
theory on the cigar is not well understood, and we 
will not discuss it further here.

\subsec{Bulk poles}

So far we have restricted our attention to poles of the
$n$-point function \sltwocon\ that occur as we tune 
a particular $j$ (or $p_\mu^2$ \masssh) to specific 
values. We identified the origin of such poles
in the $SL(2,\IR)/U(1)$ CFT and interpreted them as 
associated with the analog of LSZ reduction in 
string theory on the cigar. 

The correlation functions \sltwocon\ have another 
class of singularities that plays an important role 
in understanding string dynamics on spaces of the 
form \rrpphh, to which we turn next. These 
singularities are due to the infinite length of the 
cigar and are associated with processes that can
occur arbitrarily far from the tip.

To explain the origin of these singularities, we
next recall some features of CFT on asymptotically
linear dilaton spaces, such as Liouville theory
and $SL(2,\IR)/U(1)$ (see \eg\ \DiFrancescoUD\ for a
more detailed discussion), and contrast them with 
the more familiar case of CFT on flat space. 

Let $y$ be a non-compact scalar field on the worldsheet,
\eg\ one of the spatial directions in
$\IR^{d-1}$. The analog of the correlator \sltwocon\
for it is the  $n$-point function
$\langle e^{ip_1y}\cdots e^{ip_ny}\rangle$.
Translation invariance implies that this correlator 
is proportional to $\delta(p_1+\cdots+p_n)$, \ie\
it vanishes when the sum of momenta is non-zero,
and is infinite when it is zero. The infinity
is interpreted as the volume (or length) of 
$y$-space. Physically, it appears since the
process computed by this correlator can occur
at any $y$, with an amplitude that does not
depend on $y$. We are interested in the amplitude
per unit volume (for example, this is what enters
the calculation of the S-matrix in flat spacetime);
hence, this infinity is usually factored out.

Replacing $y$ by an asymptotically linear dilaton
direction $\phi$ \ggss, such as the radial coordinate
along the cigar, leads to a different picture.
The dynamics is no longer translationally invariant,
both because of the non-trivial dilaton and due to
whatever effects resolve the strong coupling singularity
at $\phi=-\infty$. In the case of the cigar, this is
the metric, which depends non-trivially on $\phi$. 

Thus, there is no longer any reason for the amplitudes
to be proportional to the length of $\phi$-space, and
indeed, in general they are not. In fact, since the
string coupling goes to zero far from the tip of the 
cigar, correlation functions are typically dominated by 
the vicinity of the tip.

By tuning the ``momenta along the cigar'' $j_i$ \sltwocon\ 
one can reach resonances, or bulk amplitudes, which are
processes that {\it can} occur anywhere along the cigar 
with uniform amplitude and are thus enhanced by the length
of $\phi$-space. For example, keeping only the leading 
terms in the large $\phi$ expansion of the $V_{j;m,\bar m}$'s 
\largecos, and focusing on the $\phi$ dependence, we have
\eqn\bulkres{
\langle V_{j_1;m_1,\bar m_1}\cdots V_{j_n;m_n,\bar m_n}\rangle
\sim \langle e^{Qj_1\phi}\cdots e^{Qj_n\phi}\rangle.}
If the $j_i$ satisfy the ``anomalous momentum conservation
condition''
\eqn\anomcons{\sum_i j_i=-1~,}
the amplitude can be shown to diverge like the volume of
$\phi$-space, as in the flat space case mentioned above. 
One can think of this divergence as coming from the
integral over the zero mode of the field $\phi$. However,
unlike the flat space case, here the amplitude \bulkres\ is
non-zero in the vicinity of the surface \anomcons; it 
behaves like
\eqn\nearpole{
\langle V_{j_1;m_1,\bar m_1}\cdots V_{j_n;m_n,\bar m_n}\rangle
\sim {F_0(j_i;m_i,\bar m_i)\over 1+\sum_i j_i}}
as $\sum j_i\to -1$. $F_0(j_i;m_i,\bar m_i)$ is finite
at $\sum j_i=-1$. 

The poles \nearpole\ are different in many ways from
the external leg poles corresponding to the LSZ reduction,
that were discussed in the previous subsections:
\item{(1)} While the LSZ poles are associated with
states living near the tip of the cigar, the bulk
poles are due to the presence of the semi-infinite
linear dilaton direction and are supported infinitely
far from the tip. In particular, introducing a physical
cutoff which stops $\phi$ from going to $+\infty$
regularizes the bulk poles, replacing them by large but finite
contributions proportional to the length of the cutoff cigar;
the LSZ poles are insensitive to the presence of such a cutoff. 
The fact that the bulk poles are associated with the region 
far from the tip of the cigar, where the worldsheet theory
simplifies, also allows one to compute the residue of the poles  
\nearpole\ using free field techniques,  as reviewed in
\DiFrancescoUD.
\item{(2)} The location of the bulk poles depends on the
genus of the worldsheet, $g$. On the sphere, there is a 
bulk pole at \anomcons, but the same amplitude is finite 
and dominated by the region near the tip of the cigar for 
$g\ge 1$. The higher genus analog of the condition for the
bulk pole \anomcons\ is
\eqn\highgen{\sum_i j_i=g-1.}
Thus, in different orders in string perturbation theory,
one finds different sets of bulk poles; the LSZ
poles remain the same.
\item{(3)} The poles \nearpole\ (and their generalizations described
below) depend on all the $\{j_i\}$ appearing in a correlation function,
while the LSZ poles depend only on a single $j$. 
\item{(4)} The locations of the bulk poles only depend on the 
$\phi$-momentum $j$, and not on the other quantum numbers on the cigar
$(m,\bar m)$ \foot{The residues of the poles,
$F_g(j_i;m_i,\bar m_i)$, do in general depend on all the 
other parameters.}.
In contrast, the locations of the LSZ poles
\poles, \polesinf\ involve in a non-trivial way $(j;m,\bar m)$.

\noindent
The discussion above can be generalized in two different
directions. First, we have only considered the bulk poles 
due to the leading terms in the large $\phi$ expansion of 
the vertex operators \largecos. It turns out that the only 
other term in this expansion that we need to consider is 
the leading normalizable term (the last term in \largecos)
-- the other terms in the expansion \largecos\ do not
give any new poles.

Consider, for example, the contribution to the $n$-point function
\sltwocon\ in which we keep the leading non-normalizable term
in the large $\phi$ expansion of the first $l$ operators
$V_{j_i;m_i,\bar m_i}$, $\exp(Qj_i\phi)$, and the leading
normalizable term, $\exp(-Q(j_i+1)\phi)$, for the other $n-l$.
The consideration after equation \bulkres\ lead in this case to poles
located at
\eqn\mixedpoles{\sum_{i=1}^l j_i-\sum_{i=l+1}^n(j_i+1)=g-1~.}
Another way of deducing the presence of the poles \mixedpoles\ 
is to use the reflection property satisfied by the observables
$V_{j;m,\bar m}$, \reflrel. Applying \reflrel\ to $n-l$ of the
operators in \sltwocon\ takes \highgen\ to \mixedpoles.

The second generalization of the discussion above involves 
perturbative interactions with the background. The results \highgen,
\mixedpoles\ were obtained in the leading approximation, where
we replace the cigar by an infinite flat cylinder, and are due to
interactions among the $V_{j;m,\bar m}$ in the bulk of this cylinder.
It is known from studies of CFT in this and related backgrounds
that a more general class of bulk amplitudes involves interactions
which include both the $V_{j;m,\bar m}$ and the metric. To study
such interactions in practice, one expands the metric on the cigar
around its large $\phi$ limit (the metric on a cylinder). This leads 
to an effective interaction term on the cylinder labeled by $(\phi,Y)$ 
(see \largecos) of the form
\eqn\lwakim{\CL_1=\lambda\partial Y\bar\partial Y e^{-Q\phi}.}
This interaction can be related to the Wakimoto description of
$SL(2,\IR)$ CFT in terms of free fields (see \eg\
\refs{\BershadskyIN,\GiveonUP}). Bulk interactions involving the
metric deformation \lwakim\ give rise to poles that generalize
\highgen, \mixedpoles. Roughly speaking, we can differentiate 
the correlation functions \sltwocon\ $n$ times with respect to 
$\lambda$, bringing down $n$ vertex operators \lwakim, and 
repeat the discussion above. It is easy to see that this leads to
poles at the same locations as in \highgen, \mixedpoles, but with
$g$ replaced by $g+n$.

Another class of perturbative interactions involves the dual 
description of the $SL(2,\IR)/U(1)$ CFT due to V. Fateev, 
A. Zamolodchikov and Al. Zamolodchikov \refs{\fzz,\KazakovPM} 
in terms of Sine-Liouville theory. For our purposes,
the main consequence of this duality is that in addition to the
metric perturbation \lwakim\ one should think of the CFT on the cigar
as containing in the Lagrangian the Sine-Liouville interaction
\eqn\lsineliouv{\CL_2=\mu e^{-{1\over Q}\phi}\cos\left[\sqrt{k+2\over2}
(Y-\bar Y)\right].}
This interaction carries one unit of winding, and thus explicitly 
breaks winding number conservation, in agreement with what one 
expects from the cigar picture. The precise relation between 
$\lambda$ \lwakim\ and $\mu$ \lsineliouv\ can be found in 
\GiveonUP. 

The most general bulk pole is obtained by expanding the correlation
function \sltwocon\ to order $n_1$ in $\lambda$ \lwakim\ and to
order $n_2$ in $\mu$ \lsineliouv. It is not difficult
to show that the general form of \mixedpoles\ is thus
\eqn\finalpoles{\sum_{i=1}^l j_i-\sum_{i=l+1}^n(j_i+1)=g+n_1+n_2{k\over 2}-1,}
where $n_1$ and $n_2$ are non-negative integers.

A natural question is whether one can ever confuse LSZ poles
in correlation functions with bulk poles. As discussed previously,
in general the answer is no, due to the different kinematic structure 
associated with the two kinds of poles. While the former give poles
that depend on the individual $j_i$, the latter occur as a function
of a particular linear combination of all the $j_i$ in a correlation
function.

One situation in which the two
can be confused is in two-point functions, since then there is only
one independent $j$. We will discuss examples of this in \S5.

Another case where one might worry about this issue is the following.
Consider the $n$-point function \sltwocon\ near the LSZ poles
corresponding to the external legs with indices 
$2,\cdots, n$. The residue of these
LSZ poles is a correlator of $n-1$ normalizable operators which behave
as $\phi\to\infty$ like $\exp(-Q(j_i+1)\phi)$, $(i=2,3,\cdots,n)$, and one
non-normalizable operator,  $V_{j_1;m_1,\bar m_1}$. Suppose we now
want to take this last operator to an LSZ pole \poles, \polesinf\ as well,
in order to study the S-matrix \smatr.

A natural question is whether the residue of this pole also receives 
contributions from a bulk amplitude which satisfies \finalpoles\ with 
$l=1$. This appears to be possible if the LSZ pole corresponding to
$j_1$ occurs at
\eqn\confpoles{j_1=\sum_{i=2}^n(j_i+1)+n_1+n_2{k\over2}-1}
for some non-negative integers $n_1$, $n_2$. In particular,
\confpoles\ implies that a necessary condition for a bulk pole
to give a contribution with the same kinematic structure as
an S-matrix element is
\eqn\jineq{j_1\geq n-2+\sum_{i=2}^nj_i.}

For any given S-matrix element \smatr, there are now two
possibilities. One is that \jineq\ is not satisfied. In 
this case, there is clearly no bulk contribution to the
residue of the LSZ poles. 

The second possibility is that \jineq\ is satisfied, 
in which case at least at this level of analysis it 
looks like a bulk contribution might exist. But in fact,
since $(j_1,j_2,\cdots, j_n)$ appear symmetrically in
\smatr, we can in that case exchange $j_1$ with one of
the other $j_i$, and repeat the analysis. It is easy to
show that since all $j_i$ are larger than $-1/2$, for all
$n>2$ there must exist a choice for which the analog of 
\jineq\ is no longer satisfied. For this choice, it is 
manifest that there is no bulk contribution that can mix 
with the LSZ pole.

Thus, we conclude that in $n\ge 3$ point functions,
we can always perform the calculation in a way that
makes it manifest that the only contribution to the residue
of the poles in \smatr\ comes from the S-matrix of the
normalizable states localized near the tip of the cigar.

Another natural question is how bulk poles of the sort discussed 
in this section arise when we compute the correlation functions 
\sltwocon\ using equation \densga, by first computing the correlator of
$\Phi_j(x,\bar x)$ on $AdS_3$ and then integrating over
the $x_i$. There are two ways this might happen.
One is that the $AdS_3$ correlation function
$\langle\Phi_{j_1}(x_1)\cdots \Phi_{j_n}(x_n)\rangle$
has such poles even before we integrate over the $\{x_i\}$.
Examples of such poles were studied in 
\refs{\GiribetFT,\GiveonUP,\MaldacenaKM}.

The other is a contribution from regions in the integrals over the $\{x_i\}$
\densga\ in which two or more of them approach each other. A
useful way of thinking about these regions is from the point
of view of the $AdS_3/CFT_2$ correspondence. The behavior of
correlation functions in the spacetime $CFT_2$ as some of the 
$x_i$ approach each other is mapped by the correspondence to 
physics that occurs near the boundary of $AdS_3$. After modding 
out by $U(1)$ to go from $AdS_3$ to $SL(2,\IR)/U(1)$, the boundary
of $AdS_3$ becomes the region far from the tip of the cigar. 
Thus, it is natural to expect that singularities of the $x_i$
integrals that come from these regions correspond to bulk processes
on the cigar.

The analysis of the regions $x_i\to x_j$ to the integral \densga\
is familiar from studies of string amplitudes, where the
role of the $x_i$ is played by the string worldsheet coordinates.
We will present an example of both sources of bulk poles in a particular
correlation function in the next subsection.

\subsec{The three-point function}

The discussion of the previous subsections can be made quite
explicit for the case of the three-point function \sltwocon, 
since the relevant correlation function in $AdS_3$ is known 
\TeschnerFT. In this subsection we will describe how various 
elements of that discussion manifest themselves in this case. 
For simplicity we will restrict to pure winding modes, 
$m_i=\bar m_i$, with $m_1+m_2+m_3=0$, and (without loss of 
generality) take $m_1,m_2<0$, $m_3>0$.

The three-point function \densga\ takes in this case the form
\eqn\threept{\eqalign{
\langle V_{j_1;m_1,m_1}V_{j_2;m_2,m_2}&V_{j_3;m_3,m_3}\rangle
=D(j_1,j_2,j_3;k)\times\cr
\int d^2 x_1  d^2 x_2 & d^2 x_3
|x_1|^{2(j_1+m_1)}|x_2|^{2(j_2+m_2)}|x_3|^{2(j_3+m_3)}
\times\cr
 &|x_1-x_2|^{2(j_3-j_1-j_2-1)}|x_1-x_3|^{2(j_2-j_1-j_3-1)}
 |x_2-x_3|^{2(j_1-j_2-j_3-1)},\cr
 }}
where 
\eqn\formdd{\eqalign{
D(j_1,j_2,&j_3;k)={k\over 2\pi^3}\nu(k)^{j_1+j_2+j_3+1}\times\cr
&{G(-j_1-j_2-j_3-2)G(j_3-j_1-j_2-1)G(j_2-j_1-j_3-1)G(j_1-j_2-j_3-1)
\over G(-1)G(-2j_1-1)G(-2j_2-1)G(-2j_3-1)}.\cr
}}
$\nu(k)$ is given by \nukk; $G(x)$ is a known function, 
whose analytic properties are described in appendix A. 
The factor $C_{123}$ in \densga\ can be omitted in this case.

To study the LSZ poles, one is interested in the contribution
to the integrals \threept\ from $x_i\to0,\infty$. As in the general
discussion leading to equation \polezinf, the leading divergence as
$x_1\to 0$, $x_3\to\infty$ gives rise to a pole at $j_1=-m_1-1$,
$j_3=m_3-1$. Expanding \threept\ near this pole we find that it behaves as
\eqn\nearpole{
\langle V_{j_1;m_1,m_1}V_{j_2;m_2,m_2}V_{j_3;m_3,m_3}\rangle
\simeq
{\pi^2D(j_1,j_2,j_3;k)\over (m_1+j_1+1)(m_3-j_3-1)}
\int {d^2 x_2\over x_2^2}.}
The infinite factor $\int{d^2 x_2\over x_2^2}$ imposes the momentum
and winding conservation conditions \windcons\ \GiveonTQ. In the
decomposition of $SL(2,\IR)$ as ${SL(2,\IR)\over U(1)}\times U(1)$, it
belongs to the $U(1)$ part.  Therefore, it appears uniformly in all
correlators (as we will exhibit explicitly below) and should be ignored.

At first sight it seems that since the $x_2$ integral only imposes momentum
and winding conservation, we are missing the LSZ poles of the correlation 
function \threept\ that involve $j_2$. In fact, this is not the case -- 
these poles come from the ``structure function'' $D(j_1,j_2,j_3;k)$. 
To see that, recall that since $m_2$ is negative, we expect to find 
poles when $m_2+j_2$ is a negative integer. When computing the residue 
of the poles in \nearpole\ we can set $m_1+j_1+1=m_3-j_3-1=0$. Hence,
\eqn\mjtwo{j_2+m_2=j_2-m_1-m_3=j_2+j_1-j_3.}
The function $D(j_1,j_2,j_3;k)$ \formdd\ has a factor,
$G(j_3-j_1-j_2-1)$, which has a pole whenever its argument,
$j_3-j_1-j_2-1$, is a non-negative integer. These poles correspond, via
\mjtwo, to poles as a function of $m_2+j_2$ which occur at precisely
the right places to correspond to contributions to the S-matrix.

While the pole in $j_2$ appears in a different way than those in $j_1$
and $j_3$, it is an LSZ pole due to 
normalizable states and does not receive a bulk contribution. This
follows from the general discussion in the previous subsection. Indeed, 
suppose we wanted to compute the bulk contribution to the residue of 
the LSZ poles in $j_1$ and $j_3$ \nearpole. Since the residue of the 
poles involves normalizable vertex operators (see \largecos), the
leading large $\phi$ behavior of the correlator computing this residue 
is $\langle e^{-Q(j_1+1)\phi} e^{Qj_2\phi} e^{-Q(j_3+1)\phi}\rangle$. 
The total power of $e^{-Q\phi}$ coming from the vertex operators
is $j_1+j_3+2-j_2$. Near a pole at 
$m_2+j_2=-n$ (where $n$ is a positive integer), we can use 
equation \mjtwo\ to rewrite this as $j_1+j_3-j_2+2=2j_1+n+2$. Thus, the 
$\phi$ zero mode integral goes at large $\phi$ as 
$\int d\phi_0 e^{-Q(2j_1+1+n)\phi_0}$. Using the fact that 
$2j_1+1\ge 0$ and $n\ge 1$, we see that the $\phi_0$ integral 
rapidly converges at infinity, and therefore there
cannot be a bulk contribution to the pole at $m_2+j_2=-n$.
Insertions of the interactions \lwakim, \lsineliouv\ suppress the
$\phi_0$ integral even more. Hence, we conclude that this pole
is entirely due to the contribution of an on-shell normalizable
state -- it is an LSZ pole.

The following question might seem puzzling at this point. 
The factor $G(j_3-j_1-j_2-1)$ from which we got the $j_2$
LSZ pole is an overall factor in the three-point function
\threept. It gives rise to poles of the three-point
function whenever $j_3-j_1-j_2-1$ is a non-negative integer,
irrespective of the values of the $\{m_i\}$. Hence, the
analysis of the previous subsections leads one to believe
that these poles can be interpreted as bulk poles. As we 
will see next, this is indeed the case for generic values of the $j_i$, but it 
does not contradict the discussion above, which interpreted
the same poles as LSZ poles.

Looking back at equation \finalpoles\ we see that in order
to interpret the poles of the factor $G(j_3-j_1-j_2-1)$ as bulk
poles we have to focus on the contribution to the
three-point function in which we take the leading
non-normalizable contribution from $V_{j_3}$, and
the leading normalizable terms from $V_{j_1}$, 
$V_{j_2}$ (see \largecos). For generic $j_1$ and $j_3$, 
the poles which appear when $j_3-j_1-j_2=n=1,2,\cdots$ are indeed
interpreted as bulk poles.

To compare to the preceding analysis, in which 
the same poles were interpreted as LSZ poles,
we need to take the external legs on-shell. In
the discussion around equation \nearpole\ this was done
by first taking $j_1\to -m_1-1$ and $j_3\to m_3-1$
and then studying the origin of the third pole
of the three-point function \threept\ associated
with $j_2$. 

In the present way of computing the amplitude
it is convenient to first take the external legs
$j_1$ and $j_2$ on-shell and then study the 
singularity structure as a function of $j_3$.
Indeed, the residue of the poles at (say)
$j_1=-m_1-1$, $j_2=-m_2-1$ is proportional
to the three-point function of two normalizable
operators, corresponding to $j_1$ and $j_2$,
and a third operator $V_{j_3}$, which has
both normalizable and non-normalizable parts.
Thus, in this way of calculating the correlator,
one would say that there are two contributions
to the pole at $j_3=m_3-1$. One is an LSZ pole,
whose residue is the three-point function of normalizable
operators. The other is a bulk pole; its residue is
the three-point function of two normalizable ($j_1$
and $j_2$) and one non-normalizable ($j_3$) operators. 

This situation is an example of the general discussion
around equation \confpoles. If we first go to the LSZ poles
in $j_1$ and $j_2$, the LSZ pole in $j_3$ seems to mix
with a bulk pole. As argued there, we can always perform
the calculation in such a way that the bulk contribution
is manifestly absent. Here, one way to do that is to first go to the
LSZ poles in $j_1$ and $j_3$, as we have done in \nearpole.
In that way of doing the calculation it is clear that the 
bulk contribution to the last pole (in $j_2$) vanishes.

Thus, we see that the same pole in the three-point function
can sometimes be interpreted as a bulk pole, and sometimes
as an LSZ pole. There is clearly more to be said on this
subject; in particular, it would be interesting to understand
how the apparent bulk contributions to the three-point
function cancel near the LSZ poles in the second way of
calculating it. We will not pursue  this and other similar
questions here.

So far we have mainly studied the way that LSZ poles
appear in the three-point function \threept. We next
discuss briefly the different bulk poles exhibited
by these amplitudes\foot{We will think of $k$ as large,
and study the bulk poles at $j_i$ of order one. There 
are additional poles at which some or all of the $j_i$
must be of order $k$. They can easily be studied
in the same way. We will also assume that $k$ is 
irrational. For rational $k$ (the case of interest 
in applications) there are some further issues that
need to be discussed.}. We have already seen one source
of such poles, corresponding to singularities of the 
function $G(j_3-j_1-j_2-1)$ in the $AdS_3$ form factor 
\formdd. Clearly, there are similar poles that are due 
to the other two factors of $G$ in \formdd\ that are 
obtained by permutations of $(j_1,j_2,j_3)$. 

The remaining factor in the numerator of \formdd,
$G(-j_1-j_2-j_3-2)$, has poles when 
$j_1+j_2+j_3+2=0,-1,-2,\cdots$. In the general 
analysis \finalpoles\ such poles can in principle
arise in a contribution with $l=0$. However, 
due to the constraint $j_i>-1/2$, these poles
are actually out of the physical range. 

As mentioned in the previous subsection, an 
additional source of bulk poles is the regions
in the integrals over $x_i$ where two or more
of them approach each other. Consider, for example,
the contribution to the three-point function \threept\
from the region $x_1\to x_2$. This region leads to
poles located at
\eqn\locpoleonetwo{j_1+j_2-j_3=0,1,2,\cdots.}
The residues of these poles are easy to compute
using \threept. Comparing to \finalpoles\ we see that 
these poles are bulk poles occurring in the contribution to the three-point 
function where we take the leading non-normalizable 
terms in $j_1$ and $j_2$, and the leading normalizable
term in $j_3$. 

Note also that the poles \locpoleonetwo\ do not overlap
those due to the factor $G(j_3-j_1-j_2-1)$ in the
$AdS_3$ structure constant. This is a necessary condition
for the interpretation of these poles as bulk contributions,
since the integral over the zero mode of $\phi$ can only
give single poles, and not double poles. 

Similarly, one can study the poles of \threept\ that come from
the region where all three of the $x_i$ approach each other.
Changing variables from $(x_2,x_3)$ to $(\epsilon,y)$ where
$x_2-x_1=\epsilon$, $x_3-x_1=\epsilon y$, and studying the
contribution of the region $\epsilon\to 0$ to \threept,
it is not difficult to see that this region gives poles 
when 
\eqn\threecoin{j_1+j_2+j_3=-1,0,1,2,3,\cdots}
These poles are due to the leading term in the
three-point function \threept, in which we replace 
all three vertex operators by their  leading, 
non-normalizable  contributions as $\phi\to\infty$. 
Indeed, \threecoin\ is in agreement with \finalpoles,
for the relevant case $l=0$.

Many elements of the discussion above generalize to $n>3$
point functions. The fact that we can see only $n-1$ of
the $n$ LSZ poles in \smatr\ by studying the region 
$x_i\to 0,\infty$ follows from the properties 
of the $n$-point function of the $\Phi_{j_i}(x_i)$ \densga\
under rescaling of the $x_i$. Using the fact that 
\eqn\rescxx{
\langle\Phi_{j_1}(\lambda x_1)\cdots\Phi_{j_n}(\lambda x_n)\rangle=
|\lambda|^{-2\sum_i(j_i+1)}
\langle\Phi_{j_1}(x_1)\cdots\Phi_{j_n}(x_n)\rangle,
}
and changing variables from $(x_1,\cdots, x_n)$ to 
$(y_1,\cdots, y_{n-1},x_n)$, with $y_i=x_i/x_n$, 
one finds that
\eqn\densgaaa{\eqalign{
\prod_{i=1}^n &\int d^2 x_i x_i^{j_i+m_i}\bar x_i^{j_i+\bar m_i}
\langle\Phi_{j_1}(x_1)\cdots\Phi_{j_n}(x_n)\rangle =\cr
&\prod_{i=1}^{n-1} \int d^2 y_i y_i^{j_i+m_i}
\bar y_i^{j_i+\bar m_i}
\langle\Phi_{j_1}(y_1)\cdots
\Phi_{j_{n-1}}(y_{n-1})
\Phi_{j_n}(1)\rangle
\int d^2 x_n x_n^{\sum_i m_i - 1}
{\bar x}_n^{\sum_i {\bar m}_i - 1}.\cr
}}
Thus, one of the $n$ integrals over $x_i$ (which
in \densgaaa\ has been chosen to be the one over 
$x_n$) can be thought of as imposing momentum
and winding conservation, just like we have found 
for the three-point function (see \nearpole). Presumably,
the last LSZ pole arises in all $n$-point functions from
the behavior of the unintegrated $n$-point function on
$AdS_3$, like we saw it does in the three-point function.
It would be interesting to understand this in more detail.

The fact that the last pole appears in this case
at the same value of $j_n$ as a bulk pole is general
as well. Suppose, for example, that $m_1,\cdots, m_l>0$
and $m_{l+1},\cdots, m_n<0$. Take the first
$n-1$ external legs to LSZ poles with
$j_i=m_i-1-s_i$ for $i=1,\cdots, l$ and
$j_i=-m_i-1-s_i$ for $i=l+1,\cdots, n-1$,
with $s_i\in \Z_+$. The location of the 
last LSZ pole, which is expected to occur 
at $j_n=-m_n-1-s_n$, can be rewritten using
momentum and winding conservation \windcons\
as
\eqn\lastpole{\sum_{i=1}^l j_i
-\sum_{i=l+1}^n j_i=n-1-2l 
+\sum_{i=l+1}^n s_i-\sum_{i=1}^l s_i.
}
This has the general form \finalpoles\
and hence corresponds to a position of a
bulk pole. However, as discussed in the previous
subsection, one can always perform the calculation
in such a way that it is manifest that the S-matrix 
\smatr\ does not receive contributions from bulk poles.

\subsec{Winding non-conserving correlators}

So far in this section we focused on $SL(2,\IR)/U(1)$ 
correlation functions \sltwocon\ that preserve both 
momentum and winding \windcons. As mentioned in the
beginning of \S2.3, winding conservation can actually 
be violated on the cigar. In this subsection we briefly 
comment on the generalization of the analysis above to 
winding violating correlation functions, leaving a more
detailed discussion to future work.

A nice way to study winding number violating amplitudes, which
is reminiscent of an analogous construction in the $SU(2)/U(1)$ CFT,
was proposed by \fzz. Consider the $AdS_3$ operator\foot{Note that
the reflection formula \reflrel\ naively relates this operator to an
operator with $j={k\over2}$, $m=\bar m={k+2\over2}$, which violates the
bound mentioned at the end of \S2.3. This is not necessarily a
problem since the reflection relation \reflrel\ does not obviously 
apply when one of the operators it relates violates this bound;
it originates in CFT on Euclidean $AdS_3$, where such operators
are not expected to exist.} 
$\Phi_{j;m,\bar m}$ \furtrans, with $m=\bar m=-j={k+2\over2}$.
This operator belongs to a degenerate representation of the $SL(2,\IR)$
affine Lie algebra (see \eg\ \S4.2 in \GiveonUP).
Comparing to \momwin, we see that this operator carries one
unit of winding number, $w$. At the same time, \dimvjm\ implies that
the $SL(2,\IR)/U(1)$ part of this operator has dimension 
$\Delta=\bar\Delta=0$.

The basic idea of \fzz\ is that this operator should be
interpreted as a product of an operator in the $U(1)$ CFT,
and the identity operator in the coset. Thus, to study 
correlation functions in the coset, \sltwocon, that
violate winding number by $n$ units, we can start with 
the $AdS_3$ correlation function \densga\ with $n$
extra insertions of the operator
$\Phi_{-{k+2\over2};{k+2\over2},{k+2\over2}}$
(or its complex conjugate),
and compute the resulting $AdS_3$ correlation function,
which satisfies \windcons. Stripping off the
$U(1)$ part (which can be easily computed)
we find the winding number violating 
$SL(2,\IR)/U(1)$ correlation function we are after.

Another (related) point that can be made about winding
number violating amplitudes on the cigar is the following.
Suppose we want to compute the residue of an LSZ pole
in one or more of the external legs (see \eg\ \nptfact).
Then, at least one of the operators in \sltwocon\ is 
normalizable, \eg\ for $m<-1/2$ it could be:
\eqn\normfirst{V_{-m-1;m,m}^{({\rm norm})}
=\lim_{j+m+1\to 0}(j+m+1)V_{j;m,m}.}
Then, a reflection symmetry\foot{This is not to be confused
with the different reflection property \reflrel.} relating
naively different
states in $SL(2,\IR)/U(1)$ implies that, in a suitable
normalization of the operators,
\eqn\refsltwo{V_{-m-1;m,m}^{({\rm norm})}=
V_{{k\over2}+m;m+{k+2\over2},m+{k+2\over2}}^{({\rm norm})}.}
In other words, different non-normalizable operators, in this
case the operators $V_{j;m,m}$ and 
$V_{{k-2\over2}-j; m+{k+2\over2},m+{k+2\over2}}$,
can create the same state from the vacuum. This phenomenon was
noticed in \refs{\MaldacenaHW,\ParnachevGW}, and given 
this interpretation in \AharonyVK. In field theory it is not
surprising that the same state may be created by more than one
operator, and it was verified in \AharonyVK\ in a specific
example that \refsltwo\
is consistent with the low-energy field theory content
of these operators.

Since the winding number of the right and left hand sides of
\refsltwo\ differs by one unit, we can use it to relate correlation
functions (involving normalizable operators) that violate winding number 
to those that preserve it. Examples of winding-non-conserving correlators
will be given in \S5.6.

\newsec{Superstring theory on $SL(2,\IR)/U(1)$}

Bosonic string theory on spacetimes of the form \rrpphh\ 
is in general IR unstable (for $d>0$ or non-trivial $\tilde\MM$), 
a fact that typically manifests itself
in the presence of tachyons in the spectrum of normalizable 
states, and IR divergences in loop amplitudes. To avoid 
these instabilities we turn in this section to the 
superstring. This is the case which is relevant for the applications of our
formalism to the decoupled theories of fivebranes and  
Calabi-Yau singularities
in type II string theory, which we will discuss below.

There are two steps involved in generalizing the discussion of \S2
to the superstring. First, we need to supersymmetrize the worldsheet
theory and enlarge the gauge principle from $\NN=0$ to
$\NN=1$ supergravity. This leads to type 0 string theory and in itself
does not solve the IR problems of the bosonic string. To achieve that,
one needs to also perform a chiral GSO projection 
\refs{\deAlwisPR,\KutasovUA}. 

The supersymmetric level $k$ $SL(2,\IR)/U(1)$ CFT can be constructed
as follows. We start with a bosonic $SL(2,\IR)$ WZW model with
central charge $c=3(k+2)/k$, as in \cbossl, and add to it three free
fermions $\lambda^a$, $a=3,\pm$ (and their anti-chiral analogs).
Associated with the fermions $\lambda^a$ is an $SL(2,\IR)$ current
algebra of level $(-2)$. The full theory is $\NN=1$ superconformal; we
review the structure of its superconformal algebra in
appendix A. The fermions $\lambda^a$ are bottom components of 
superfields whose top components are the total $SL(2,\IR)$
currents $J_a^{\rm (total)}$. The level of this current algebra
is $k+2+(-2)=k$, where the two contributions come from the bosonic
$SL(2,\IR)$ WZW model and from the fermions.

To describe the coset, we would like to gauge the $U(1)$ superfield
whose bottom component is $\lambda_3$ and whose top component
is the total $U(1)$ current $J_3^{\rm (total)}=\{G_{-\half},\lambda_3\}$.
This leads to a SCFT with central charge
\eqn\cferm{c_{sl}={3(k+2)\over k}+{3\over2}-{3\over2}={3(k+2)\over k},}
where the $+{3\over2}$ is the contribution of the fermions $\lambda^a$,
and the $-{3\over2}$ is due to the gauging. While the underlying CFT on
$SL(2,\IR)$ is $\NN=1$ superconformal, the coset $SL(2,\IR)/U(1)$ is 
actually invariant under an $\NN=2$ 
superconformal symmetry. 
The generators of the $\NN=2$ algebra are given in appendix A.

The requirement that \rrpphh\ corresponds to a solution of the classical
equations of motion of the superstring leads to the constraint
(compare to \critbos)
\eqn\critfer{{3\over2}d+c_{sl}+c_{\tilde\MM}=15.}
To study perturbative string theory on \rrpphh\ we need to construct
vertex operators on $SL(2,\IR)/U(1)$. This can be done as in
section 2, by studying vertex operators in the underlying
SCFT on $AdS_3$, and then removing from them the $U(1)$ part.
We will describe this here for the Neveu-Schwarz (NS) sector. 
The analysis of the Ramond sector is similar;
for a recent discussion, see \AharonyVK.

We can again take as a starting point the $SL(2,\IR)$ vertex operators
$\Phi_{j;m,\bar m}$ \furtrans. While these operators do not contain
the worldsheet fermions $\lambda^a$, removing their $U(1)$ part does
{\it not} lead to the  operators $V_{j;m,\bar m}$ that we encountered
in section 2. The reason
is that the $U(1)$ current $J_3^{\rm (total)}$ that is being gauged
includes contributions from both the bosonic WZW model and from the
fermions. Removing the $U(1)$ part of the operators $\Phi_{j;m,\bar m}$
in the superstring 
gives operators that we will denote by $V_{j;m,\bar m}^{(sl,susy)}$,
whose dimension is (compare to \dimvjm)
\eqn\susydimvjm{\eqalign{
\Delta_{j;m}=&{m^2-j(j+1)\over k},\cr
\bar\Delta_{j;\bar m}=&{\bar m^2-j(j+1)\over k}.\cr
}}
The operators $V_{j;m,\bar m}^{(sl,susy)}$ are in fact primaries of the
full $\NN=2$ superconformal symmetry. Their R-charges are given by
\eqn\rvjm{\eqalign{
R_m=&{2m\over k},\cr
\bar R_{\bar m}=&{2\bar m\over k}.\cr
}}
Far from the tip, the cigar looks like a semi-infinite cylinder. As in the
bosonic case, we will parametrize the semi-infinite direction along this
cylinder by $\phi$. The  direction around the cylinder, $Y$, now lives on a
circle of radius $\sqrt{2k}$. The momentum and winding in the $Y$
direction, $(n,w)$, are related to $(m,\bar m)$ in \susydimvjm\ in a way
similar to \momwin:
\eqn\susymomwin{\eqalign{
m=&\half(wk+n),\cr
\bar m=&\half(wk-n).\cr
}}
The vertex operator $V_{j;m,\bar m}^{(sl,susy)}$ has the asymptotic form
\eqn\asymsusy{
V_{j;m,\bar m}^{(sl,susy)}\simeq {e^{iQ(mY-\bar m\bar Y)}
\over 2j+1}\left[ e^{Qj\phi}+R(j,m,\bar m;k)e^{-Q(j+1)\phi}+\cdots\right],}
where the reflection coefficient $R(j,m,\bar m;k)$ is given by \reflrel,
and $Q$ is related to $k$ by \kkqq.

Other observables on the cigar can be obtained by acting on
$V_{j;m,\bar m}^{(sl,susy)}$ by the $\NN=2$ superconformal symmetry
generators. For example, at the first excited level we find the operators
$[G_{-\half}^\pm,V_{j;m,\bar m}^{(sl,susy)}]$ (and similarly for the other
worldsheet chirality). Far from the tip of the cigar the theory simplifies
into that of two free superfields $(\phi,\psi_\phi)$ and $(Y, \psi_Y)$, and
these excited operators can be written in terms of the $\NN=2$ descendants
$\psi_\phi e^{ipY+\beta\phi}$ and $\psi_Y e^{ipY+\beta\phi}$. At higher
excited levels the situation is similar.

A class of physical observables in superstring theory on
\rrpphh\ can be constructed as follows. Let $\WW$ be an $\NN=1$ superconformal
primary on $\tilde\MM$ with scaling dimension $(\Delta_L,\Delta_R)$. An (NS,NS)
sector physical operator can be formed by ``dressing'' $\WW$ as in \physw:
\eqn\physwsusy{\CO_\WW(p)=
e^{-\varphi-\bar\varphi}\WW V_{j;m,\bar m}^{(sl,susy)}e^{ip\cdot x}.}
Here, $(\varphi,\bar\varphi)$ are bosonized superconformal ghosts, and
the factor $e^{-\varphi-\bar\varphi}$ indicates that the vertex
operator \physwsusy\ is written in the $(-1,-1)$ picture. The physical
state condition for \physwsusy\ is
\eqn\massshsusy{\half p_\mu^2+\Delta_{j;m}+\Delta_L=
\half p_\mu^2+\bar\Delta_{j;\bar m}+\Delta_R=\half~.}
As mentioned earlier, in order to study superstring propagation on
\rrpphh\ we need to perform a chiral GSO projection. A sufficient
condition for being able to do that is that the full background
\rrpphh\ is $\NN=2$ superconformal. This is the case, \eg, if $d$ is
even and the conformal theory on $\tilde\MM$ is $\NN=2$ superconformal. 
In that case, the chiral GSO projection amounts to the requirement that
the total R-charge of the vertex operator \physwsusy\ must be an odd
integer. The GSO projection thus acts as an orbifold projection on the
background \rrpphh. We will see below that an important consequence
of this orbifold is the appearance of states with non-integer winding $w$
\susymomwin\ in the twisted sectors.

Since the observables in the superconformal field theory on 
the cigar can be lifted to those on (supersymmetric) $SL(2,\IR)$, we can use 
the results of the analysis of \S2 to study the analytic 
structure of their correlation functions. There are again LSZ 
poles at the locations \poles, \polesinf\ and bulk poles at 
the locations \finalpoles. The Wakimoto and Sine-Liouville
perturbations \lwakim, \lsineliouv\ have a slightly different
form in this case (see \eg\ \GiveonUP) but this does not alter 
the analysis leading to \finalpoles. For example, the analog
of the Sine-Liouville perturbation \lsineliouv\ is in this case
the $\NN=2$ Liouville perturbation,
\eqn\ntwoliouv{\CL_2=\mu G_{-\half}^-\bar G_{-\half}^-
e^{-{1\over Q}(\phi-i(Y-\bar Y))}+{\rm c.c.}}
Here $\mu$ is the (generally complex) $\NN=2$ Liouville coupling.

In the next sections we will use the general formalism of \S2 and this
section to study the dynamics associated with $NS5$-branes and
singularities of Calabi-Yau manifolds in superstring theory.

\newsec{Six dimensional $\NN=(1,1)$ supersymmetric LST}

\subsec{Review}

An important application of the formalism developed above
is to the dynamics of Neveu-Schwarz fivebranes (or, equivalently,
ADE singularities of $K3$ surfaces). The main example which we will
discuss in detail in this paper is
a system of $k$ parallel $NS5$-branes, extended in the directions
$(x^0,x^1,\cdots, x^5)$, in type IIB string theory. At low energies
the dynamics of this system includes a supersymmetric Yang-Mills (SYM) 
theory with $\NN=(1,1)$ 
supersymmetry in six dimensions (\ie\  sixteen supercharges), and
gauge group\foot{There are also versions of the construction with other
simply laced (ADE) gauge groups. We will not discuss them here.}
$SU(k)$ (there is also a decoupled $U(1)$ gauge field which will play no role
in this paper). The full theory on the fivebranes can be thought of as a UV
completion of this non-renormalizable gauge theory. We will see later
that the Green functions of this theory exhibit some unexpected features
at low energies.

On general grounds, one expects the decoupled theory on the fivebranes to be
holographically related to string theory in the near-horizon geometry
of the branes. As discussed in section 1, the near-horizon geometry of  
$k$ coincident fivebranes is \CallanAT
\eqn\chsmetr{\IR^{5,1}\times\IR_\phi\times SU(2)_k,}
where $\phi$ is related to the radial direction away from the branes,
and the supersymmetric $SU(2)_k$ WZW model describes the angular 
three-sphere. The $SU(2)_L\times SU(2)_R$ symmetry of the WZW model 
is identified with the $SO(4)$ rotation symmetry about the fivebranes, 
and gives rise to an R-symmetry of the LST. The dilaton depends linearly
on $\phi$, as in \ggss; the slope $Q$ is related to the number of
fivebranes $k$ by the relation \kkqq.

The six dimensional LST has a moduli space $\IR^{4k}/S_k$, corresponding
to the positions of the $k$ $NS5$-branes in the transverse $\IR^4$.
To avoid the strong coupling singularity of \chsmetr\ it is convenient
to separate the branes in the transverse $\IR^4$. Parametrizing this
$\IR^4$ by the complex coordinates
\eqn\defab{\eqalign{
a=&x^6+ix^7,\cr
b=&x^8+ix^9,\cr
}}
we will consider the point in moduli space at which the  $l$'th fivebrane
($l=1,\cdots, k$) is located at
\eqn\Bj{(a,b)=r_0(0,e^{2\pi il\over k}).}
This corresponds to a configuration in which the fivebranes are 
evenly spaced on a circle of radius $r_0$ in the $(x^8,x^9)$ 
plane. The $SO(4)$ rotation symmetry around the fivebranes is 
broken in the background \Bj\ to $SO(2)_{67}\times \Z_k$. Thus, 
this is a rather  symmetric point in moduli space -- generically, 
the rotation symmetry is broken completely. It should be possible 
to generalize the discussion below to generic points in the moduli 
space; we will briefly comment on this problem below (in section 11.2).

In the low energy gauge theory on the fivebranes, the
coordinates $x^i$, $i=6,7,8,9$, or $a$ and $b$ \defab, 
are promoted to scalar fields in the adjoint representation 
of $SU(k)$, which we will denote by $X^i$, $A$ and $B$, 
respectively. The displacement of the fivebranes \Bj\ 
corresponds to giving an expectation value to $B$, of the form
\eqn\bexpv{\langle A\rangle=0;\;\;\;
\langle B\rangle=M_WM_s{\rm diag}(e^{2\pi i\over k},
e^{4\pi i\over k},\cdots,e^{2\pi i(k-1)\over k},1).}
The expectation value \bexpv\ breaks the gauge symmetry
from $SU(k)$ to $U(1)^{k-1}$. The off-diagonal components of
the matrices $A$, $B$, etc, which correspond to D-strings
stretched between the fivebranes, get a mass proportional to
\eqn\mmww{M_W=r_0M_s^2/g_s^{({\rm far})},}
where $g_s^{({\rm far})}$ is the value of the string coupling far
from the fivebranes.

To decouple the fivebranes from the bulk, we consider the double scaling
limit \GiveonPX
\eqn\doublesc{r_0, g_s^{({\rm far})}\to 0;\;\;\;
{r_0\over g_s^{({\rm far})}l_s}= {M_W\over M_s}={\rm fixed}.}
In this limit, fundamental strings propagating in the vicinity of the
fivebranes see the ``near-horizon geometry''
\eqn\slsu{ \IR^{5,1}\times \left({SL(2,\IR)_k\over U(1)} \times
{SU(2)_k\over U(1)}\right)/\Z_k.}
The geometry \slsu\ is a cut-off version of \chsmetr; it reduces
to \chsmetr\ far from the tip of the cigar. It was mentioned
above that the brane configuration \Bj\ spontaneously breaks
the rotation symmetry about the fivebranes from $SO(4)$ to
$SO(2)\times \Z_k$. In the background \slsu\ this corresponds
to the fact that while the asymptotic large $\phi$ geometry has
an $SU(2)_L\times SU(2)_R$ symmetry, the full background
\slsu\ is only invariant under an $SO(2)\times \Z_k$ subgroup.
The $SO(2)$ corresponds to translations of $Y$, the angular coordinate
around the cigar. The $\Z_k$ is a ``quantum symmetry'' of the $\Z_k$
orbifold in \slsu. Note that, as usual in holography, a spontaneously
broken  symmetry in the gauge theory corresponds in the bulk description
to a symmetry that is preserved near the boundary at $\phi=\infty$, but is
broken at finite $\phi$.

The string coupling at the tip of the cigar in \slsu\ is:
\eqn\gstip{g_s^{({\rm tip})}\simeq M_s/M_W.}
In order for perturbative string theory in the background \slsu\
to be reliable, $g_s^{({\rm tip})}$ has to be small. Indeed, the
non-perturbative states corresponding to D-strings stretched
between the fivebranes correspond in the geometry \slsu\ to
D-branes localized near the tip of the cigar. Their mass is thus
proportional to $M_s/g_s^{({\rm tip})}=M_W$, and requiring
that they are much heavier than perturbative string excitations
leads to the condition
\eqn\mws{M_W\gg M_s.}
Our computations below will be done in this limit, which was
referred to in \refs{\GiveonPX,\GiveonTQ} as Double Scaled
Little String Theory (DSLST).

The $\Z_k$ orbifold acts on the SCFTs on the cigar 
and on $SU(2)/U(1)$ (which is an $\NN=2$ minimal model) in
\slsu\ as follows. The compact coordinate on the cigar, $Y$, lives on
a circle whose asymptotic radius is $R=\sqrt{2k}$ (see \S3). The orbifold
\slsu\ acts as a translation, $Y\to Y+{2\pi R\over k}$ (or, alternatively,
a rotation of the cigar by the angle $2\pi/k$). Clearly this
generates a $\Z_k$ action on the cigar. The $\NN=2$ minimal model
$SU(2)_k/U(1)$ has a discrete $\Z_k$ symmetry, and it is the product 
of this symmetry with the $Y$-translation mentioned above
that is gauged in \slsu. One important effect of orbifolding by $\Z_k$
is that the twisted sectors contain states with fractional winding
around the cigar, $w\in \Z/k$. The fractional part of $w$ is the
$\Z_k$ charge.

Another way of presenting the double scaling limit \doublesc\ corresponds
to the description of the near horizon geometry of the separated fivebranes
in terms of $\NN=2$ Liouville theory \ntwoliouv. The radius of the circle
on which the fivebranes are placed is related to the $\NN=2$ Liouville
coupling via the relation \GiveonPX\ $r_0/l_s=|\mu|^{1\over k}$. 
Thus, we see that $M_W$ scales like
\eqn\mwsc{M_W\sim |\mu|^{1\over k}.}
This relation will be used below for determining the dependence of amplitudes
on $M_W$, via KPZ scaling.

Since the DSLST background \slsu\ is a special case of \rrpphh, we can
use the results of \S2, \S3 to analyze it. In particular, we can construct
physical vertex operators as in \S3, and study the analytic structure of
Green functions using the results of \S2. The main focus of our discussion
will be on the behavior of these Green functions at low energies, \ie\ in
situations where all the kinematic invariants $p_i\cdot p_j$ (including those
with $i=j$) are much smaller than $M_s^2$.

The low energy limit of the fivebrane theory at the point \bexpv\ on its
moduli space is expected to be a
$U(1)^{k-1}$ gauge theory with sixteen supercharges. Thus, one might 
expect the physical vertex operators in string theory on \slsu\ to 
reduce at low energies to local operators in that gauge theory, 
and the correlation functions of these vertex operators to reduce 
to off-shell Green functions in the gauge theory. Of course, since 
the gauge theory is non-renormalizable, we expect its Lagrangian to 
include  high dimension operators, such as ${\rm tr}(F^{2n})$ with $n>1$ 
(appropriately supersymmetrized).

We will see that the actual situation is more subtle. The spectrum of
massless states is indeed in one to one correspondence with states in
the $U(1)^{k-1}$ gauge theory with sixteen supercharges. The residues
of LSZ poles as a function of the external momenta agree with low
energy expectations as well, \ie\ they can be obtained from a
gauge-invariant effective action which has a power series expansion in
local gauge-invariant fields, when we assume that the string
theory vertex operators in the background \slsu\ reduce at low
energies to particular gauge-invariant operators in the field theory.

However, as we saw in \S2, the off-shell Green functions obtained from
string theory have some additional singularities associated with bulk
dynamics. We will see that some of these singularities influence the
low energy behavior of correlation functions and they cannot be
described by an effective action written purely in terms of the low
energy interpolating fields. Thus, the statement that the low energy
theory on $k$ $NS5$-brane in type IIB string theory along its Coulomb
branch is a $U(1)^{k-1}$ gauge theory is not quite accurate off-shell,
even at arbitrarily large distances.

To make the above discussion more concrete, we will study below a
particular class of string theory observables, and try to match their
correlation functions to those of the corresponding operators in the
low energy gauge theory.

The following non-normalizable vertex operators in the fivebrane
background \chsmetr\ were identified (at low energies) 
with operators in the low energy gauge theory in \refs{\AharonyUB,\AharonyVK}:
\eqn\scalars{e^{-\varphi-\bar\varphi}(\psi\bar\psi\Phi_j^{(su)})_{j+1;m,\bar m}
e^{Q\tilde j\phi}e^{ip\cdot x}\leftrightarrow
\ttr(X^{i_1}X^{i_2}\cdots X^{i_{2j+2}}).}  
The notation in \scalars\ is as follows. $\psi^a$, with $a=3,\pm$, are three
free fermions which transform in the adjoint of $SU(2)_L$ (and
similarly for $\bar\psi^{\bar a}$). The supersymmetric level $k$ $SU(2)$
WZW theory 
includes these three fermions plus a bosonic WZW model of level $(k-2)$.
The operators $\Phi_{j;m,\bar m}^{(su)}$
(with $2j=0,1,\cdots, k-2; m,\bar m=-j,-j+1,\cdots,j$) are primaries of the
bosonic $SU(2)_{k-2}$ WZW model, whose dimension is
\eqn\dimsutwo{\Delta^{(su)}_j=\bar\Delta^{(su)}_j={j(j+1)\over k}~.}
The notation $(\psi\bar\psi\Phi_j^{(su)})_{j+1;m,\bar m}$
means that we are coupling the fermions and bosons into a primary
of spin $j+1$ and $(J_3^{\rm (tot)},\bar J_3^{\rm (tot)})=(m,\bar m)$ in
the supersymmetric $SU(2)_k$ WZW model.
For example, for the special case $m=j+1$ that will be useful below, one
has (in a natural overall normalization):
\eqn\explexp{\eqalign{
(\psi\bar\psi\Phi_j^{(su)})_{j+1;j+1,\bar m}
={1\over\sqrt{(2j+1)(2j+2)}}\psi^+[
&\sqrt{(j+\bar m)(j+\bar m+1)}\bar\psi^+\Phi_{j;j,\bar m-1}^{(su)}+\cr
\sqrt{2(j+\bar m+1)(j-\bar m+1)}\bar\psi^3\Phi_{j;j,\bar m}^{(su)}+
&\sqrt{(j-\bar m)(j-\bar m+1)}\bar\psi^-\Phi_{j;j,\bar m+1}^{(su)}].
\cr}}
The physical state condition \massshsusy\ requires that the vertex
operator \scalars\ satisfy
\eqn\mamama{Q^2(\tilde j-j)(\tilde j+j+1)=p^2~.}
On the right hand side of \scalars, $X^i$ with $i=6,7,8,9$
are the four scalar fields in the adjoint of $SU(k)$ which 
parametrize the locations of the fivebranes in the transverse
directions. To match to the representation of $SO(4)\simeq
SU(2)_L\times SU(2)_R$ that appears on the left hand side, 
one must restrict to symmetric traceless tensors in $(i_1,i_2,
\cdots, i_{2j+2})$.

The notation $\ttr$ refers to the fact that the gauge theory operator
which appears on the right-hand side of \scalars\ is a combination of
single and multi-trace operators.  We will normalize $\ttr$ such that
the single-trace component has a coefficient equal to one.  The relative
normalization between the left and right hand sides of \scalars\ will
be discussed in section 5.

The fact that single string vertex operators correspond to 
combinations of single and multi-trace operators is expected 
to be generic in holographic dualities, but the precise combinations
are in general unknown. The identification \scalars\ is based
on the fact that the two sides transform in the same chiral
representation of the supersymmetry algebra, but this does not
enable us to distinguish the single-trace from the multi-trace
operators (which transform in the same way under supersymmetry).
Interestingly, in the case of $NS5$-branes, we will be able to
determine the precise combination of multi-trace operators in
the gauge theory that corresponds to the string theory vertex
operators (in \S6).

We will now further restrict the discussion to the operators
\eqn\aba{ \OO_{j+1-\bar m,j+1+\bar m}(p_{\mu})\equiv
e^{-\varphi-\bar\varphi}(\psi\bar\psi\Phi_j^{(su)})_{j+1;j+1,\bar m}
e^{Q{\tilde j}\phi}e^{i p_\mu x^\mu}
\leftrightarrow\ttr(A^{j+1-\bar m}B^{j+1+\bar m})(p_\mu)~.
}
The identification with the gauge theory operators in \aba\ relies on an
identification of the rotation groups $SO(2)_A$, $SO(2)_B$ with particular
subgroups of $SU(2)_L\times SU(2)_R$ in the geometry \chsmetr. This
identification and other aspects of the operator maps are discussed in
more detail in \AharonyVK.

So far we discussed the form of the vertex operators in the unresolved CHS
geometry \chsmetr, or equivalently in the full resolved geometry \slsu\
but far from the tip of the cigar. In order to compute correlation functions,
we will need the form of the full vertex operators in the coset theory. It is
convenient to discuss separately the cases $|\bar m|=j+1$ and
$|\bar m|\le j$.

For $\bar m=j+1$ we expect \aba\ that
$\OO_{0,2j+2}\sim \ttr(B^{2j+2})$, and we have
\eqn\formb{\OO_{0,2j+2}(p_{\mu})=e^{-\varphi-\bar\varphi}\psi^+\bar\psi^+
\Phi_{j;j,j}^{(su)}e^{Q{\tilde j}\phi}e^{i p_\mu x^\mu}.}
In order to write $\OO_{0,2j+2}$ as a vertex operator in the full
geometry \slsu, we would like to decompose the $SU(2)$ WZW part of
the vertex operator \formb\ into its $U(1)$ and  $SU(2)\over U(1)$
components. It is not difficult to show that
\eqn\decompsu{\psi^+\bar\psi^+\Phi_{j;j,j}^{(su)}=e^{iQ(j+1)(Y-\bar Y)}
V^{(su,susy)}_{{k\over2}-j-1;-{k\over2}+j+1,-{k\over2}+j+1}.}
Here, $Y$, $\bar Y$ are the left and right moving parts of the worldsheet
field corresponding to the compact coordinate around the cigar. In particular,
we see that for generic $j$, the vertex operator \decompsu\ carries fractional
winding around the circle labeled by $Y$. As mentioned earlier, this is
possible due to the $\Z_k$ orbifold in \slsu.

The operators $V^{(su,susy)}_{j;m,\bar m}$ can be defined 
(as in the $SL(2)$ discussion above) by starting 
with the (supersymmetric) $SU(2)$ vertex operator $\Phi^{(su)}_{j;m,\bar m}$ 
and removing from it the $U(1)$ part. In particular, 
$V^{(su,susy)}_{j;m,\bar m}$ has dimension and $R$-charge
\eqn\drsu{\eqalign{
&\Delta_{j;m}^{(su)}={j(j+1)-m^2\over k};\;\;\;
\bar\Delta_{j;\bar m}^{(su)}={j(j+1)-\bar m^2\over k},\cr
&R_m^{(su)}=-{2m\over k};\;\;\;\bar R_{\bar m}^{(su)}=-{2\bar m\over k}.\cr
}}
Plugging \decompsu\ in \formb, and using \asymsusy, we find that
\eqn\dsbt{\OO_{0,2j+2}(p_\mu)=e^{-\varphi-\bar\varphi}
V^{(su,susy)}_{{k\over2}-j-1;-{k\over2}+j+1,-{k\over2}+j+1}
V^{(sl,susy)}_{\tilde j;j+1,j+1}e^{ip_\mu x^\mu}.}
For the case $\bar m=-(j+1)$ a similar discussion leads to the result
 \eqn\dsat{\OO_{2j+2,0}(p_\mu)=e^{-\varphi-\bar\varphi}
V^{(su,susy)}_{{k\over2}-j-1;-{k\over2}+j+1,{k\over2}-j-1}
V^{(sl,susy)}_{\tilde j;j+1,-(j+1)}e^{ip_\mu x^\mu}.}
In the gauge theory, the operator \dsat\ should correspond to
$\ttr(A^{2j+2})$ \aba.

Next we turn to vertex operators  \aba\ with $-j\le \bar m\le j$. In general,
there are now three terms in the expansion \explexp. The terms proportional
to $\bar\psi^\pm$ in \explexp\ can be described as follows. 
Consider the operator
\eqn\opslsufull{\psi^+\bar\psi^\pm \Phi^{(su)}_{j;j,\bar m\mp1}
\Phi^{(sl)}_{\tilde j;j+1,\bar m}}
in SCFT on $SU(2)\times SL(2,\IR)$. This operator  has the property
that if  we decompose it under $[{SU(2)\over U(1)}\times {SL(2,\IR)\over U(1)}]
\times [U(1)^2]$, the ${SU(2)\over U(1)}\times {SL(2,\IR)\over U(1)}$ component
is precisely the operator we are interested in, 
whose asymptotic form in the CHS geometry is
\eqn\asymff{\psi^+\bar\psi^\pm \Phi^{(su)}_{j;j,\bar m\mp1}e^{Q\tilde j\phi},}
while the $U(1)^2$ component has dimension zero and has the form
$e^{iQmZ}$ with $Z$ a null scalar field (a combination of the scalar 
fields associated with the two $U(1)$'s). The only effect of the 
$U(1)^2$ component is to impose the conservation of $U(1)^2$ charges, 
which we will require in our calculations anyway. Thus, the $U(1)^2$ 
part of the vertex operator \opslsufull\ plays no role in the calculations.

As we saw in \S2, the analytic structure of amplitudes is largely determined
by the $SL(2)/U(1)$ part of the vertex operators. We will see later (in \S5.3)
that the $SL(2)/U(1)$ part of \opslsufull\ does not lead to singularities at 
low energies.
Thus, the contributions \opslsufull, \asymff\ to the vertex operator \aba\
can in fact be neglected at low energies.

The remaining contribution to the vertex operator \aba\ is proportional to
\eqn\lastcont{\psi^+\bar\psi^3\Phi^{(su)}_{j;j,\bar m}e^{Q\tilde j\phi}.}
In order to write it in a form similar to that given above for the other terms,
we would like to decompose \lastcont\ into its $SL(2,\IR)/U(1)$ and
$SU(2)/U(1)$ components, and then lift the results to $SL(2,\IR)\times SU(2)$.
The first thing to note is that $\bar\psi^3$ belongs to the $SL(2,\IR)$
component. It is in fact identical to $\bar\psi_Y$ discussed after equation
\asymsusy. In the notation of \S3, we can write the right-moving,
$SL(2,\IR)/U(1)$ part of the vertex operator \lastcont\ as
\eqn\rightvv{\bar\psi^3 e^{Q(\tilde j\phi-i\bar m \bar Y)}=
\bar\psi_Y e^{Q(\tilde j\phi-i\bar m \bar Y)}.}
As explained in \S3, this operator is a descendant of the $\NN=2$
primary $e^{Q(\tilde j\phi-i\bar m \bar Y)}$. Using the form
of the $\NN=2$ superconformal generators far from the tip of the cigar,
\eqn\gpm{\eqalign{
\bar G^+=&(\bar\psi_\phi+i\bar\psi_Y)\partial(\phi+i\bar Y)+
Q\partial(\bar\psi_\phi+i\bar\psi_Y),\cr
\bar G^-=&(\bar \psi_\phi-i\bar \psi_Y)\partial(\phi-i\bar Y)+
Q\partial(\bar\psi_\phi-i\bar\psi_Y),\cr
}}
one can show that
\eqn\respsi{\bar\psi_Y e^{Q(\tilde j\phi+i(mY-\bar m \bar Y))}\sim
\left(
{1\over \tilde j-\bar m} \bar G_{-{1\over2}}^+-
{1\over \tilde j+\bar m} \bar G_{-{1\over2}}^-
\right)V^{(sl,susy)}_{\tilde j;m,\bar m}.}
A natural question at this point is how can the operator on the left hand side
of \respsi, which seems to be regular as $\tilde j\to \pm {\bar m}$, 
be the same as
the operator on the right hand side, which diverges in this limit. The answer
is that on the left hand side we only wrote the leading, non-normalizable,
contribution to the operator which is defined algebraically on the right 
hand side. That contribution is indeed finite in the limit 
$\tilde j\to \pm {\bar m}$ also on the right-hand side (one can show that
${\bar G}_{-{1\over 2}}^+$ annihilates the non-normalizable component of
$V_{\bar m;m,\bar m}$ and that ${\bar G}_{-{1\over 2}}^-$ annihilates the
non-normalizable component of $V_{-{\bar m};m,\bar m}$). 
The divergence of the operator is reflected in the  divergence of the leading
normalizable contribution, \ie\ the poles as $\tilde j\to \pm {\bar m}$
are potential LSZ poles. We will see later (in \S5.3)
that a more careful analysis of
these poles leads to a picture which is in agreement with expectations.

One can also  write an operator in SCFT on $SL(2,\IR)\times SU(2)$
which has the property that removing from it the $U(1)^2$ components
leads to the operator \lastcont:
\eqn\lifttt{\psi^+
\left({1\over \tilde j-\bar m} \bar G_{-{1\over2}}^+-
{1\over \tilde j+\bar m} \bar G_{-{1\over2}}^-\right)
\Phi^{(su)}_{j;j,\bar m}\Phi^{(sl)}_{\tilde j;j+1,\bar m}.}
Here $\bar G_{-{1\over2}}^\pm$ are operators defined in the
full $SL(2,\IR)$ CFT, but they commute with the $U(1)$. Their
description in terms of $SL(2)$ currents is reviewed in appendix A.

This concludes our brief review of six dimensional DSLST. In the
next subsection we will describe the LSZ poles in this theory and
compare the resulting pattern with the low-energy gauge theory.
In the following two sections, we will use our results to
analyze some correlation functions in this theory, and in particular
their analytic structure at small momenta.

\subsec{LSZ poles}

In this subsection we analyze the LSZ poles associated with the operators
\aba, and compare the low-energy pole structure with our expectations from
the gauge theory. In the gauge theory, operators are expected to exhibit
poles only if they can create single-particle states. The operators
\scalars, \aba\ obviously couple to a state with $2j+2$ massless particles.
However, if $2j+1$ of the $2j+2$ fields appearing in the operator are
$B$'s (or $B^*$'s), then at the point \bexpv\ in the moduli space we
can replace these $2j+1$ fields by their vacuum expectation value (VEV), 
and obtain an operator that can create
from the vacuum a single-particle state involving the remaining field.
Thus, we expect to find LSZ poles for the operators \aba\ with
${\bar m}=j,j+1$, corresponding to the gauge theory operators
$\ttr(A B^{2j+1})$, $\ttr(B^{2j+2})$, respectively, but not for the other
values of $\bar m$.

Let us start with the case $\bar m=j+1$, corresponding to the
vertex operator \dsbt. The analysis of \S2 and \S3 shows
that this operator has LSZ poles at $\tj=j,j-1,j-2,\cdots>-1/2$.
The mass-shell condition \mamama\ maps these to
momentum-space poles occurring at (for $\alpha'=2$)
\eqn\mompoles{p^2=0,\qquad p^2=-{4j\over k},\qquad 
p^2=-{4(2j-1)\over k},\qquad \cdots,}
respectively.
Thus, we find that this operator has a massless LSZ pole for all
$j\geq 0$, in agreement with the field theory expectations described
above.

For the case $\bar m=-(j+1)$, it is clear from \dsat\ and from the discussion
of \S2, \S3 that the operator $\OO_{2j+2,0}$ has no LSZ poles.
Again, the absence of a massless LSZ pole is consistent with the fact
that the corresponding gauge theory operator, $\ttr(A^{2j+2})$, does not
couple to single particle massless states. Interestingly, we find that it does
not couple to massive single particle states localized near the tip of 
the cigar either.

It remains to discuss the case $|\bar m|\le j$, for which the operators
in question are obtained from \lifttt\ by removing the $U(1)^2$ part.
For $|\bar m|<j$ the situation is simple. Massless poles would again
have to appear at $\tilde j=j$ \mamama, but this case does not belong
to the set \poles, \polesinf, so these operators do not create massless
states when acting on the vacuum, although some of them do create
massive states. Again, this is compatible with the gauge theory expectations,
since all these operators contain at least two $A$'s and thus do not couple
to single particle massless states.

For $|\bar m|=j$ the situation is slightly more subtle, because of the
explicit factors of $\tilde j\pm \bar m$ in the denominator of \lifttt,
which as we explained above, can potentially lead to (massless) LSZ
poles. Repeating the analysis of \S2 we find\foot{There are two ways to
repeat this analysis here. One possibility is to decompose the operators
of the SCFT on $SL(2)/U(1)$ in terms of operators in a {\it bosonic} $SL(2)$
theory, and then the analysis of \S2 may be applied directly and 
it is straightforward to see that a massless pole originating by taking
$x\to 0,\infty$ in an $n$ point function appears for an operator in
\aba\ iff it should appear in the low energy QFT via the LSZ reduction.
Alternatively, we can continue working using the supersymmetric LST,
by writing the operators $\bar G^{\pm}$ in terms of $SL(2)$ currents
${\bar J}^{\pm}$ which act on the $x$-variables in a known way (as an
$\bar x$ derivative, see \KutasovXU\ and references therein), and
analyzing whether the regions $x\to 0,\infty$ give poles or not. 
This leads, of course, to the same answers.} 
that while the pole at
$\tilde j=\bar m=j$ does create a massless particle from the vacuum,
and thus is an LSZ pole, the pole at $\tilde j=-\bar m=j$ does not correspond
to an LSZ pole, and is analogous to the poles associated with $\Gamma(-2j-1)$
in the reflection coefficient \reflrel. This is again consistent with the
gauge theory expectations, since the operators \aba\ with $\bar m=j$
correspond to $\ttr(AB^{2j+1})$ and thus should couple to single particle
states, while those with $\bar m=-j$ correspond to $\ttr(A^{2j+1}B)$, and should not.

Thus, for all the operators discussed here we find a precise agreement
between the LSZ poles found in string theory and our low-energy
field theory expectations. We have verified that this is true also for
some additional operators (including Ramond-Ramond (RR) 
sector operators \AharonyVK).

\newsec{Some examples of correlation functions}

In this section we will study some simple examples of correlation
functions of operators in six dimensional DSLST, focusing
on their low energy behavior. Some puzzles regarding these
correlation functions were encountered in \GiveonTQ. We
will use the discussion of the previous sections to clarify 
their analytic structure, and in particular
to resolve the aforementioned puzzles.

We will see that, even at low momenta, some of the poles exhibited
by these amplitudes are due to dynamics that is not captured by the
$U(1)^{k-1}$ IR free gauge theory that one expects to describe $k$
separated type IIB NS fivebranes at long distances. In the first
three subsections we discuss examples of two-point functions, and
show that they generally receive contributions both from bulk poles
and from the LSZ poles discussed in \S4.2. In \S5.4 we discuss
the double pole terms in a specific non-trivial three-point function.
In \S5.5 we review an S-matrix computation performed in \AharonyVK,
and in \S5.6 we comment on some winding-number-violating correlation 
functions.

\subsec{$\langle \ttr(B^n)  \ttr((B^*)^n) \rangle$}

The string theory vertex operator with the quantum numbers of
${\rm tr}(B^n)$ is $\OO_{0,n}$ \formb, \dsbt. As discussed
earlier, $\OO_{0,n}$ actually corresponds (in the low-energy gauge
theory) to a mixture
of single and multi-trace operators,
\eqn\onbtilde{\OO_{0,n}=C_{n,k}{1\over n}
\ttr(B^n),}
where $C_{n,k}$ is a normalization constant to be determined, and the
operator $\ttr(B^n)$ is a specific linear combination of
$\tr(B^n)$ and multi-trace operators, which will be determined in
the next section.

In the $U(1)^{k-1}$ gauge theory at the point \bexpv\ in its moduli
space, the leading (in $1/M_W$)
contribution to the two-point function of $\ttr(B^n)$ comes
from the single trace term:
\eqn\leadtwob{\langle {1\over n}\ttr(B^n) {1\over n}\ttr((B^*)^n)\rangle
=\langle {1\over n}{\rm tr}(B^n) {1\over n}{\rm tr}((B^*)^n)\rangle \simeq
{k M_W^{2n-2}\over p_\mu^2},}
where we have contracted a single $B$ with a single $B^*$,
and replaced the remaining $B$'s and $B^*$'s by their
VEV \bexpv\ (setting $M_s=1$ in the process). The multi-trace
terms in $\ttr(B^n)$ do not contribute at this order in the $1/M_W$
expansion, since they involve factors of $\vev{\tr(B^l)}$
with $0<l<n$, which vanish.

The string theory analog of \leadtwob\ involves a two-point
function of the operators $\OO_{0,2j+2}$, \formb, \dsbt. For small
$p_\mu^2$, the mass shell condition \mamama\ implies that
\eqn\masssmall{\tilde j-j\simeq {k\over2} {p_\mu^2\over 2j+1}.}
The two-point function of $\OO_{0,2j+2}$ is thus given by the
product of the $SU(2)/U(1)$ and $SL(2,\IR)/U(1)$ contributions (the
$\IR^{5,1}$ contribution is equal to one). We will normalize the
$SU(2)/U(1)$ vertex operators $V^{(su,susy)}$ in \dsbt\ such that 
their two-point function is also equal to one. Thus, we have\foot{The
operator corresponding to the complex conjugate of $\ttr B^{2j+2}$ is 
$\OO_{0,-(2j+2)}\sim \ttr(B^*)^{2j+2}$.}
\eqn\twoptb{\langle \OO_{0,2j+2}(p_\mu) \OO_{0,-(2j+2)}(-p_\mu)\rangle=
{1\over g_s^2}
\langle V^{(sl,susy)}_{\tilde j;j+1,j+1}
V^{(sl,susy)}_{\tilde j;-(j+1),-(j+1)}\rangle.
}
We have inserted a factor of ${1\over g_s^2}$ to account for the fact that
we are computing an amplitude on the sphere. 
$g_s$ is proportional to the string coupling
at the tip of the cigar, \gstip.
The general formula for the two-point function in $SL(2,\IR)/U(1)$ appears
in appendix A. It indeed has a pole at $\tilde j=j$ (or
$p_\mu^2=0$, \masssmall), near which it behaves as
\eqn\smallpbt{\langle \OO_{0,2j+2}(p_\mu) \OO_{0,-(2j+2)}(-p_\mu)\rangle
          \simeq {1\over g_s^2}\left({2j+1\over k}\right)^2{1\over p_\mu^2}.}
Combining \onbtilde, \leadtwob\ and \smallpbt\ we seem to conclude that
\eqn\cnk{C_{2j+2,k}^2 k M_W^{4j+2}={1\over g_s^2}\left({2j+1\over k}\right)^2.}
Comparing to \gstip\ we see that the 
solution of this equation is 
\eqn\gscnk{\eqalign{
{1\over g_s^2}=&~ C(k) M_W^2,\cr
C_{2j+2,k}=&~{2j+1\over {k M_W^{2j}}} \sqrt{C(k)\over k}\cr
}}
for some function $C(k)$. This function can be determined by a careful
computation of three-point functions, but we will not need its explicit form
here.

Actually, the discussion above (and a similar discussion which appeared
in \S5.1 of \AharonyVK) contains a subtle flaw, which turns out to be
relatively benign in this case, but plays an important role in understanding
other correlation functions. The issue is whether it is really the case that
the behavior of the correlation function \smallpbt\ near the pole at 
$p_\mu^2=0$
is entirely due to the low energy gauge theory contribution described above,
or whether it receives additional contributions.

To answer this question using the techniques of the previous sections
we need to take a closer look at the two-point function \twoptb. The
analytic structure as a function of $\tilde j$ is due to the behavior of
the integral (see \useint)
\eqn\xinteg{\int d^2 x |x|^{2(\tilde j+j+1)}|1-x|^{-4(\tilde j+1)}=
\pi \gamma(-2\tilde j-1)\gamma(\tilde j-j)\gamma(\tilde j+j+2),}
where $\gamma(x) \equiv \Gamma(x) / \Gamma(1-x)$.
Taking $\tilde j=j+\epsilon$ and studying the right-hand side of \xinteg\
in the limit $\epsilon\to 0$, one finds that the integral behaves
like $\pi/2\epsilon$; this was used to arrive at \smallpbt.
As we saw in the previous sections, the contribution to the residue
of the pole at   $\epsilon\to 0$ due to a normalizable
state created from the vacuum by the operator \dsbt\ comes
from the region near $x=0$ or $x=\infty$ in the integral \xinteg. In our
case, the relevant region is $x\to\infty$, where one has (as $\epsilon \to 0$)
\eqn\intinf{\int^{\infty} d^2x |x|^{2(j-\tilde j-1)}\simeq {\pi\over\epsilon}.}
Thus, we see that the contribution of the normalizable discrete
state leads to a pole with a residue that is twice as large as the 
one computed for the full integral.

What cancels half of the contribution of this normalizable state?
On the level of the integral \xinteg, the answer is clear. As
$\tilde j\to j$, in addition to the contribution from $x\to\infty$,
there is a further divergent contribution coming from $x\to 1$, which
also gives a pole (since $2j$ is a non-negative integer). This contribution
can be computed by methods similar to the ones we used in \S2,
and gives $-\pi/2\epsilon$. Together, the two contributions
account for the behavior of the full integral.

The physical interpretation of the two terms is clear as well, following
the discussion of the previous sections. The contribution to the residue of
the pole from $x\to\infty$ should be interpreted as due to the overlap of
the string theory vertex operator \dsbt\ with the state created by the
operator $\ttr(B^{2j+2})$
in the low energy gauge theory. This contribution is always positive, as
required by unitarity. The contribution from $x\to 1$ (corresponding to
$x_1 \to x_2$ before we rescale the $x$'s, as in \densgaaa) has a non-gauge
theoretic origin. In the deformed CHS geometry \slsu\ it is associated with
processes that occur in the bulk of the cigar, very far from the tip. Indeed,
since the pole occurs when $2j$ is an integer, the two-point function in
question satisfies the condition for bulk poles \finalpoles\ (with $l=n=2$
and $g=n_2=0$).

We see that the correlation function \twoptb\ receives two kinds of large
contributions at low energies. One can be interpreted as due to the dynamics
of the low energy $U(1)^{k-1}$ gauge theory, and comes from the region
near the tip of the cigar; the other is due to dynamics in the bulk of
the cigar, and has a non-gauge theoretic origin. It is interesting that 
the bulk of the cigar can give a massless pole in the correlation function, 
even though all the physical states that live there have masses
obeying $m^2 \geq 1/k\alpha'$.
Obviously, such an effect could not happen in a standard field theory,
where a pole in the two-point function would necessarily be associated 
with creating a particle of the appropriate mass. The non-local nature
of LST seems to play an important role in making this possible here.

The above discussion implies that our analysis of the relation between
the vertex operators $\OO_{0,2j+2}$ and the gauge theory operators
$\ttr(B^{2j+2})$ must be modified slightly. In particular, the
right-hand side of equation \smallpbt\ must be multiplied by a factor
of two, since it should only contain the contribution to the residue
of the pole from $x\to\infty$. This factor of two will not play an
important role below, since we will not attempt to keep track of
numerical coefficients (but only of the $j$ dependence). It is
nevertheless interesting that the behavior of this amplitude at low
momenta is not fully accounted for by the $U(1)^{k-1}$ gauge theory
that lives on the separated fivebranes. We next turn to a more
dramatic manifestation of this phenomenon.

\subsec{$\langle \ttr(A^n) \ttr((A^*)^n)\rangle$}

In this subsection we will repeat the discussion of the previous
subsection for the operator $\OO_{n,0}$ \dsat, which has the quantum
numbers of ${\rm tr}(A^n)$ in the low energy gauge theory. Again, the
vertex operator $\OO_{n,0}$ actually corresponds to a mixture of
single and multi-trace operators in the gauge theory, but like in the
previous subsection, this will not play an important role in our
discussion.

Since the scalar field $A$ does not have an expectation value at the point
in moduli space where we are working, \bexpv, the gauge theory calculation
is even simpler than \leadtwob\ in this case. At order $M_W^{2n-2}$, the
two-point function $\langle \ttr(A^n)\ttr((A^*)^n)\rangle$ is exactly zero.
So are all other contributions to this two-point function that go like
$M_W^{2l}$, with $l>0$.

It is thus interesting to compute the two-point function of
$\OO_{n,0}$ in string theory. The fact that in the $U(1)^{k-1}$
gauge theory, the
contributions that go like $M_W^{2(n-1-l)}$ with $l=0,1,2,\cdots, n-2$
vanish, would lead one to expect that in string theory in the
background \slsu\ the genus $l<n-1$ contributions to the two-point
function of $\OO_{n,0}$ (at low energies) vanish as well. Any non-vanishing
contributions at low genus would have to have a non-gauge theoretic
origin.

The contribution of $\IR^{5,1}$ and $SU(2)/U(1)$ to the two-point function
of the operators \dsat\ is again equal to one in the conventional
choice of normalizations on the compact coset. The two-point function in
$SL(2,\IR)/U(1)$ can be read off the general formula in appendix A,
\eqn\sltwopt{
\langle V^{(sl,susy)}_{\tilde j;j+1,-(j+1)}
V^{(sl,susy)}_{\tilde j;-(j+1),j+1}\rangle
={2\tilde j+1\over k}\gamma(-2\tilde j-1){\Gamma^2(\tilde j-j)\over
\Gamma^2(-\tilde j-j-1)}.}
Using the physical state condition \masssmall\ we find that not only is
the tree level contribution to the two-point function \sltwopt\ non-zero,
it in fact has a pole at $p_\mu^2=0$ \GiveonTQ ,
\eqn\twopt{\langle \OO_{2j+2,0}(p_\mu) \OO_{-(2j+2),0}(-p_\mu)\rangle
\simeq {1\over g_s^2}(-1)^{2j}\left({2j+1\over k}\right)^2{1\over p_\mu^2}.}
The alternating sign of the residue makes it clear that this pole
cannot in general be interpreted as due to an on-shell one particle
state, which is just as well, since we know that the operator ${\rm
tr}(A^n)$ should not create such states. Indeed, 
the analysis of \S2 shows that
vertex operators such as \dsat\ do not have LSZ poles 
in their correlation functions.

To understand the physical interpretation of the pole \twopt\ we
go back to the representation of the two-point function \sltwopt\ in terms
of an integral over $x$,
\eqn\momxint{\int d^2x x^{\tilde j+j+1}\bar x^{\tilde j-j-1}
|1-x|^{-4(\tilde j+1)} =\pi\gamma(-2\tilde j-1){\Gamma^2(\tilde
j-j)\over\Gamma^2(-\tilde j-j-1)}.}  
This integral exhibits a pole as $\tilde j\to j$, and we would like to
understand its origin.  It is easy to see that the integral is well
behaved as $x\to 0,\infty$.  The divergence is in this case entirely
due to the behavior as $x\to 1$. As explained in the previous
sections, such divergences are due to bulk processes.

We conclude that for the operators $\OO_{n,0}$, which have the quantum
numbers of $\ttr(A^n)$, the full string theory tree-level two-point
function, which has a pole at vanishing $p_\mu^2$, is non-zero due to
effects that cannot be seen in the low energy $U(1)^{k-1}$ gauge theory 
of $k$ separated fivebranes.

\subsec{$\langle \ttr(A^nB^{2j+2-n}) \ttr ((A^*)^n(B^*)^{2j+2-n})\rangle$}

The vertex operators corresponding to gauge theory operators that
include both $A$ and $B$ are given by \aba, \explexp, \lifttt. For
$|\bar m|< j$ it is easy to check that the two-point function is finite
as $p^2\to 0$ (or $\tilde j\to j$). This leaves the cases $\bar m=\pm j$  
corresponding to the gauge theory operators $\OO_{1,2j+1}\sim\ttr(A B^{2j+1})$ 
and $\OO_{2j+1,1}\sim\ttr(A^{2j+1}B)$. 

Consider the case $\bar m=j$. As we saw in \S4, the vertex operator
$\OO_{1,2j+1}$ has two terms. One, whose asymptotic form far from
the tip of the cigar is given by \asymff, has the property that its 
$SL(2)/U(1)$ part is proportional to $V^{(sl,susy)}_{\tilde j,j+1;j}$. 
Using the results of appendix A for the two-point function, it is not 
difficult to see that this term has a two-point function that is 
regular as $p^2\to 0$. The other term, whose asymptotic form looks like
\lastcont, can be obtained from the $SU(2)\times SL(2)$ operator \lifttt\
by removing the $U(1)^2$ part, as explained in \S4. 

The two-point function of this operator is proportional to
\eqn\twolift{{1\over (\tilde j-j)(\tilde j+j)}
\langle V^{(sl,susy)}_{\tilde j,-j-1;-j}
\{\bar G^-_{-\half},\bar G^+_{-\half}\}
V^{(sl,susy)}_{\tilde j,j+1;j}\rangle.}
Using the fact that the anti-commutator 
$\{\bar G^-_{-\half},\bar G^+_{-\half}\}$
is proportional to ${\bar L}_{-1}$, which acts as a worldsheet derivative on
$\langle V^{(sl,susy)}_{\tilde j,-j-1;-j}
V^{(sl,susy)}_{\tilde j,j+1;j}\rangle$,
we find that \twolift\ is proportional to 
\eqn\finmmww{{\Delta_{\tilde j;j}\over (\tilde j-j)(\tilde j+j)}
\langle V^{(sl,susy)}_{\tilde j,-j-1;-j}
V^{(sl,susy)}_{\tilde j,j+1;j}\rangle,}
where $\Delta_{\tilde j;j}$ is the scaling dimension \susydimvjm.
In the limit $\tilde j\to j$, which corresponds via \masssmall\ to
$p^2\to 0$, we find a single pole, whose residue is readily computable.

In the case $\bar m=-j$ one again finds a pole in the two
point function, however, as we saw in \S4.2, it should be thought
of as a bulk pole, and not an LSZ pole.

\subsec{$\langle \ttr(F_{\mu\nu}B^{n_1})
\ttr(F_{\mu\nu}B^{n_2}) \ttr((B^*)^{n_1+n_2})\rangle$}

As another test of our techniques we next turn to a correlation
function involving the Ramond-Ramond vertex operators $\OO_n^+$
discussed in \AharonyVK. These operators have the quantum numbers
of
\eqn\fbop{\OO_n^+\sim \xi^{\mu\nu}\ttr(F_{\mu\nu} B^n),}
where, again, the left-hand side is a vertex operator in the 
background \slsu, while the right-hand side is a combination
of single and multi-trace operators in the low-energy gauge theory.
As in the previous subsections, the multi-trace
components will not play a role in our calculations.

Consider first the leading contribution to the connected three-point
function in the low energy gauge theory:
\eqn\threefree{\eqalign{
\xi_{\mu\nu}^{(1)} & \xi_{\mu'\nu'}^{(2)}
\langle {\rm tr}(F^{\mu\nu} B^{n_1})(p_1)
{\rm tr}(F^{\mu'\nu'} B^{n_2})(p_2)
{1\over n_1+n_2}{\rm tr}((B^*)^{n_1+n_2})(p_3)\rangle=\cr
&\delta^6(p_1+p_2+p_3)k M_W^{2(n_1+n_2-1)}
{\xi_{\mu\nu}^{(1)}\xi_{\mu'\nu'}^{(2)}
\over p_3^2}
\left[n_2{p_1^\mu p_1^{\mu'}\eta^{\nu\nu'}
\pm (\mu\leftrightarrow\nu,\mu'\leftrightarrow\nu')
\over p_1^2}+(1\leftrightarrow 2)\right].\cr}}
It comes from a contraction of the two $F$'s, and a second contraction
of one of the $B$'s with one of the $(B^*)$'s.

We would like to compute the same object in string theory,
and in particular reproduce the pole structure and the dependence
on $n_1$, $n_2$ in \threefree. The vertex operators 
$\OO_n^+$ in the $(-1/2,-1/2)$ picture are given in equation (5.4) in
\AharonyVK. Substituting them into the three
point function 
\eqn\threeooo{\langle \OO_{2j_1+1}^+(p_1) \OO_{2j_2+1}^+(p_2)
\OO_{0,-2(j_1+j_2)-2}(p_3)\rangle}
one finds two terms. One goes like $p_1^\rho p_1^\sigma$;
the other, like $p_2^\rho p_2^\sigma$. This is consistent
with the gauge theory answer \threefree, which has two terms
related by interchanging $1\leftrightarrow 2$.
Thus, we focus on the term that goes like $p_1^\rho p_1^\sigma$.

In this term one has \AharonyVK\ asymptotically
\eqn\formver{\eqalign{
&\OO_{2j_1+1}^+  \simeq
{\xi^{(1)}_{\mu\nu}\gamma_{a\dot a}^{\mu\nu}
e^{-{1\over2}(\varphi+\bar\varphi)+{i\over2}(H+\bar H)}\over 
Q^2(j_1+\tilde j_1+1)^2}
(\gamma_\rho)^{\dot c}_a p_1^\rho S_{\dot c}
(\gamma_\sigma)^{c}_{\dot a} p_1^\sigma \bar S_c
e^{-{i\over2}(H'+\bar H')}\Phi^{(su)}_{j_1;j_1,j_1}
e^{Q\tilde j_1\phi+ip_1\cdot x},\cr
&\OO_{2j_2+1}^+ \simeq \xi^{(2)}_{\mu'\nu'}\gamma_{b\dot b}^{\mu'\nu'}
e^{-{1\over2}(\varphi+\bar\varphi)+{i\over2}(H+\bar H+H'+\bar H')}
S_b\bar S_{\dot b}\Phi^{(su)}_{j_2;j_2,j_2}
e^{Q\tilde j_2\phi+ip_2\cdot x},\cr
&\OO_{0,-2(j_1+j_2)-2}\simeq e^{-\varphi-\bar\varphi-i(H+\bar H)}
\Phi^{(su)}_{j_1+j_2;-j_1-j_2,-j_1-j_2}
e^{Q\tilde j_3\phi+ip_3\cdot x}.\cr
}}
The expectation value \threeooo\ factorizes into the
contributions of the ghosts, the $\IR^{5,1}$ CFT and the 
$\left({SU(2)\over U(1)}\times {SL(2,\IR)\over U(1)} \right)/\Z_k$ CFT.
The ghost contribution is equal to one. The 
$\IR^{5,1}$ contribution is similar to a computation performed 
in equation (5.9) of
\AharonyVK. It gives the correct
kinematic structure (compare to \threefree)\foot{As in \AharonyVK,
there is another
term in this contribution proportional to $p_1^2$, which turns out to
be analytic in all the momenta so it is interpreted as a contact-term
in space-time.}
\eqn\kinstr{{8\over Q^2(j_1+\tilde j_1+1)^2}
\xi_{\mu\nu}^{(1)}\xi_{\mu'\nu'}^{(2)}
\left(p_1^\mu p_1^{\mu'}\eta^{\nu\nu'} \pm
(\mu\leftrightarrow\nu,\mu'\leftrightarrow\nu')\right).
}
What remains is the three-point function
\eqn\remthree{\eqalign{
&\langle e^{{i\over2}(H+\bar H-H'-\bar H')}
\Phi_{j_1;j_1,j_1}^{(su)} e^{Q\tilde j_1\phi} \cdot
e^{{i\over2}(H+\bar H+H'+\bar H')}
\Phi_{j_2;j_2,j_2}^{(su)} e^{Q\tilde j_2\phi} \cdot\cr
&\qquad e^{-i(H+\bar H)}
\Phi_{j_1+j_2;-j_1-j_2,-j_1-j_2}^{(su)} e^{Q\tilde j_3\phi}
\rangle.\cr}}
The RR vertex operators that enter equation \remthree\ correspond
in the exact background \slsu\ to (see section 4.3 of \AharonyVK\ for a
discussion of these formulae and definitions of the operators appearing
in them)
\eqn\corrr{\eqalign{
&e^{{i\over2}(H+\bar H-H'-\bar H')}\Phi_{j_1;j_1,j_1}^{(su)}
e^{Q\tilde j_1\phi}\leftrightarrow
V^{(su,susy)}_{j_1;j_1,j_1}(RR,+)
V^{(sl,susy)}_{\tilde j_1;j_1+1,j_1+1}(RR,-),\cr
&e^{{i\over2}(H+\bar H+H'+\bar H')}\Phi_{j_2;j_2,j_2}^{(su)}
e^{Q\tilde j_2\phi}\leftrightarrow
V^{(su,susy)}_{j_2;j_2,j_2}(RR,+)V^{(sl,susy)}_{\tilde j_2;j_2,j_2}(RR,+).\cr
}}
Together with the form of the third operator in \remthree, given in \dsbt,
we conclude that the three-point function \remthree\ is given by the following
product of $SU(2)/U(1)$ and $SL(2,\IR)/U(1)$ three-point functions:
\eqn\exactthree{\eqalign{
&\langle V^{(su,susy)}_{j_1;j_1,j_1}(RR,+)V^{(su,susy)}_{j_2;j_2,j_2}(RR,+)
V^{(su,susy)}_{{k-2\over2}-j_1-j_2;{k-2\over2}-j_1-j_2,{k-2\over2}-j_1-j_2}
\rangle\times\cr
&\langle V^{(sl,susy)}_{\tilde j_1;j_1+1,j_1+1}(RR,-) 
V^{(sl,susy)}_{\tilde j_2;j_2,j_2}(RR,+)
V^{(sl,susy)}_{\tilde j_3;-j_1-j_2-1,-j_1-j_2-1}\rangle.\cr}}
The $SU(2)/U(1)$ correlator can be simplified by using the 
reflection relation (see \S4.3 in \AharonyVK)
\eqn\refsu{V^{(su,susy)}_{j;m,m}(RR,+)=
V^{(su,susy)}_{{k-2\over2}-j;-{k-2\over2}+m,-{k-2\over2}+m}(RR,-).}
Applying \refsu\ to (say) $j_1$ in \exactthree, we find the $SU(2)/U(1)$
amplitude
\eqn\reflectsu{
\langle V^{(su,susy)}_{{k-2\over2}-j_1;-{k-2\over2}+j_1,-{k-2\over2}+j_1}(RR,-)
V^{(su,susy)}_{j_2;j_2,j_2}(RR,+)
V^{(su,susy)}_{{k-2\over2}-j_1-j_2;{k-2\over2}-j_1-j_2,{k-2\over2}-j_1-j_2}
\rangle.}
Since this is an amplitude that preserves $U(1)$, we can calculate it in the 
underlying bosonic $SU(2)_{k-2}$ CFT, where it is given by
\eqn\sucorfin{\eqalign{
\langle \Phi^{(su)}_{{k-2\over2}-j_1;-{k-2\over2}+j_1,-{k-2\over2}+j_1}
& \Phi^{(su)}_{j_2;j_2,j_2}
\Phi^{(su)}_{{k-2\over2}-j_1-j_2;{k-2\over2}-j_1-j_2,{k-2\over2}-j_1-j_2}
\rangle=\cr
&\left[\gamma({1\over k})\gamma(1-{2j_1+1\over k})\gamma(1-{2j_2+1\over k})
\gamma({2(j_1+j_2)+1\over k})\right]^{1\over2}.\cr
}}
In the last step we used equation (B.11) in \AharonyVK, which is taken from
\ZamolodchikovBD.

It remains to compute the $SL(2,\IR)/U(1)$ three-point function on
the second line of \exactthree. As before, since this correlation
function preserves $U(1)$, we can compute it in the underlying
$SL(2,\IR)_{k+2}$ CFT. 
Thus, we have to compute (in notations
which are described in appendix A)
\eqn\slthree{\eqalign{
\langle \tilde\Phi_{\tilde j_1;j_1+1,j_1+1}
\tilde\Phi_{\tilde j_2;j_2,j_2} \tilde
\Phi_{\tilde j_3;-j_1-j_2-1,-j_1-j_2-1}& \rangle=\cr
\tilde D(\tilde j_1,\tilde j_2,\tilde j_3)
\int d^2x_1d^2x_2 & |x_1|^{2(\tilde j_1+j_1+1)}
|x_2|^{2(\tilde j_2+j_2)} |1-x_1|^{2(\tilde j_2-\tilde j_1-\tilde j_3-1)}
\cdot\cr
&
|1-x_2|^{2(\tilde j_1-\tilde j_2-\tilde j_3-1)}
|x_1-x_2|^{2(\tilde j_3-\tilde j_1-\tilde j_2-1)}.\cr
}}
This is precisely the computation described in \S4.3 of \AharonyVK, 
after performing
the change of variables described in \densgaaa, and for our current purposes
we are interested in a double-pole contribution to this expression.
As $\tilde j_1\to j_1$, \slthree\ exhibits a pole
coming from $x_1\to\infty$,
\eqn\poleone{\eqalign{
&\langle \tilde\Phi_{\tilde j_1;j_1+1,j_1+1}
\tilde\Phi_{\tilde j_2;j_2,j_2}\tilde
\Phi_{\tilde j_3;-j_1-j_2-1,-j_1-j_2-1}\rangle
\simeq \cr
&{\pi^2\over \tilde j_1-j_1}
\tilde D(j_1,\tilde j_2,\tilde j_3)
\gamma(\tilde j_3-j_1-j_2)\gamma(\tilde j_2+j_2+1)
\gamma(j_1-\tilde j_2-\tilde j_3).\cr}}
As $\tilde j_3\to j_1+j_2$, $\tilde j_2\to j_2$, this
behaves as
\eqn\polesonethree{\langle \tilde\Phi_{\tilde j_1;j_1+1,j_1+1}
\tilde\Phi_{\tilde j_2;j_2,j_2}\tilde
\Phi_{\tilde j_3;-j_1-j_2-1,-j_1-j_2-1}\rangle
\simeq{\pi^2\tilde D(j_1,j_2, j_1+j_2)
\over ( \tilde j_1-j_1)( \tilde j_3-j_1-j_2)}.
}
Evaluating $\tilde D(j_1,j_2, j_1+j_2)$ using the formulae
in appendix A we finally find 
\eqn\slfinform{\eqalign{
&\langle \tilde\Phi_{\tilde j_1;j_1+1,j_1+1}
\tilde\Phi_{\tilde j_2;j_2,j_2}\tilde
\Phi_{\tilde j_3;-j_1-j_2-1,-j_1-j_2-1}\rangle\simeq\cr
&{\pi^2\over ( \tilde j_1-j_1)( \tilde j_3-j_1-j_2)}
\left[\gamma({1\over k})\gamma(1-{2j_1+1\over k})
\gamma(1-{2j_2+1\over k})
\gamma({2(j_1+j_2)+1\over k})\right]^{-{1\over2}}.\cr}}
Multiplying by the $SU(2)/U(1)$ contribution \sucorfin,
we find that the correlator \exactthree\ is given by 
\eqn\ppoott{\eqalign{
\langle V^{(su,susy)}_{j_1;j_1,j_1}(RR,+) V^{(su,susy)}_{j_2;j_2,j_2}(RR,+)
V^{(su,susy)}_{{k-2\over2}-j_1-j_2;{k-2\over2}-j_1-j_2,{k-2\over2}-j_1-j_2}
\rangle &\times\cr
\langle V^{(sl,susy)}_{\tilde j_1;j_1+1,j_1+1}(RR,-)
V^{(sl,susy)}_{\tilde j_2;j_2,j_2}(RR,+)
V^{(sl,susy)}_{\tilde j_3;-j_1-j_2-1,-j_1-j_2-1}\rangle & \simeq\cr
{\pi^2\over ( \tilde j_1-j_1)( \tilde j_3-j_1-j_2)} \simeq 
{4\pi^2\over k^2p_1^2p_3^2}(2j_1+1)&[2(j_1+j_2)+1].
\cr}}
We see that the string calculation \kinstr, \ppoott\ gives rise to 
the correct kinematics, reproducing the pole structure of the
first term in \threefree. It remains to check that the dependence
on $j_1$, $j_2$ comes out correctly as well. 

To do that we first need to assemble all the factors in the string
calculation, and then take into account the relation between the
string theory vertex operators and the field theory operators. 
Bringing together the results \kinstr, \ppoott, we find that the
residue of the double pole in
the string theory three-point function \threeooo\ is proportional 
to 
\eqn\ththth{\eqalign{
\langle \OO_{2j_1+1}^+&\OO_{2j_2+1}^+
\OO_{0,-2(j_1+j_2)-2}\rangle\sim \cr
&{8k\over 2(2j_1+1)^2}4\pi^2 {(2j_1+1)[2(j_1+j_2)+1]\over k^2}=
{16\pi^2\over k}{2(j_1+j_2)+1\over 2j_1+1}.\cr}}
The relation between the vertex operator $\OO_{2j+1}^+$
and the corresponding gauge theory observable was determined in\foot{Here
and below we omit the powers of $M_W$.} equation (5.12) in
\AharonyVK, up to $j$-independent 
constants which were not carefully followed there,
\eqn\ooppmap{\OO^+_{2j+1}\sim {1\over {(2j+1) 
}} \xi^{\mu\nu}\ttr(F_{\mu\nu} B^{2j+1}).}
Together with the relation \onbtilde, \gscnk, the string theory
predicts that the gauge theory amplitude should scale with
$j_1$, $j_2$ like (up to $k$-dependent constants):
\eqn\scstring{{1\over 2(j_1+j_2)+1}
(2j_1+1)(2j_2+1)
{2(j_1+j_2)+1\over 2j_1+1}=
(2j_2+1).}
Thus, we see that the dependence on $n_1=2j_1+1$, $n_2=2j_2+1$ comes
out correctly as well: the first term in \threefree\ is indeed
proportional to $n_2$ and is independent of $n_1$. Thus, we find
precise agreement of the behavior near the poles
between the string theory and low-energy gauge
theory results, consistent with the fact that the string theory
contribution \poleone\ comes purely from the region $x\to 0,\infty$ and with
our general discussion.
Matching the constants (which we would need to do in order to determine the
precise value of $g_s^2$ appearing in \gscnk) 
requires more work, and we will not attempt
to do this here.

\subsec{$\vev{\ttr(F_{\mu_1 \nu_1} B^{n_1}) \ttr(F_{\mu_2 \nu_2} B^{n_2})
\ttr(F_{\mu_3 \nu_3} (B^*)^{n_3}) \ttr(F_{\mu_4 \nu_4} (B^*)^{n_4})}$}

The contributions to the correlation functions we discussed until now
from the low-energy gauge theory all came simply from free-field
contractions, and they do not teach us anything about non-trivial
S-matrix elements in this theory. This is due to the particularly
simple examples we have chosen, and is not a general property of our
formalism. An example of a non-trivial S-matrix computation arises
from the 4-point function of the operators \fbop.  This 4-point
function was computed (at tree-level) in \AharonyVK, where it was
shown that it has a quadruple pole as a function of all the momenta when
$p_{\mu}^2=0$, and that the coefficient of this pole may be interpreted
as a non-trivial S-matrix element, arising from an $F^4$ interaction in
the low-energy field theory. The result of this computation was
verified by comparing it with predictions from the duality of type IIA
string theory on K3 with heterotic string theory on $T^4$. The
analysis of the present paper makes it clear that this agreement is
based on the fact that the $SL(2,\IR)/U(1)$ contribution to the residue of
the pole is localized near $x=0,\infty$, so (as assumed in \AharonyVK)
it arises purely from the
low-energy gauge theory. Other non-trivial S-matrix elements may be
similarly computed by using the methods of \AharonyVK\ and this paper.

\subsec{$\langle \ttr(F_{\mu \nu} B^n)\,  \ttr(F_{\mu' \nu'} B^{k-n}) \rangle$}

As a final example in this section let us discuss a two-point function
which does not conserve winding number, involving operators of the form
\fbop. As we discussed in \S2.6, one way to compute a correlation function
like $\vev{\OO_n^+ \OO_{k-n}^+}$, which violates winding number by one
unit, is to insert into the correlation function an additional
degenerate vertex operator $\Phi_{-(k+2)/2;-(k+2)/2,-(k+2)/2}$. The
resulting correlation function conserves winding number, so it can be
analyzed by the same methods we used above, and as
discussed in \S2.6, its $SL(2)/U(1)$ part is precisely the same as
that of the two-point function that we are interested in.

Another way to compute this two-point function is to use the
reflection property \refsltwo. As discussed in \AharonyVK, this
reflection property relates the normalizable part of the operator 
$\OO_n^+\simeq \xi^{\mu \nu} \ttr(F_{\mu \nu} B^n)$ 
(coming from its massless pole) to that of $\OO_{k-n}^-
\simeq \xi^{\mu \nu}
\ttr(F_{\mu \nu} (B^*)^{k-n})$. Thus, we can use it to relate 
the two-point function we are interested in to the two-point function
$\vev{\OO_{k-n}^- \OO_{k-n}^+}$ which conserves winding number. From the 
point of view of the 
low-energy field theory this relation between the two operators
$\ttr(F_{\mu \nu} B^n)$ and $\ttr(F_{\mu \nu} (B^*)^{k-n})$ (of
different winding numbers) is obvious, since the one-particle states
that these operators create are the same due to the relation
$\vev{B^n} = M_W^{2n-k} \vev{(B^*)^{k-n}}$ which follows from \bexpv.
In this way of computing, the final two-point function that we end up
with is similar to the one discussed in \S5.1, and again it receives
contributions both from an LSZ pole (related to the low-energy
field theory) and from a bulk pole.

Note that the simplest example of a winding-number non-conserving
correlation function is just the one-point function of the operator
$\ttr(B^k)$: $\vev{\ttr(B^k)} = k M_W^k$ \bexpv. 
This can be computed by similar methods,
or by computing its derivative with respect to the $\NN=2$ Liouville
coupling \ntwoliouv, which gives a winding-conserving two-point function.

\newsec{Fixing the mixing}

In this section we discuss another class of  correlation functions in 
the six dimensional DSLST,
whose low energy behavior is expected from the analysis of \S2-\S4
to be dominated by the low energy gauge theory. By comparing the
string calculation of these correlation functions to the gauge theory
one we determine the precise form of the gauge theory operators on
the right hand side of \scalars, including all single-trace and multi-trace
contributions.

We wish to study correlation functions of the operators\foot{The overall
normalization of these operators, which can be determined as discussed
in \S5.1, will not be important in this section.}
\eqn\defbn{\BB_n \equiv {1\over n} \ttr(B^n) \simeq \OO_{0,n}~,}
corresponding to the vertex operators \formb, \dsbt\ (obeying \mamama) in six
dimensional $\NN=(1,1)$ supersymmetric LST. 
Consider a general correlation function of these operators,
\eqn\corrsecsixa{\vev{\BB_{n_1}(p^1) \BB_{n_2}(p^2) \cdots \BB_{n_r}(p^r)
\bBB_{{\hat n}_1}({\hat p}^1) \cdots \bBB_{{\hat n}_{\hat r}}
({\hat p}^{\hat r})},}
where $\bBB_n(p)$ is the complex conjugate of $\BB_{n}(-p)$. As in \S2,
we will impose winding conservation \windcons. This leads to the
constraint\foot{Note that the $\Z_k$ symmetry $B\to e^{2\pi i\over k} B$,
which is preserved (up to a gauge transformation)
by the background \bexpv, implies that in general
$\sum_{i=1}^r n_i - \sum_{i=1}^{\hat r} {\hat n}_i\in k\Z$.}
$\sum_{i=1}^r n_i = \sum_{i=1}^{\hat r} {\hat n}_i$.
We will also restrict to $r+{\hat r}> 2$; the case of the
two-point function was already discussed in \S5.1.

In order to understand the structure of  \corrsecsixa\ at low momenta,
consider the behavior of the correlation function
\eqn\loweb{\langle {\rm tr}(B^{n_1})(p^1)\cdots {\rm tr}(B^{n_r})(p^r)
{\rm tr}({B^*}^{n_1})({\hat p}^1)\cdots
{\rm tr}({B^*}^{{\hat n}_{\hat r}})({\hat p}^{\hat r})\rangle}
in the low energy field theory. The leading contribution to this correlation
function is due to standard free field contractions in which $B$'s from the
operators ${\rm tr}(B^{n_i})$ are contracted with $B^*$'s from the operators
${\rm tr}({B^*}^{\hat n_{\hat i}})$. The minimal number of contractions needed
to get a connected diagram is $r+{\hat r}-1$. Replacing the uncontracted
$B$'s and $B^*$'s by their expectation value \bexpv, we find that the
correlation function \loweb\ behaves at large $M_W$ like  $M_W^x$ with
\eqn\powergauge{x=\sum_{i=1}^r n_i + \sum_{{\hat i}=1}^{\hat r}
{\hat n}_{\hat i} - 2(r + {\hat r} -1)=2\left[\sum_{i=1}^r n_i
-(r+\hat r-1)\right].}
Terms with more than $r+{\hat r}-1$ contractions scale like lower powers
of $M_W$ and hence are subleading in the $1/M_W$ expansion. Recalling
that in string theory on \slsu\ the $1/M_W$ expansion is equivalent to
the string loop expansion (see \gstip), we expect these terms to come 
from higher loop diagrams in string theory.

Why is there no contribution to the correlation function
\loweb\ from higher order terms in the Lagrangian of the 
six dimensional gauge theory? As mentioned above, since 
this theory is non-renormalizable, we must allow interaction 
terms in the Lagrangian that go like arbitrarily high 
powers of the fields. For example, an interaction of the 
form ${\rm tr}(B^n{B^*}^n)$ with $n\ge r, \hat r$ would 
contribute to the correlation function \loweb\ terms that 
go like a power of $M_W$ that increases linearly with $n$.

The answer to this question is that we know that the 
eigenvalues of $B$ are exact moduli in the full theory 
on the fivebranes, and thus the Lagrangian for $B$ must
have the symmetry $B\to B+C$, with $C$ an arbitrary 
diagonal matrix. Taking this into account, it is easy 
to convince oneself that interactions of this form cannot exist,
and the free contractions indeed 
give the leading contribution in the $1/M_W$ expansion.

A useful special case is $\hat r=1$, for which one has
\eqn\specialb{
\langle {\rm tr}(B^{n_1})(p^1)\cdots {\rm tr}(B^{n_r})(p^r)
{\rm tr}({B^*}^{n})({\hat p})\rangle\simeq
M_W^{2(n-r)}\prod_{i=1}^r{n_i\over (p^i)^2}~,} 
where $\sum_{i=1}^r n_i=n$.  In this case, the leading contribution
comes from contracting one $B$ from each of the first $r$ operators
with a $B^*$ from the last operator.  This leads to the structure
indicated on the right hand side of \specialb.  Note that there are
poles corresponding to the first $r$ momenta, but not to the last
one. This is very natural, since if there was a pole associated with
the last external leg as well, the S-matrix for scattering $B$
particles would be non-zero at zero momentum, which would be
inconsistent with the expected absence of a potential for the 
eigenvalues of $B$.

Equations \loweb, \specialb\ are field theory correlation functions of single
trace operators. For comparison with string theory we have to generalize
the discussion to more general operators,
\eqn\multit{{\rm tr}(B^n)\to \ttr(B^n)={\rm tr}(B^n)+
\sum_{l_1+l_2=n} c_{l_1,l_2}{\rm tr}(B^{l_1}){\rm tr}(B^{l_2})+\cdots.}
The multi-trace terms in \multit\ contribute to some of the amplitudes
\loweb\ at various orders in the $1/M_W$ expansion. Consider for example
the special case \specialb. It is easy to see that at the order in $M_W$ indicated
on the right hand side, multi-trace contributions to the first $r$ operators,
${\rm tr}(B^{n_i})$, vanish. The reason is that there can only be one free
field contraction that involves a given operator ${\rm tr}(B^{n})$. Thus,
contributions of multi-trace operators necessarily include factors of
$\langle{\rm tr} (B^{l_i})\rangle$, which vanish for all $l_i<k$ at
the point in moduli space at which we are working, \bexpv.

Multi-trace contributions to the last operator,
${\rm tr}({B^*}^{n})$, {\it do} contribute to 
\specialb. For example, terms of the form 
${\rm tr}({B^*}^{n_1}){\rm tr}({B^*}^{n_2})\cdots{\rm tr}({B^*}^{n_r})$,
${\rm tr}({B^*}^{n_1+n_2}){\rm tr}({B^*}^{n_3})\cdots{\rm tr}({B^*}^{n_r})$,
${\rm tr}({B^*}^{n_1+n_2+\cdots+n_{r-1}}){\rm tr}({B^*}^{n_r})$,
etc, all contribute at the same order as ${\rm tr}({B^*}^{n})$, 
and give rise to the same analytic structure as that on the 
right hand side of \specialb. This fact will prove useful below.

Having understood the field theory amplitudes \corrsecsixa\ --
\specialb, we now turn to the string theory calculation of the
corresponding correlation function of the vertex operators \dsbt.
Our first task is to reproduce the scaling of the correlation
function with $M_W$. To do that, note that the vertex operator 
$\BB_n$ \formb\ goes for small $p_\mu^2$ as $e^{Q(n-2)\phi/2}$,
due to the physical state condition \mamama. The sphere 
correlation function \corrsecsixa\  scales with the $\NN=2$ 
Liouville coupling $\mu$ \ntwoliouv\ like $\mu^y$, with  
\eqn\kpzone{
{Q\over 2} (n_1 - 2 + n_2 - 2 + \cdots + n_r - 2 + {\hat n}_1 - 2
+ \cdots + {\hat n}_{\hat r} - 2) - {1\over Q} y = -Q,} 
or $y=(\sum_i n_i + \sum_i {\hat n}_i - 2 r - 2 {\hat r} + 2)/k$. 
Using the relation between $\mu$ and $M_W$ \mwsc\ we conclude that
the correlator scales like $M_W^x$ with $x=ky$. Comparing to the
field theory result \powergauge, we see that the string and field
calculations give the same answer for $x$. 

Our next task is to calculate the string theory correlation 
function \corrsecsixa. In general, this correlation function 
is complicated; however, the case $\hat r=1$, in which 
\corrsecsixa\ reduces to 
\eqn\corrsecsix{\vev{\BB_{n_1}(p^1) \BB_{n_2}(p^2) \cdots \BB_{n_r}(p^r)
{\bar \BB}_n({\hat p})} {\rm\ \ with\ \ } \sum_{i=1}^r n_i = n~,}
turns out to be much more tractable, and we will restrict to it here.

To calculate \corrsecsix\ we recall an important feature
of the vertex operators $\BB_n$ \dsbt: their $SU(2)/U(1)$ 
component is chiral. Indeed, from \drsu\ we see that any operator 
of the form $V^{(su,susy)}_{m;-m,-m}$ has the property that its 
dimension and R-charge are related: 
$\Delta=\bar\Delta=\half R=\half \bar R=m/k$.
Thus, it is a chiral operator (annihilated by  $G_{-1/2}^+$ and
${\bar G}_{-1/2}^+$). Defining 
$V^{(su,susy)}_{\half;-\half,-\half}\equiv\chi$, one has 
$V^{(su,susy)}_{m;-m,-m}=\chi^{2m}$. One can think of $\chi$ as the
bottom component of a Landau-Ginzburg superfield, in terms of which
the $\NN=2$ minimal model is naturally formulated. In terms of $\chi$, 
the vertex operator \dsbt\ can be written as
\eqn\oondecomp{\OO_{0,2j+2}(p_{\mu})=
e^{-\varphi-{\bar \varphi}} \chi^{k-2j-2} V^{(sl,susy)}_{\tj;j+1,j+1}
e^{ip\cdot x} \leftrightarrow \BB_{2j+2}~.}
To compute the $(r+1)$-point function \corrsecsix\ in
tree-level string theory we need to put two of the 
vertex operators in the $(-1,-1)$ picture as
in \oondecomp, and the other $(r-1)$ in the $(0,0)$
picture, where they are given by
\eqn\oonidentn{\BB_{2j+2}(p_{\mu}) \leftrightarrow 
G_{-1/2} {\bar G}_{-1/2}\left( \chi^{k-2j-2} V^{(sl,susy)}_{\tj;j+1,j+1}
e^{ip\cdot x}\right).}
Each of the superconformal generators $G$ and ${\bar G}$ can be written
as a sum of contributions from the $SU(2)/U(1)$, $SL(2,\IR)/U(1)$ and 
$\IR^{5,1}$ CFTs. The three component CFTs  are $\NN=(2,2)$ worldsheet 
supersymmetric; therefore, $G_{-1/2}$ may be further decomposed
in terms of the two $\NN=2$ superconformal generators as $G_{-1/2} = 
G_{-1/2}^+ + G_{-1/2}^-$, where $G^+$ ( $G^-$ ) raises (lowers) the
$U(1)_R$-charge by one unit. Similar comments apply to the other
worldsheet chirality.

The result of the calculation cannot depend on which two operators
we take to be in the $(-1,-1)$ picture. A convenient choice is to
pick the first two operators (say) in \corrsecsix. We will next show 
that the correlation function \corrsecsix\ vanishes for this choice, 
point by point in the worldsheet moduli space. 

To see that, focus on the $SU(2)/U(1)$ part of the unintegrated 
correlation function \corrsecsix. In order to get a non-vanishing 
result, the total $U(1)_R$ charge must vanish. The first two operators 
have $U(1)_R$ charges $(1-{n_1\over k})$ and $(1-{n_2\over k})$.
The next $(r-2)$ operators have components with two different 
values of the R-charge. Contributions to $G_{-\half}$ in \oonidentn\ 
from $\IR^{5,1}\times SL(2)/U(1)$ give operators with $SU(2)/U(1)$ 
R-charge $(1-{n_i\over k})$. In the contribution to $G_{-\half}$
of the $SU(2)/U(1)$ CFT, only the $G^-$ part acts (recall that $G^+$
annihilates the chiral operators $\chi^l$), and gives an operator with
R-charge $(-{n_i\over k})$. Similarly, for the last operator in \corrsecsix,
$\bBB_n$, we find contributions with R-charge $({n\over k}-1)$ or 
$({n\over k})$. 

Remembering that $n=\sum_i n_i$, it is easy to see that all possible
contributions to the correlation function \corrsecsix\ have a total 
R-charge which is a positive integer. Since the total R-charge 
does not vanish, we conclude that this correlation 
function vanishes in string theory for any $r \geq 2$. This result 
is true for the unintegrated correlation function of the vertex
operators in \corrsecsix; therefore, it obviously holds after integrating
over the moduli as well.

Two comments:
\item{(1)} The derivation above is valid for arbitrary momenta
$(p^1,\cdots,p^r,\hat p)$. Below we will see that at low momenta,
this vanishing has interesting consequences for the gauge-string
correspondence. It would be interesting to understand the significance
and implications of the vanishing of \corrsecsix\ more generally.
\item{(2)}If both $r$ and $\hat r$ in \corrsecsixa\ are larger than one,
the argument above does not imply the vanishing of the correlation
function, but it leads to a simplification of its structure. It would be
interesting to explore this further.

\noindent
The vanishing of \corrsecsix\ derived from string theory is quite
surprising from the point of view of the low-energy field theory,
where the free field theory diagrams do not seem to vanish.
Indeed, if the operators $\BB_n$ were precisely identified 
with ${1\over n}\tr(B^n)$ in the low-energy field theory 
we would get a contradiction, since the relevant field theory 
correlation function \specialb\ does not vanish. One might ask 
whether it is possible that this is a case where non-gauge 
theoretic contributions to the amplitude are important, but 
it is not difficult to see that the residue of the poles in 
\specialb\ should not receive bulk contributions. Hence,
the vanishing of \corrsecsix\ should have an interpretation
purely in the low energy gauge theory.

The resolution of the apparent paradox is that in general
the LST operators $\BB_n$ corresponding to the single
string vertex operators \oondecomp\ reduce at low energies
to linear combinations of single-trace and multi-trace
operators\foot{It is important to distinguish this mixing from
the mixings between single-trace and multi-trace operators computed in the
context of the AdS/CFT correspondence. The mixings computed there
usually involve non-chiral operators, for which specific combinations
of the single-trace and multi-trace operators are eigenfunctions of
the dilatation operator. On the other hand,
for chiral operators there is no intrinsic
preference for a particular linear combination over any other combination
from the field theory point of view. However, it is still true in all
examples of the AdS/CFT correspondence that the
string theory vertex operators map to specific linear combinations of
operators. It would be interesting to work out the generalization
of our results to other examples. Of course,
generally this is difficult due to the presence of RR backgrounds, but it
should be possible to do this at least in the BMN limit \BerensteinJQ\ 
of string
theory on $AdS_5\times S^5$.}
\eqn\genident{\BB_n \simeq
{1\over n} \tr(B^n) + \sum_{n_1} a_{n_1} \tr(B^{n_1})
\tr(B^{n-n_1}) + \sum_{n_1,n_2} a_{n_1,n_2} \tr(B^{n_1}) \tr(B^{n_2})
\tr(B^{n-n_1-n_2}) + \cdots}
with some coefficients $a$. As mentioned above, at
the particular point in moduli space \bexpv\ that
we are studying, where $\vev{\tr(B^l)}=0$ for all
$l<k$, only the first term in the expansion of the
$\BB_{n_i}$ operators in \corrsecsix\ can contribute
(at leading order in $M_W$). However, many different
terms in the expansion \genident\ of the operator
$\bBB_n$ can contribute.
The correlation function \corrsecsix\ is a linear combination
of all of these contributions, and its vanishing may be used to determine the
value of the relevant coefficients in \genident. By looking at all correlation
functions of the type \corrsecsix\ we can determine all the coefficients
appearing in \genident; this computation is detailed in appendix B. The result
of appendix B may be written as\foot{In appendix B we prove this formula using
a certain plausible assumption. A complete proof is still missing.}:
\eqn\ourident{\BB_n \simeq
\sum_{l=1}^{\infty} \sum_{n_i=2}^n
{1\over {l!}} \left( {1-n\over k} \right)^{l-1} \delta(\sum_i n_i,n)
\left({1\over n_1}
\tr(B^{n_1})\right) \left({1\over n_2} \tr(B^{n_2})\right) \cdots
\left({1\over n_l} \tr(B^{n_l})\right)}
for $n=2,3,\cdots,k$. Note that the sum over the $n_i$ in \ourident\ goes
over all possible values of the $n_i$ without any ordering between them,
so that (for instance) a term involving $l$ specific traces with all the $n_i$
different from each other will appear $l!$ times in the sum, canceling the
factor of $1/l!$ appearing explicitly in the formula.

We can test whether the result \ourident\ is reasonable by considering
two different limits. In a general $SU(k)$ gauge theory with fields in the
adjoint representation, 't Hooft has argued
\thooft\ that in the limit of large $k$ with fixed $g_{YM}^2 k$ the
gauge theory may be described as a string theory, with single-trace
operators (involving a product of a finite number of adjoint fields) mapping
to string vertex operators. We expect that the same should be true
also in the LST (even though it is not simply a gauge theory), namely
that for large $k$ and finite $n$ we should have $\BB_n \simeq {1\over
n} \tr(B^n)$.  And indeed, in this limit only the $l=1$ term in
\ourident\ contributes, and the other terms are suppressed by powers
of $1/k$, as expected. Note that we have not taken any large $k$ limit
in deriving \ourident\ (or elsewhere in our computations).

The opposite limit of large $n$ and $k$ with small $(k-n)$ 
corresponds to operators
with a large angular momentum on the $S^3$ (in the asymptotic region
\chsmetr).  It was argued in \McGreevyCW\ that in AdS/CFT examples
such operators are not described by perturbative string states but
rather by ``giant graviton'' wrapped D-brane states, since the
Ramond-Ramond (RR) flux in these backgrounds causes states with large
angular momentum to expand into D-branes. It was further argued in
\refs{\BalasubramanianNH\CorleyZK\AharonyND\BerensteinAH-\BerensteinKK}
(by examining various correlation functions of such states) that these
states should be identified with subdeterminant operators of the form
\eqn\subdet{{\rm subdet}_n(B) \equiv {(-1)^n \over {n! (k-n)!}}
\epsilon^{a_1 a_2\cdots
a_n a_{n+1} \cdots a_k} \epsilon_{b_1 b_2 \cdots b_n a_{n+1} \cdots a_k}
B_{a_1}^{b_1} B_{a_2}^{b_2} \cdots B_{a_n}^{b_n}}
rather than with single-trace states\foot{The subdeterminant operators may
be written in terms of the eigenvalues ${\hat b}_i$ ($i=1,\cdots,k$) of the
matrix $B$ as ${\rm subdet}_n(B) = (-1)^n \sum_{i_1 < i_2 < \cdots < i_n}
{\hat b}_{i_1} {\hat b}_{i_2} \cdots {\hat b}_{i_n}$.}.
In our theory we do not have an RR flux
but we have an NS-NS three-form flux on the $S^3$, so similar arguments suggest
that states with large angular momentum should again correspond to ``giant
gravitons,'' but now these ``giant gravitons'' are fundamental strings wrapped
on an $S^1$ inside the $S^3$ rather than D-branes, explaining why we are
able to describe them (and see the bound on their angular momentum)
in string perturbation theory. Most of the
arguments that the ``giant gravitons'' should map to subdeterminants may also
be generalized to our case, so we expect that for $k-n \ll k$ the operators 
\ourident\ should be approximately equal to the operators \subdet.

This is indeed true, since the operators \subdet\ may be written as
\eqn\subdetform{{\rm subdet}_n(B) = \sum_{l=1}^{\infty}
\sum_{n_i=2}^n {(-1)^l \over {l!}} \delta(\sum_i n_i,n)
\left({1\over n_1}
\tr(B^{n_1})\right) \left({1\over n_2} \tr(B^{n_2})\right) \cdots 
\left({1\over n_l} \tr(B^{n_l})\right);}
this formula\foot{We have not found the formula \subdetform\
for subdeterminants in the literature, but at least for the case $n=k$
it is easy to show that it is equivalent to existing formulae for the
determinant. For a general matrix whose trace does not necessarily vanish
one has the same formula with $n_i \geq 1$.} may be easily seen to be a
solution to the recursion relation
\eqn\recur{n \cdot {\rm subdet}_n(B) + \sum_{i=2}^n {\rm subdet}_{n-i}(B)
\cdot \tr(B^i) = 0}
which the subdeterminant operators obey (with ${\rm subdet}_0(B)=1$,
${\rm subdet}_1(B)=0$). The expression for the subdeterminant \subdetform\
is almost the same as the formula \ourident\ describing the dual of single 
string vertex operators, except for an (unimportant) overall sign, and the
factor of $((n-1)/k)^{l-1}$ in \ourident. However, for large $k$ and
small $k-n$ this factor goes to one for any finite $l$, so the operators
we find indeed approach\foot{More precisely, the coefficients of the
terms in \ourident\ with a finite number of traces for large $k$ approach
those of the subdeterminant operators. This it not true for the coefficients
of terms with a large number of traces (of order $k$). However, this difference
between the operators does not contribute for large $k$ to the correlation
functions of ``giant gravitons'' that motivate their identification with
subdeterminants, so it is not inconsistent. It is amusing that the two
formulae coincide exactly for $n=k+1$, outside the range in which the
string theory operators exist, where both expressions give operators which
vanish (at least classically).} subdeterminants as $n \to k$, though they
are never precisely equal to them.  The formula
\ourident\ provides (for large $k$) a smooth interpolation between the two
limits where the operators look like single traces and like subdeterminants,
and based on the arguments above it is exact for all values of $n$ and $k$.

As mentioned above, our argument for the vanishing of \corrsecsix\
does not apply to general correlation functions of the form
\corrsecsixa\ with $r,{\hat r} > 1$, for which one has contributions to the
correlation function with vanishing total R-charge in
$SU(2)/U(1)$. Indeed, using \ourident\ one can compute the low-energy
limit of such correlation functions and find that it is not
vanishing. It would be interesting to verify that the non-zero
tree-level correlation functions of the form
\corrsecsixa\ are consistent with \ourident\ at low energies.

In \ourident\ we identified precisely the low-energy field theory
content of the operators $\BB_n$, by using a computation performed at
a particular point in the moduli space of the LST. However, since this
identification involves non-normalizable operators it should be independent
of the moduli. The operators $\BB_n$ are related by the 
global $SO(4)$ symmetry of the LST (at the origin of its moduli space) and by
supersymmetry to all chiral operators in the LST, so from \ourident\
we can read off the precise low-energy field theory content of all of these
chiral operators. For example, for the operators $\FF_n =\ttr(F_{\mu \nu} B^{n-1})$
discussed in \AharonyVK\ we find\foot{The chiral operator $\FF_n$ includes
also fermionic contributions, which we omit here.}
\eqn\ouridentf{\FF_n \simeq 
\sum_{l=1}^{\infty} \sum_{n_i=2}^n
{\delta(\sum_i n_i,n)\over {(l-1)!}} \left( {1-n\over k} \right)^{l-1}
\left(
\tr(F_{\mu \nu} B^{n_1-1})\right) 
\left({\tr(B^{n_2})\over n_2}\right) \cdots 
\left({\tr(B^{n_l})\over n_l}\right),}
and for the operators $\OO_{1,n-1} = \ttr(A B^{n-1})$ we find
\eqn\ouridenta{\OO_{1,n-1} \simeq
\sum_{l=1}^{\infty} \sum_{n_i=2}^n
{\delta(\sum_i n_i,n)\over {(l-1)!}} \left( {1-n\over k} \right)^{l-1}
\left(
\tr(A B^{n_1-1})\right)
\left({\tr(B^{n_2})\over n_2}\right) \cdots 
\left({\tr(B^{n_l})\over n_l}\right).}
We can use equation \ouridenta\ to perform another consistency check
of our analysis. As we saw before (see \eg\ equation \aba), the left-moving
component of the operator $\OO_{1,n-1}$ is identical to that of the
operator $\OO_{0,n}\sim\BB_n$; in particular, it also
involves (in the $(-1,-1)$ picture) a chiral primary operator in the
$SU(2)/U(1)$ CFT of R-charge $(1-{n\over k})$. Consider the
correlation function
\eqn\corrsecsixab{\vev{\OO_{1,n_1-1}(p^1) \BB_{n_2}(p^2) \cdots \BB_{n_r}(p^r)
{\bar \OO}_{1,n-1}({\hat p})} {\rm\ \ with\ \ } \sum_{i=1}^r n_i = n.}
As far as the left-movers are concerned, it is identical to \corrsecsix.
Thus, we can repeat the arguments used in proving that \corrsecsix\
vanishes to show that \corrsecsixab\ vanishes (at leading order in
$1/M_W$, corresponding to string tree-level) as well.

In the low-energy gauge theory, there are various contributions to this
correlation function, involving (at leading order in $1/M_W$)
contracting the $A$ from the first operator with the $A^*$ from the last 
operator, and performing the other contractions as before. Using \ourident\ and
\ouridenta\ we can compute the coefficients of all of these different
contributions and sum them up. We find that the result is  zero,
consistent with the string theory computation of \corrsecsixab. This
provides another consistency check of our analysis.

\newsec{Four dimensional $\NN=2$ supersymmetric LST}

Our discussion above focussed on the case of $d=6$, but the same considerations
apply to all other LSTs. As an additional example we discuss here the case
of $d=4$ LSTs with $\NN=2$ supersymmetry which arise as decoupled theories
on generalized conifold singularities of the form 
\eqn\gencon{z_1^n + z_2^2 + z_3^2 + z_4^2 = \mu} 
in type II string theory (or alternatively on wrapped $NS5$-branes).
The low-energy theory in this case is a $d=4$ $\NN=2$ $U(1)^{n-1}$ gauge
theory, which can be thought of as describing an $SU(n)$ gauge theory
near an Argyres-Douglas type point \refs{\ArgyresJJ\ArgyresXN-\EguchiVU}
in its moduli space (the point $\mu=0$, which corresponds to a non-trivial
superconformal field theory for $n>2$).
Denoting the complex scalar field in the $SU(n)$ gauge multiplet by $\Phi$,
the gauge-invariant chiral operators are $\ttr(\Phi^l)$ ($l=2,3,\cdots,n$) and
their descendants. We can choose a basis of operators such that their
vacuum expectation values vanish for $\mu=0$, and moving away from this point
is achieved by giving a non-zero expectation value to $\Phi$ (which plays the
same role as $B$ in our discussion of the $d=6$ case above).

The worldsheet description of this decoupled theory is given by an orbifold
of the supersymmetric $\IR^{3,1}\times SL(2)/U(1)\times SU(2)_n/U(1)$ CFT,
similar to \slsu, where the $SL(2)/U(1)$ theory has $Q^2 = 1 + 2 / n$. 
Using the same notations as in \S4, the vertex operators
for the chiral operators on the worldsheet take the form \GiveonZM\ 
\eqn\fourdver{\OO_i(p_{\mu}) = e^{-\varphi -{\bar \varphi}} 
V^{(su,susy)}_{i/2;-i/2,-i/2} V^{(sl,susy)}_{j;m,m} 
e^{ip_{\mu} x^{\mu}}}
with $m=(n-i)/(n+2)$, for $i=0,1,\cdots,n-2$. The mass-shell condition for
this operator is
\eqn\fourdmass{p^2 = {{(i+2)(n-i) + j(j+1)(n+2)^2}\over {n(n+2)}}.}
Our analysis in \S2,\S3 
suggests that for the values of $i$ which give $1>m>1/2$,
namely $i < n/2-1$, the operator \fourdver\ has an LSZ pole for $j=m-1$.
Using \fourdmass, we find that this pole is at $p^2=0$. On the other hand, for
$i > n/2-1$ we find no LSZ poles\foot{The case $i=n/2-1$ ($m=1/2$), 
which can occur for even values of $n$, 
is a special case for which the mass gap to the bulk continuum
goes to zero, so we cannot tell whether a massless pole exists in this case
or not.}. In particular, the appropriate solution of the mass-shell condition
for these operators when $p^2=0$ is $j=-m > -1/2$, and there is no LSZ
pole at this value of $j$.
Note that (unlike in $d=6$ LST) in the two-point-functions of $\OO_i$
\fourdver\ there is no bulk contribution to the low energy poles.

The analytic structure we found in string theory at small momenta agrees 
precisely with the mapping of these operators to the low-energy
field theory which was suggested in \GiveonZM. The operators with $i < n/2-1$
were argued to correspond to bottom components of space-time anti-chiral
superfields, of the form $\ttr((\Phi^*)^{n-i})$. Since $\Phi$ has a non-zero
VEV, these operators are expected to exhibit massless LSZ poles, as
we found above. On the other hand, the operators with $i > n/2-1$ were
argued to correspond to the 
top components of superfields whose bottom component
is $\ttr(\Phi^{i+2})$. These operators do not include any component of the
form $\ttr(X \Phi^l)$ (for a field $X$ whose two-point function has a pole;
they include, for instance, components of the form
$\ttr(\del_{\mu} \Phi \del^{\mu} \Phi \Phi^i)$), so they are not expected to
exhibit massless poles, in agreement with our string theory results.

A similar analysis may be performed for RR operators in this case. These
map in space-time to operators of the form $\ttr(F_{\mu \nu} \Phi^l)$, and
again we find precise agreement between the string theory results and the
low-energy field theory expectations. By computing the S-matrix related
to these operators we can reconstruct the $F^{2n}$ terms in the effective
action of this LST, as in \AharonyVK.

\newsec{Two dimensional string theory}

In this paper we focused on spacetimes of the form \rrpphh,
but much of the discussion is more general, and should apply
to other asymptotically linear dilaton backgrounds in string
theory. A famous example of such a background is Liouville
theory, and in particular its application to two dimensional
string theory (for reviews see \eg\
\refs{\KlebanovQA,\GinspargIS,\PolchinskiMB}).
In this section we  briefly discuss two dimensional string theory
from the point of view of our analysis.

Consider string theory in a two dimensional Euclidean
space labeled by the coordinates $(\phi,x)$. $\phi$ is a
Liouville coordinate, which is described asymptotically by
a free field with a linear dilaton (as in \ggss) whose slope $Q$
is related to its central charge by the relation\foot{In this section
we are using conventions that are different from those of the
previous sections, but are standard in studies of Liouville theory
and two dimensional string theory. In particular, we take $\alpha'=1$
and take $\phi\to-\phi$, such that the weakly coupled region is at
$\phi\to-\infty$.} $c_L=1+6Q^2$. The strong coupling singularity
at $\phi\to\infty$ is resolved by adding to the worldsheet Lagrangian
the cosmological term
\eqn\cosmterm{\delta \CL=\mu_0 e^{2b\phi},}
where $b$ is related to $Q$ by $Q=b^{-1}+b$. This interaction
creates a wall repelling the worldsheet fields from the strong
coupling region $\phi\to\infty$ and thus regularizes the theory.

The worldsheet cosmological constant $\mu_0$ plays in this theory
a role similar to that of the Sine-Liouville coupling \lsineliouv\
and the $\NN=2$ Liouville coupling \ntwoliouv\ in the $SL(2,\IR)/U(1)$
backgrounds  studied earlier in this paper. In particular, one can again
think of $\mu_0$ as setting an energy scale (an analog of $M_W$ \mwsc\
in our previous discussion) and of string perturbation theory as an 
expansion in powers of the inverse of this scale.

A set of natural observables in Liouville theory is
\eqn\vvaall{V_\alpha=e^{2\alpha\phi}~.}
The scaling dimension of $V_\alpha$ is
$\Delta_\alpha=\bar\Delta_\alpha=\alpha (Q-\alpha)$. These operators
play a role similar to that of the $SL(2,\IR)/U(1)$ observables
$V_{j;m,\bar m}$ \physw, \dimvjm, and their supersymmetric analogs
\susydimvjm. We saw that the observables on the cigar satisfy a
reflection relation, \reflrel, which played an important role in our
discussion. In particular, singularities of the reflection coefficient
$R$ provided important information on the possible singularities of
correlation functions in that case.  The Liouville observables satisfy
a similar reflection property (see \eg\
\TeschnerRV):
\eqn\liouvrefl{\eqalign{
V_\alpha=&R_\alpha V_{Q-\alpha},\cr
R_\alpha=&-\mu^{Q-2\alpha\over b}
{\Gamma(-{1\over b}(Q-2\alpha))\Gamma(-b(Q-2\alpha))
\over
\Gamma({1\over b}(Q-2\alpha))\Gamma(b(Q-2\alpha))}~,\cr
}}
where $\mu=\pi\mu_0\gamma(b^2)$.

We are interested here in the case  $b=1$,  $Q=2$, for which
the Liouville central charge is $c_L=25$. Together with the
second space-like coordinate $x$, we have $c=26$, so we can
study bosonic string theory on this space.

Bosonic string theory in the two dimensional Euclidean space
labeled by $(\phi,x)$ has one field theoretic degree of freedom,
the massless ``tachyon'' whose vertex operator is given by
\eqn\tachver{T_k=e^{ikx+(2-|k|)\phi}~.}
In addition to the tachyon, the theory has some discrete states
that occur only at some particular values of the momenta, $k\in \Z$.

The dynamics of the tachyon is rather well understood, both in the
continuum formulation, and in the dual matrix quantum mechanics
in the double scaling limit. The operators \tachver\ are non-normalizable
observables in the theory, and we can compute their correlation functions.
One finds that these correlation functions have the following structure:
\eqn\corrtach{\langle T_{k_1}\cdots T_{k_n}\rangle=
\left[\prod_{i=1}^n\gamma(1-|k_i|)\right] F(k_1,\cdots, k_n)~,}
where the term in square brackets is usually referred to as the ``leg pole''
contribution, and $F(k_1,\cdots, k_n)$ 
does not have poles as a function of the $k_i$.
$F$ is a continuous function of the $k_i$, which is bounded for finite
values of the momenta; its singularities are discontinuities 
of the first derivative
with respect to the $k_i$, 
arising from structures like $|\sum_i k_i|$ with the sum
running over some subset of the momenta.

The discrete states have always been more mysterious; in particular, correlation
functions involving discrete states seemed to be 
divergent since they were related
to those of tachyons with the special momenta $k\in \Z$, 
which are infinite according to \corrtach.

The discussion of our paper is useful for elucidating the analytic
structure of the tachyon correlation functions \corrtach, and for
studying the role of the discrete states.  Consider first the
reflection relation for the special case of two dimensional string
theory:
\eqn\reflcone{V_\alpha=-\mu^{2(1-\alpha)}
{\Gamma^2(-2(1-\alpha))\over\Gamma^2(2(1-\alpha))}V_{2-\alpha}~.}
The reflection relation exhibits second order poles for
$2(1-\alpha)\in \Z_+$. Comparing to \tachver, we see that these double poles
occur at momenta corresponding to discrete states, $k\in \Z$, and moreover the
power of $\mu$ on the right hand side of \reflcone\ is a positive
integer in this case.

In \S2 we saw that poles of the reflection coefficient have one of two
origins.  They either signal the presence of normalizable states
living in the strong coupling region, or are due to bulk interactions
that occur very far from the wall. Each of these gives rise to first
order poles at the appropriate locations.  In our case we find that
the reflection coefficient has double poles for integer momenta, so
both of those mechanisms must be operating at the same time.  It must
be that for $k\in \Z$ there are normalizable states bound to the
Liouville wall, and we see from the integer power of $\mu$ appearing in 
\reflcone\ that bulk interactions are
possible in that case as well. Each of these effects gives one of the
two poles for $k\in \Z$ in \reflcone.

This picture is compatible with the structure of the $n$ point
functions of tachyons
\corrtach, which exhibit single poles as a function 
of each external momentum $k_i$.
These ``leg poles'' are nothing but LSZ poles for the particular case
of Liouville theory.  For $k\in \Z$ the non-normalizable tachyons
$T_k$ do not exist; one has to study instead the normalizable operators
obtained by taking the residues of the LSZ poles, as in the general
discussion of \S2 (see \eg\ \smatr).

There is also an analog of the discussion at the end of \S2.4 of the
question whether the same poles can sometimes be thought of as bulk poles
and sometimes as LSZ poles. The leg poles in \corrtach\ were first found in
bulk correlation functions
\refs{\DiFrancescoSS,\DiFrancescoUD} where one takes $n-1$ of the momenta to
be positive (say) and one negative. In that case, it was shown in
\refs{\DiFrancescoSS,\DiFrancescoUD} that the $n-1$ poles corresponding to
the positive momentum tachyons are LSZ poles, while the last pole,
corresponding to the negative momentum tachyon, was a bulk pole. As we
saw in \S2.4, it is a rather general phenomenon that the same poles
have seemingly different interpretations in different ways of doing
the calculation. We expect that the situation should be the same here,
namely, for generic kinematics there should be a way to do the
calculation in such a way that it is manifest that all $n$ leg poles
in \corrtach\ are LSZ poles.

Finally, the discussion above clarifies the status of the discrete
states in two dimensional string theory. Unlike tachyons with generic
(non-integer) momenta, there are no non-normalizable vertex operators
corresponding to these observables. They must be described by
normalizable vertex operators, which can be formally defined by the
procedure of going to the poles outlined in \S2. 

It is also useful to note that the discussion of this section generalizes
in a simple way to the case of two dimensional type 0 string theory, since
the structure of reflection coefficients and correlation functions in that
theory is almost identical to the bosonic case.

\newsec{The Thermodynamics of Little String Theories}

In this section we use our improved understanding of the physics
of normalizable states on the cigar to study the thermodynamics of 
LSTs, and in particular the possible existence of a Hagedorn phase 
transition in these theories.

The thermodynamics of LSTs was studied in 
\refs{\MaldacenaCG\AharonyTT\HarmarkHW\BerkoozMZ
\KutasovJP\RangamaniIR\BuchelDG\NarayanDR-\DeBoerDD}. 
In the bulk description, LST at a large energy density (in string units)
is described by the near-horizon limit of near-extremal $NS5$-branes (or
singularities), in which the background \typform\ is replaced by\foot{The
discussion of this section applies only to LSTs with $k>1$, since then 
the $SL(2,\IR)/U(1)$ black hole corresponds to a normalizable state in 
the theory ($k>1$ is required so that in equations \lsineliouv,\ntwoliouv,
$\beta=-1/Q<-Q/2$). In particular, among the examples we discussed above, it 
applies to the six dimensional example and to the four dimensional 
examples with $n > 2$, but not to two dimensional string theory, where
the black hole is non-normalizable \KazakovPM. We thank J. Maldacena for
a discussion on this issue.}
\eqn\rrpphht{SL(2,\IR)_k/U(1) \times \IR^{d-1} \times 
{\cal M}.}
For the theory at finite energy density, 
the $SL(2,\IR)/U(1)$ in \rrpphht\ is Lorentzian and replaces the time
direction and the linear dilaton direction of \typform; this should
not be confused with the unrelated appearance of the Euclidean
$SL(2,\IR)/U(1)$ space in our discussions of the double scaling 
limit\foot{In the case of the six dimensional LSTs discussed in
sections 4-6, 
the finite energy-density background \rrpphht\ takes the form
$SL(2,\IR)_k/U(1)\times \IR^5\times SU(2)_k$.} \doublesc. 
The behavior at finite temperatures may
be analyzed by the Euclidean continuation of \rrpphht, which involves
the same Euclidean $SL(2)/U(1)$ whose correlation functions we
discussed in detail above, although its interpretation now is quite
different (since the direction around the cigar is now the Euclidean 
time direction). The canonical partition function of LST 
also receives contributions from the bulk geometry
in which we just compactify the time direction of \typform\ on a
circle, but (as in the AdS/CFT correspondence \WittenZW) 
this is sub-dominant near the Hagedorn temperature.

In classical string theory, the radius of the (asymptotic) circle in
the $SL(2)/U(1)$ theory at level $k$ is
$\sqrt{k\alpha'}$. Thermodynamically, this corresponds to having
$\beta = 2\pi \sqrt{k\alpha'}$, so the temperature is fixed at a value
\eqn\hagtemp{T = T_H \equiv 1 / 2\pi
\sqrt{k\alpha'}.}  
This temperature is independent of the energy
density, which in \rrpphht\ is a function of the string
coupling $g_s^{(tip)}$ at the horizon of the Lorentzian $SL(2)/U(1)$
space,
\eqn\enerden{{E\over V} \simeq {k M_s^d \over {(g_s^{(tip)})^2}}.}
This is a Hagedorn-like behavior, corresponding in the
micro-canonical ensemble to a density of states $\rho(E) \simeq e^{E /
T_H}$ (we will assume in this section that the $(d-1)$-dimensional space
of the LST has been compactified on some finite large volume $V$, and we
will not explicitly write the volume dependence).

Quantum corrections can change this behavior; they were analyzed
at one-loop in string theory in \KutasovJP. 
It was found there that the classical density
of states is corrected to 
\eqn\coorbeh{\rho(E) \simeq E^{\alpha} e^{E / T_H}} 
with a
negative value of $\alpha$ (smaller than $(-1)$). 
This leads to a temperature which is slightly
above the Hagedorn temperature (at large energy densities), and to a
negative specific heat, meaning that the canonical ensemble is ill-defined.

The behavior \coorbeh\ implies that as one gradually increases the
temperature of LST, the Hagedorn temperature is reached at a finite
energy density (below the energy densities for which \coorbeh\ is a good
approximation).  Due to the negative specific heat, higher
temperatures cannot be accommodated in the background
\rrpphht. However, the fact that the Hagedorn temperature is reached at a
finite energy density suggests that perhaps at this temperature there
could be a transition to a different phase, as one finds
(for instance) in $SU(N)$ gauge theories with large $N$.  As usual,
the thermodynamic instability described above is reflected in the
canonical ensemble by a winding mode of the string around the thermal
time direction which is (classically) massless \KutasovJP, and the
phase transition would correspond to a condensation of this mode (as
in \AtickSI). We would like to argue that such a phase transition actually
does not occur, and that the Hagedorn temperature \hagtemp\ really is a 
maximal temperature for LST.

The simplest argument for this comes from the micro-canonical
ensemble.  In this ensemble the black hole states \rrpphht\ 
give rise to a density of states $\rho(E) \sim
e^{E/T_H}$ (at finite volume).  These states are not exactly stable
(due to Hawking radiation), but their life-time increases as $E$
becomes larger. So, it seems that one can trust this prediction for
the density of states at asymptotically high energies (at least, it is
a lower bound on the density of states -- there could be other types
of states that would increase it further).  The existence of such a
density of states implies that the theory should not make sense at
temperatures above the Hagedorn temperature, since all thermodynamic
quantities would diverge there.

In order to study the same question in the canonical ensemble, we
need to analyze the dynamics of the (classically massless) winding
mode mentioned above in the Euclidean background \rrpphht, to compute
its effective potential and
to see if it condenses or not. The vertex
operator for this mode is \KutasovJP\ 
\eqn\winding{V_W = e^{-\varphi -\bar\varphi} V^{(sl,susy)}_{j;m,\bar m} e^{ip_{\mu}
x^{\mu}},} 
with $m={\bar m}=k/2$, and the massless mode corresponds to the residue of
the pole in this operator at $j=m-1=(k-2)/2$ (which exists only for $k>1$). 
As mentioned above, this normalizable mode could become unstable at one-loop 
order and destabilize the background. In the string theory
the condensation of the corresponding normalizable state
(with zero momentum) would be described by adding to the worldsheet Lagrangian
the deformation 
\eqn\taccond{G_{-1/2} {\bar G}_{-1/2} V^{(susy,norm)}_{j;m,\bar m} + c.c.}
with $m={\bar m} =k/2, j=(k-2)/2$.

The winding operator $V_W$ in the Euclidean $SL(2)/U(1)$ theory is
precisely the same as the operator that corresponds to $\ttr(B^k)$ 
in the six dimensional $\NN=(1,1)$ DSLST, discussed in the previous
sections (see equations \aba, \dsbt). 
In the DSLST we saw that condensing this mode 
is equivalent to changing the string coupling at the tip of the
$SL(2,\IR)/U(1)$ cigar\foot{Note that this is true even though naively the
vertex operator \taccond\ is not the same as the vertex operator for a
change in the string coupling at the tip of the ``throat,'' which does
not carry any winding number. The two different non-normalizable vertex 
operators create the same normalizable state, as can be seen by
applying equation
\refsltwo\ to the supersymmetric case, so adding \taccond\ to the
worldsheet action is the same as 
adding the deformation changing the string coupling at the tip of the 
cigar. This is related to the (worldsheet supersymmetric version of the) 
FZZ duality \refs{\fzz,\KazakovPM} between $SL(2)/U(1)$ and sine-Liouville 
theory.}. Thus, the same must be
true in the background \rrpphht, where this is interpreted as changing
the energy density. This is not surprising -- in the thermodynamic
context one may expect the instability of the thermal LST to be
towards lowering or raising the energy of the system. This means that
the tachyon condensation here would either increase the energy
density, perhaps driving it to infinity (and the string coupling at
the tip to zero), or decrease the energy density (where for small
enough energy densities the perturbative analysis of \rrpphht\ would
break down).

It is important to note that the full spectrum of the string theory
involving the Euclidean $SL(2)/U(1)$ is
quite different in the two cases \rrpphht\ and \rrpphh, with different
GSO projections, orbifolds and so on -- but the particular
operator $V_W$ \winding\ appears in both cases. 
In particular, this operator has
the same tree-level correlation functions in DSLST and in
\rrpphht. This means that we can compute the tree-level potential
for the would-be thermal tachyon $V_W$ by computing zero-momentum
correlation functions of the normalizable mode of $\ttr(B^k)$
in the six dimensional DSLST.

However, these all vanish, since we know that there is no potential for $B$
in this (maximally supersymmetric) theory, so the S-matrix must vanish
at zero momentum. This means that the potential for the would-be
tachyon vanishes at tree-level. This again should not be surprising
given the interpretation of this tachyon as the energy density, since
the tree-level partition function of \rrpphht\ vanishes at all
energies.

Thus, the potential for the ``winding tachyon'' $V_W$
first arises at the one-loop level. Moreover, since we interpreted this
mode as the energy density, we already know what this potential is --
it can simply be read off from the one-loop partition function of
\KutasovJP. This gives a potential of the form
$(\alpha+1)\log(E)$. This form of the potential means that the
massless mode $V_W$ does not just become tachyonic at one-loop, but
actually develops a tadpole, which (since $\alpha < -1$) drives it
towards large values of the energy.  There is no stable end-point to
this tadpole condensation process (the corrections to the one-loop
contribution become smaller and smaller as the energy increases). This
is consistent with the thermodynamics developed in \KutasovJP\ --
trying to go above the Hagedorn temperature leads to a configuration
\rrpphht\ which has negative specific heat, so one is driven towards
higher energy densities (where the temperature approaches the Hagedorn
temperature). There seems to be no stable configuration at any
temperature above the Hagedorn temperature, consistently with the
discussion of the micro-canonical ensemble above.

To summarize, we described a self-consistent picture of
the thermodynamics of LSTs with $k>1$. In the micro-canonical
ensemble the density of states is given by \coorbeh, which
implies that the canonical ensemble is only
well-defined below the Hagedorn temperature. As one goes up to this
temperature one approaches a finite average energy density \KutasovJP,
but it is not possible to achieve thermal equilibrium at any higher
temperature. The fact that
the Hagedorn temperature is a maximal temperature is
similar to the behavior in free string theory in flat space. In order
to reach this conclusion we did not need to use any of the detailed results
of the previous sections, but our arguments are based on the
understanding (described in sections 2 and 3) of the normalizable
states in LST backgrounds and their correlation functions.

\newsec{Further comments on the results}

\subsec{General structure of correlation functions}

Our discussion of perturbative string theory in backgrounds of the 
general form \rrpphh\ leads to the following qualitative picture. 
The theory has two distinct (but coupled) sectors, one associated
with the vicinity of the tip of the cigar, the other with the asymptotic
region far from the tip. This is reflected in the spectrum of normalizable
states, as well as in the structure of off-shell Green functions of the 
physical observables of the theory. 

General observables carry both momentum and winding around the cigar,
and depending on the amount of momentum and winding they are sensitive to
different aspects of the physics. Consider, for example, pure winding
modes on the cigar. Since the energy of a wound fundamental string
decreases as it moves towards the tip of the cigar, wound strings
experience an attractive potential towards the tip. It was shown
in \DijkgraafBA\ that this potential has bound states, which are
nothing but the principal discrete series states with $m=\bar m=\pm(j+n)$,
$n=1,2,\cdots$. As we have seen in \S2, the structure of the 
correlation functions of non-normalizable vertex operators (or off-shell
Green functions) corresponding to wound strings reflects the presence
of these bound states, via the appearance of LSZ poles as a 
function of the external momenta. Near such poles, the Green function 
is dominated by the contribution of the bound states and thus is 
sensitive only to dynamics near the tip of the cigar. 

An example of the above discussion in the background associated with 
$k$ type IIB $NS5$-branes \slsu\ is the operators $\ttr(B^n)$ \formb, 
\dsbt. They correspond to pure winding modes on the cigar (in general 
with fractional winding), and create massless and massive principal 
discrete series states when acting on the vacuum.

The situation with pure momentum observables is quite different. Since 
the radius of the circle decreases as we move towards the tip of the 
cigar, the potential felt by these modes is repulsive \DijkgraafBA, and 
they do not form bound states near the tip. The physical process encoded 
in the off-shell Green functions of these observables is scattering off 
the repulsive potential provided by the tip. In particular, the singularity 
structure of the Green functions of momentum operators is different from 
that of the winding modes. They do not exhibit LSZ poles associated with 
the principal discrete series states, and their singularities are instead 
due to bulk amplitudes in the cigar geometry\foot{As mentioned in \S2, 
the amplitudes also have more conventional multiparticle singularities,
that are familiar from other string theory backgrounds.}. An example of 
such operators in the fivebrane background \slsu\ is the vertex operators 
$\ttr(A^n)$ \dsat, whose correlation functions do not have LSZ poles, 
as we saw.

The momentum-carrying operators are not completely blind to
the physics associated with the principal discrete series
states. For example, one expects the operators $\ttr(A^n)$ 
in the $d=6$ example to be able to create states with $n$
$A$-particles in the low-energy theory. As we saw, such
processes do not contribute to tree-level correlation functions
in the background \slsu, but they should contribute to 
higher-genus correlation functions. 

For general observables, which carry both momentum and winding,
there is a competition between the two effects, the repulsive 
potential due to the momentum, and the attractive one due to
the winding. If the operator is ``winding dominated,'' \ie\ 
if $m,\bar m$ \momwin\ have the same sign, we saw \poles, \polesinf\
that it couples to bound states living near the tip of the cigar.
Otherwise, it behaves like a momentum mode. Note that this is consistent
with what one expects: the potential 
\eqn\potatt{V(R)=\left(n\over R\right)^2+\left( w R\over\alpha'\right)^2}
is attractive (\ie\ $V'(R)>0$) when $|w|R/\alpha'>|n|/R$, and is repulsive
otherwise.

The above general picture helps to compare the analytic structure
of the Green functions of string theory in (for example) the 
background \slsu\ to expectations based on the low energy gauge 
theory of $k$ type IIB $NS5$-branes. The massless gauge theory 
states correspond in the geometry \slsu\ to the lowest lying 
principal discrete series states on the cigar. Thus, roughly 
speaking, the low energy gauge theory lives near the tip of the cigar. 
As explained above, the physics of winding modes near the LSZ poles 
corresponding to these massless states is indeed dominated by the 
vicinity of the tip. Thus, the low energy behavior of correlation
functions of operators such as $\ttr(B^n)$ \dsbt\ is dominated by 
the gauge theory contribution\foot{Except for the two-point function,
which receives also bulk contributions, 
as discussed in \S5.1.}. On the other hand, amplitudes that 
involve momentum modes such as $\ttr(A^n)$ \dsat\ do not have this 
property. Their low energy behavior is governed by the large $\phi$
region, and therefore has a non-gauge theoretic origin.   

Thus, we see that the low energy amplitudes of string theory on the
background \slsu\ are not entirely due to the contribution of the
broken $SU(k)$ gauge theory one normally associates with the 
fivebranes. They receive another contribution from a different
source, which in the cigar description corresponds to the contribution
of the region $\phi\to\infty$. 

\subsec{Weak-weak coupling duality?}

The above discussion helps to resolve another puzzle raised by the 
construction of the double scaling limit \doublesc. The equivalence
between string theory in the background \slsu\ and the fivebrane 
theory at a point along its Coulomb branch is expected to be an 
example of a gauge-gravity duality. Such dualities usually have 
the property that the two dual descriptions of the physics are 
never simple at the same time. For example, for coincident fivebranes, 
at low energies the gauge theory is expected to be weakly coupled 
six dimensional $SU(k)$ gauge theory, while the bulk description, 
corresponding to the CHS geometry \chsmetr\ is strongly coupled as 
$\phi\to-\infty$, and thus is not useful. 

{}From this point of view, the low energy limit of the fivebrane theory
in the double scaling limit \doublesc, \mws\ is very puzzling. On the
gauge theory side, $k$ separated fivebranes at a point along the Coulomb
branch are expected to be described in the IR by a $U(1)^{k-1}$ gauge 
theory with sixteen supercharges, which of course is free in the IR. At 
the same time, the gravity description in terms of string propagation in 
the background \slsu\ is also weakly coupled in the limit \mws\ and, if 
one wishes, it is possible to make the $\alpha'$ corrections small as well 
by sending $k\to\infty$. Thus, naively we seem to conclude that DSLST at
low energies is an example of weak-weak coupling gauge-gravity duality.

The resolution is that, as we saw, the $U(1)^{k-1}$ gauge theory gives 
only part of the contributions to the DSLST correlation functions at low
energies, and there are additional contributions coming from the dynamics 
in the bulk of the cigar. Thus, while the gravity description \slsu\ is 
indeed weakly coupled, there is no alternative weakly coupled description 
which captures all of the low-energy off-shell Green functions. In fact, 
we expect that if there is a gauge theory dual of the full LST in the 
double scaling limit, it is strongly coupled, even at low energies.

It is interesting that the gauge and gravity descriptions share a weakly 
coupled sector that {\it can} be described in two different ways -- the 
broken $SU(k)$ field theory which can be described both by field theory 
methods and by focusing on the residues of the massless LSZ poles in string 
theory on \slsu. This is similar to the fact that certain correlation
functions of chiral operators in $\NN=4$ SYM can be computed either by
studying the dynamics of gravitons on $AdS_5$, or by calculations in
weakly coupled gauge theory. Presumably, the dynamics of this weakly coupled 
sector of LST is similarly constrained.

\subsec{The density of normalizable states at large energies}

In sections 2 and 3 we saw that normalizable states in DSLST correspond 
to principal discrete series states on the cigar. We would like to estimate 
the growth in the number of such states at large masses. 

A rough way to proceed is as follows. A given operator with ``more winding
than momentum'' (\ie\ the same sign of $m$ and $\bar m$ \susymomwin) can
create a number of states that is proportional to $(k|w|-|n|)$. This
grows linearly with the winding, and therefore with the mass of the state.
However, a much larger contribution to the
growth in the number of normalizable states at large mass
comes from the exponential growth in the number of different operators that
can create normalizable states\foot{In general one needs to be careful 
because the same state can be created by different operators, but we do 
not expect this to drastically change the arguments below.} in \rrpphh. 
At high excitation levels, the growth in the number of such operators is 
comparable to that of ten dimensional string theory. It is somewhat smaller, 
since it is determined by the effective central charge of the worldsheet 
CFT, which is smaller than its actual (critical) central charge because 
it partly comes from an asymptotically linear dilaton background. 

Thus, we conclude that the high energy growth in the density of normalizable
states in DSLST is exponential, 
$\rho_{\rm pert}(E)\sim\exp(\beta_{\rm pert} E)$,
with $\beta_{\rm pert}$ of order one in string units. Of course, as implied
by the notation, this only takes into account perturbative string states in
the background \rrpphh. This estimate is expected to be reliable for energies 
well above $M_s$ but well below $M_W$ (or more generally well below 
$M_s/g_s^{(\rm tip)}$). Around the scale $M_W$, non-perturbative states
such as the W-bosons in the six dimensional example start appearing, 
and the perturbative estimate of the density of states breaks down.  
For energies well above $M_W$ the density of states of
LST is similar to 
that of a free superstring theory in $4k+2$ dimensions 
\refs{\MaldacenaCG,\ItzhakiDD,\AharonyTT}. 
It is much larger than the density of perturbative states for all $k\ge 2$. 
Clearly, most of the high energy states of LST are non-perturbative,
even when we are at the DSLST point in the moduli space.

\subsec{High-energy scattering in DSLST}

As discussed in \S1, \S2, the S-matrix of normalizable states in DSLST 
is obtained by studying correlation functions of the corresponding
normalizable vertex operators. It is interesting to examine the high
energy behavior of this S-matrix (\ie\ its behavior for large values 
of the Mandelstam invariants). 

In fact, this analysis is identical to that in critical string theory.
As mentioned in \S1, the normalizable vertex operators on spacetimes
like \slsu\ behave essentially as those in the critical string. Thus, 
the momentum dependence of the S-matrix is precisely equal to that of 
the usual superstring in $\IR^{d-1,1}$ 
(recall that the $SL(2,\IR)/U(1)$ component of normalizable
vertex operators does not depend on the momentum $p_{\mu}$ in $\IR^{d-1,1}$).
This means that the high-energy 
behavior of the S-matrix in DSLST will be similar to that of standard 
string theories. It will exhibit Regge behavior in the appropriate regime, 
and will decay exponentially when all kinematic variables are large. This 
is another aspect of LSTs which is similar to that of standard string 
theories, even though they are not gravitational.

In standard string theories the behavior described above can be
trusted up to energies of the order of the Planck scale, where higher orders
in string perturbation theory and non-perturbative effects (such as black
holes) become important. Similarly, we expect that in DSLST this behavior
will persist until the scale $M_W$, but will be modified above this scale.

\newsec{Open problems}

The results of this paper lead to a number of questions. In this
section we would like to briefly discuss some of them. 

\subsec{Little string worldsheets}

In studying the analytic structure of off-shell Green functions in 
spacetimes of the general form \rrpphh, we found it convenient to 
express the $SL(2,\IR)/U(1)$ vertex operators in terms of vertex 
operators on $AdS_3$, $\Phi_j(x,\bar x)$, integrated over the variables
$(x,\bar x)$ (see \eg\ \furtrans). The correlation functions on the cigar
can then be written as integrated versions of the $AdS_3$ ones \densga. 

Even though $x$ itself is not a meaningful object in string theory on
the cigar (in contrast to the $AdS_3$ case), we saw that the analytic 
structure of the correlation functions on the cigar is usefully described
by studying various degeneration limits of the integrals over
the $x_i$. For example, the contributions of the regions $x_i\to0,\infty$
(two points that are picked arbitrarily at the outset by the definition
of the integral transform \furtrans) were seen to give rise to LSZ poles
associated with external legs going on-shell, while contributions from
regions where two or more of the $x_i$ approach each other were shown
to give rise to singularities associated with bulk interactions.

It is natural to ask whether the variables $(x,\bar x)$ are just a
convenient technical tool for analyzing the analytic structure of the
amplitudes, or whether they have a deeper physical significance. Recall
that in string theory on $AdS_3$ these variables label positions
on the boundary of spacetime, and thus describe the base space on which
the two dimensional {\it spacetime} CFT is living. It is natural
to conjecture that in LST they should be thought of as worldsheet 
variables for some sort of strings, in terms of which the dynamics
can be formulated. 

Indeed, the singularities of amplitudes that we find arise in precisely the
right way for such an interpretation. We can interpret $x=0$ as corresponding
to the far past on the worldsheet (in the sense of radial quantization,
as is standard in CFT, or by mapping the $x$ plane to a cylinder), and 
$x=\infty$ as the far future. Thus, singularities associated with these regions 
have to do with the contribution of on-shell physical states, as we have found. 
Similarly, the fact that regions in which some of the $x_i$ approach each other 
have to do with interactions is familiar from studies of Shapiro-Virasoro 
amplitudes in critical string theory. 

It would be interesting to reformulate our results in terms of the
dynamics of the strings whose worldsheet is labeled by $(x,\bar x)$. 
It is clear that this would be a very different kind of string theory from
what we are accustomed to, and the worldsheet description is bound to be
different as well. For example, this string theory is non-critical, 
off-shell amplitudes in spacetime make sense in it, and there does not 
seem to be an analog of the volume of the conformal Killing group that 
we divide by, that in the usual string theory fixes three of the $n$ 
integrals in an $n$ point function. It is also not clear whether/how 
one is supposed to sum over the genus of the worldsheet labeled by $x$, 
and if so, with what weight (\ie\ what is the value of the string coupling),
etc. As described above, the $x$ coordinates seem to naturally live on
a sphere with punctures.

It is interesting to note that some type of ``little string'' appears also
in the DLCQ description of LSTs, at least for the case of the $d=6$
$\NN=(1,1)$ LSTs which we discussed in detail above. The DLCQ of these
theories, with $N$ units of light-like momentum, 
is given by the $1+1$ dimensional SCFT which arises at low energies
on the Coulomb branch of the $\NN=(4,4)$ $U(N)^k$ gauge theory with
bifundamental hypermultiplets \refs{\SethiZZ,\GanorJX,\AharonyDW},
compactified on a circle. The
Coulomb branch includes configurations where each $U(N)$ group is broken
to an Abelian subgroup, and as in the Matrix theory description of type
IIA string theory \refs{\BanksVH,\MotlTH,\DijkgraafVV} 
one can construct ``long string'' configurations in
which the eigenvalues of the matrices are permuted around the circle, and
which could carry energies of order $1/N$. It
is tempting to conjecture that these strings could be related to the
strings mentioned in the previous paragraphs. However, there is no reason
to believe that these strings are weakly coupled, so it is hard to see why
correlation functions on their worldsheet would be meaningful.

\subsec{Other asymptotically linear dilaton spacetimes}

Throughout most of this paper we focused on backgrounds
of the form \rrpphh, which contain an $SL(2,\IR)/U(1)$
factor, and used the fact that in that case we know a
lot about the CFT. In particular, the variables $(x,\bar x)$,
which we have used extensively, appear due to the fact that
the worldsheet CFT is a coset of $SL(2,\IR)$. It is natural
to wonder how the structure of the theory changes when we
consider more general asymptotically linear dilaton spacetimes.

For example, in the six dimensional LST with $\NN=(1,1)$ supersymmetry
that we focused on in this paper, such backgrounds can be
obtained by moving away from the highly symmetric point in
moduli space \bexpv\ to more generic points, corresponding to 
other distributions of $NS5$-branes in the transverse $\IR^4$. 
For all points in the moduli space, the background looks the same
near the boundary \chsmetr; what distinguishes between
different points in the moduli space is the form of the ``wall''
that prevents $\phi$ from going to $-\infty$. The cigar, or $\NN=2$
Liouville \ntwoliouv, wall is replaced by a more general one. 

Physically, one would expect most of the qualitative conclusions
we reached to be valid in this more general setup. There should
still be normalizable states living in the vicinity of the wall.
These should include the massless gauge bosons living on the fivebranes,
which should of course be there anywhere in moduli space. Some 
of the correlation functions of non-normalizable (off-shell) operators
should exhibit LSZ poles associated with these states. There should
also be another sector of the theory that is sensitive to the structure
far from the wall, and amplitudes that involve the relevant operators
should exhibit bulk poles. The type of UV/IR mixing that we found should
also exist more generally than our examples. 

It seems that in order to verify the above claims, one has to develop
more powerful techniques for studying string theory in asymptotically 
linear dilaton spacetimes, and in particular the analytic structure
of amplitudes in such spacetimes. Of special interest is the question
whether there is an analog of the variables $(x,\bar x)$ that allows
one to analyze the analytic structure of amplitudes in a way similar
to what we have done in the case of $SL(2,\IR)/U(1)$ above. If the 
interpretation of $(x,\bar x)$ in terms of worldsheet variables for
little strings is correct, it seems that such variables should exist at 
any point in moduli space.

\subsec{Phenomenological implications}

One of the conclusions of our analysis was that some correlation 
functions in LST, such as the two-point function of \S5.2,
are enhanced in the IR (due to the bulk poles) in a way that cannot 
be understood just by studying the light degrees of freedom. 
This effect violates the usual renormalization group intuition in 
which low-energy correlators can be described just by using low-energy 
states. It is natural to ask whether such effects could possibly have 
phenomenological implications in compactifications of string theory that 
include LSTs, such as the scenario of \AntoniadisSW, where the asymptotically
linear dilaton direction is cut off by an upper bound on $\phi$, but the 
value of the string coupling there (which determines the 
four dimensional Planck scale) is very small. 

Observing the resulting IR enhancement seems to be difficult for 
the following reason. This enhancement occurs in correlation functions 
of operators whose wave-function is dominated by the weak-coupling end 
of the ``throat.'' However, in these scenarios, the standard model fields 
do not live at the (very) weakly coupled end of the ``throat,'' but rather 
near the wall (for instance, they could include the principal discrete 
series states living near the tip of the cigar in the example of this paper,
or D-branes living near the tip). It is not clear how observers made
of standard model fields can access the analytic structure encoded in
the correlation functions of operators whose wave-function is dominated 
by the other end of the ``throat''\foot{We thank S. Dimopoulos and 
E. Silverstein for discussions on this issue.}. On the other hand, 
the fact (discussed in \S9) that LSTs have a limiting temperature 
at the string scale could potentially have observable cosmological 
consequences in the scenario of \AntoniadisSW, where $M_s$ is around a TeV.

\subsec{Open-closed string duality}

It is natural to ask whether asymptotically linear dilaton
spacetimes such as \typform, \rrpphh\ have a holographically
dual open string description. In the special case of two 
dimensional string theory discussed in section 8, there is a 
well known dual -- matrix quantum mechanics in the double 
scaling limit \refs{\KlebanovQA,\GinspargIS,\PolchinskiMB}.
This is now understood as the theory on a large number of
D-branes localized deep inside the Liouville wall 
\refs{\McGreevyKB,\KlebanovKM}. A generalization of this
construction leads to a matrix model dual to two dimensional
type 0 string theory \refs{\TakayanagiSM,\DouglasUP}. Some
results on matrix models dual to two dimensional string theory
on the cigar are available as well \refs{\KazakovPM,\McGreevyDN,
\GiveonWN}.

This naturally leads to the question of whether one can similarly 
construct a matrix model dual to LST backgrounds \rrpphh\ in higher 
dimensions. One way to proceed is to consider again
the dynamics on D-branes localized in the region where the 
coupling is largest, which on the cigar is the region near 
the tip. A number of questions immediately arise. One is 
how many such branes should we take. In the AdS/CFT 
correspondence, the number of $D3$-branes is in general 
finite and related to the string coupling on $AdS_5$, 
while in the holographic duality of two dimensional string 
theory mentioned above, the matrix model has strictly 
infinite $N$ (in the double scaling limit). 

Another question is in how many dimensions of $\IR^{d-1,1}$
should the branes be extended. Naively, one might expect them
to be extended in all $d$ dimensions, so as to match the Poincar\'e
symmetry of the closed string dual, but it is possible that
other choices are allowed as well. Finally, one can ask whether 
the open string dual is expected to be the full open string 
theory on these branes, or just some low energy limit of the 
full open string theory. 

The answers to all these questions are not clear at present. 
We believe that it is likely that one should consider an 
infinite number of localized D-branes, but obviously more 
work is required to definitively decide one way or the other. 
In any case, it would be very interesting to rederive some or 
all of our results from an open string perspective.

It might also be possible to explore the physics of LST
by using its ``deconstruction'' by a large $N$ gauge theory, 
either along the lines of \ArkaniHamedIE\ or by using the new 
deconstruction of LST in \DoreyPP. In this framework the large 
$N$ gauge theory is also responsible for the construction of 
(typically two -- either compact or large) dimensions in LST.

In \DoreyPP, a certain deformation of four dimensional
${\cal N}=4$ SYM with gauge group $SU(N)$ at large $N$
was considered. This theory has a branch where it is
confined down to $SU(k)$ (or $U(1)^{k-1}$ in the Coulomb 
branch of the unconfined sector). In a certain scaling 
limit on this branch at large $N$ it is claimed that this 
gauge theory theory is equivalent to six dimensional DSLST.
It would be very interesting to make the correspondence
more precise. In particular, the construction of \DoreyPP\
raises the interesting possibility that some of the peculiar
low energy results that we found using the bulk description
reflect the physics associated with a strongly coupled sector
of the dual gauge theory.

\bigskip
\noindent{\bf Acknowledgements:}
We would like to thank Y. Antebi, M. Berkooz, S. Dimopoulos, 
N. Dorey, B. Fiol, E. Kiritsis, J. Maldacena, A. Naqvi, V. Niarchos, 
N. Seiberg, E. Silverstein and T. Volansky for useful discussions. 
O.A. would like to thank Stanford University, SLAC and
Harvard University for hospitality during the course of this project.
A.G. would like to thank the EFI and the Department of Physics at the 
University of Chicago, where this work was initiated, for its warm
hospitality. D.K. thanks the Weizmann Institute of Science, LPTHE 
Paris VI and the NHETC at Rutgers University for hospitality. 
The work of O.A. was supported 
in part by the Israel-U.S. Binational Science Foundation and by Minerva. 
O.A. is the incumbent of the Joseph and Celia Reskin career 
development chair.
The work of A.G. was supported in part by the
German-Israel Bi-National Science Foundation.
The work of O.A. and A.G. is supported in part by the Israel Academy
of Sciences and Humanities -- Centers of Excellence Program, by
the European network HPRN-CT-2000-00122, and by the Albert Einstein
Minerva Center for Theoretical Physics. Minerva is funded through the BMBF.
D.K. is supported in part by DOE grant \#DE-FG02-90ER40560.

\appendix{A}{Some results from $SL(2)/U(1)$ CFT}

\subsec{Review of two-point and three-point functions in the $SL(2)$ and
$SL(2)/U(1)$ CFTs}

We start our discussion from the bosonic $SL(2)$ WZW model of level
\eqn\levlsl{k_{SL(2)}=k+2,}
with central charge
\eqn\cntrsl{c_{SL(2)}={3(k+2)\over k}.}
The natural observables in the theory defined 
on the Euclidean version of $SL(2)$, 
$H_3^+~\equiv~SL(2,\IC)/SU(2)$, are primaries $\Phi_j(x,\bar x)$
of the $SL(2)_L\times SL(2)_R$
current algebra \refs{\TeschnerFT,\TeschnerUG} with $j>-{1\over 2}$.
The worldsheet scaling dimension of $\Phi_j(x,\bar x)$ is
\eqn\dimphij{\Delta(j)=-{j(j+1)\over k}~.}
In the papers \refs{\TeschnerFT,\TeschnerUG} the operators $\Phi_j$
are normalized as follows:
\eqn\twptsl{\langle\Phi_{j_1}(x_1,\bar x_1)
\Phi_{j_2}(x_2,\bar x_2)\rangle=
\delta(j_1-j_2){k\over\pi}\left[{1\over k\pi}\gamma
\left({1\over k}\right)\right]^{2j_1+1}
\gamma\left(1-{2j_1+1\over k}\right)|x_{12}|^{-4(j_1+1)},}
with $\gamma(x) \equiv \Gamma(x)/\Gamma(1-x)$.
In some of our computations (in particular in \S5.4) it is more convenient to
choose a different normalization
\eqn\difnorm{\tilde\Phi_j(x,\bar x)\equiv {\Phi_j(x,\bar x)\over
\sqrt{{k\over\pi}\left[{1\over k\pi}\gamma\left({1\over k}\right)\right]^{2j+1}
\gamma\left(1-{2j_1+1\over k}\right)}}.
}
In this normalization the two-point function is
\eqn\twptfn{\langle\tilde\Phi_{j_1}(x_1,\bar x_1)
\tilde\Phi_{j_2}(x_2,\bar x_2)\rangle=
\delta(j_1-j_2)|x_{12}|^{-4(j_1+1)}.}
Most of our considerations in this paper are independent of the 
normalization, since it does not affect the pole structure for
$-1/2 < j < (k-1)/2$; it is easy to translate the formulas below 
to the normalization \difnorm.

For discussing the coset $SL(2)/U(1)$ it is convenient to choose a 
different basis for the primaries $\Phi_j$ (or $\tilde\Phi_j$)
\eqn\difbas{\Phi_{j;m,\bar m}=\int d^2x \, x^{j+m}\bar x^{j+\bar m}
\Phi_j(x,\bar x).}
The two-point function in this basis was computed in \refs{\fzz,\GiveonTQ} : 
\eqn\twptmm{\eqalign{\langle\Phi_{j;m,\bar m}\Phi_{j';-m,-\bar m}\rangle=
k \left[{1\over k\pi}\gamma \left({1\over k}\right)\right]^{2j+1}
& \gamma\left(1-{2j+1\over k}\right) \delta(j-j') \times \cr
& {\Gamma(-2j-1)\Gamma(j-m+1)\Gamma(1+j+\bar m)\over\Gamma(2j+2)\Gamma(-j-m)
\Gamma(\bar m-j)}.\cr}}

The three-point function in the $SL(2)$ CFT takes the form
\eqn\thrptsl{\eqalign{\langle\Phi_{j_1}(x_1,\bar x_1)
& \Phi_{j_2}(x_2,\bar x_2)
\Phi_{j_3}(x_3,\bar x_3)\rangle=\cr
&D(j_1,j_2,j_3)|x_{12}|^{2(j_3-j_1-j_2-1)}|x_{13}|^{2(j_2-j_1-j_3-1)}
|x_{23}|^{2(j_1-j_2-j_3-1)}, }} 
where the structure constants
$D(j_1,j_2,j_3)$ were computed in
\refs{\TeschnerFT,\TeschnerUG}:
\eqn\ddef{\eqalign{D&(j_1,j_2,j_3)={k\over 2\pi^3}
\left[{1\over k\pi}\gamma \left({1\over k}\right)\right]^{j_1+j_2+j_3+1}
\times\cr
&{G(-j_1-j_2-j_3-2)G(j_3-j_1-j_2-1)G(j_2-j_1-j_3-1)G(j_1-j_2-j_3-1)\over 
G(-1)G(-2j_1-1)G(-2j_2-1)G(-2j_3-1)}.
}}
$G(j)$ is a special function which satisfies the following useful identities:
\eqn\propG{\eqalign{G(j)&=G(-j-1-k),\cr
G(j-1)&=\gamma(1+{j\over k})G(j),\cr
G(j-k)&=k^{-(2j+1)}\gamma(j+1)G(j).
}}
$G(j)$ has poles at the following values of $j$: 
$j=n+mk$, $j=-(n+1)-(m+1)k$, where $n,m=0,1,2,\cdots$. In particular, for
$j=0,1,2,\cdots < k$ it has single poles.

In the $(j,m,\bar m)$ basis the three-point function for the special case
$m={\bar m}$ is given by
\eqn\basthree{\eqalign{
\langle \Phi_{j_1;m_1, m_1} & \Phi_{j_2;m_2,m_2}
\Phi_{j_3;m_3,m_3}
\rangle=D(j_1,j_2,j_3)\times\cr
&F(j_1,m_1;j_2,m_2;j_3,m_3)\int d^2 x|x|^{2(m_1+m_2+m_3-1)}~,\cr}}
where
\eqn\jmjmjm{\eqalign{
F(j_1,m_1;j_2,m_2;&j_3,m_3)=\int d^2 x_1 d^2x_2|x_1|^{2(j_1+m_1)}
|x_2|^{2(j_2+m_2)}\times\cr
&|1-x_1|^{2(j_2-j_1-j_3-1)}|1-x_2|^{2(j_1-j_2-j_3-1)}
|x_1-x_2|^{2(j_3-j_1-j_2-1)}~.\cr}}
The integral over $x$ in \basthree\ ensures momentum conservation
$m_1+m_2+m_3=0$. The function $F$ \jmjmjm\ does not seem to be expressible
in terms of elementary functions. 

The same two-point functions and three-point functions arise also in
the coset $SL(2)/U(1)$ for the operators $V_{j;m.{\bar m}}$ (arising from
$\Phi_{j;m,{\bar m}}$) when we look at correlation functions
preserving the winding number, since the $U(1)$ part contributes
trivially.  In the coset, additional correlation functions are
non-vanishing as well.

\subsec{The superconformal algebra of $SL(2)/U(1)$}

The supersymmetric $SL(2)$ WZW model of level $k$ may be viewed as the
sum of a bosonic $SL(2)$ theory of level $k+2$, with currents $j^a$ ($a=1,2,3$),
\eqn\curralg{j^a(z) j^b(0) \sim {{1\over 2}(k+2) \eta^{ab} \over {z^2}} 
+ i \epsilon^{abc} {j_c(0) \over z},} 
($\eta = \rm{diag}(1,1,-1)$)
and of three free fermions $\lambda^a$ ($a=1,2,3$), 
which can be associated with an
$SL(2)$ current algebra of level $k=-2$. The total $SL(2)$ currents are
given by
\eqn\totcur{J_a^{\rm (total)} = j_a - {i\over 2} \epsilon_{abc}
\lambda^b \lambda^c,}
where the fermions obey
\eqn\fervev{\lambda^a(z) j^b(w) \sim 0,\qquad
\lambda^a(z) \lambda^b(w) \sim {\eta^{ab}\over {z-w}}.}
This theory has an $\NN=1$ superconformal symmetry generated by the
current
\eqn\gsltwo{G = Q (\eta_{ab} \lambda^a j^b - {i\over 6} \epsilon_{abc}
\lambda^a \lambda^b \lambda^c),}
with $Q^2 = 2 / k$.
The $\lambda^a$'s are superconformal primaries, and the $J_a^{\rm (total)}$
are the top components of the corresponding multiplets.

The supersymmetric $SL(2)/U(1)$ theory is defined by gauging the
$U(1)$ superfield including $\lambda^3$ and $J_3^{\rm (total)}$. This
theory has an $\NN=2$ superconformal algebra, generated by
\eqn\ntwogen{\eqalign{
&G^+ = Q J^- \lambda^+,\qquad G^- = Q J^+ \lambda^-, \cr
&J = (1 + Q^2) :\lambda^+ \lambda^-: + Q^2 j^3 = :\lambda^+ \lambda^-: + 
Q^2 J^{3{\rm(total)}},\cr}}
where
\eqn\defplusminus{J^{\pm} \equiv J^1 \pm i J^2,\qquad \lambda^{\pm} \equiv
{1\over \sqrt{2}}(\lambda^1 \pm i \lambda^2).}
The right-moving fields obey similar algebras.

\appendix{B}{Mixing of single and multi-trace operators}

As we saw in the text, the single string vertex operators in the background
\chsmetr\ do not correspond to single trace operators in the low energy
$SU(k)$ gauge theory, but rather to a mixture of single and multi trace
operators. In this appendix we determine the coefficients of the different
multi trace operators in this mixture.

The operators of interest to us will be
\eqn\defo{\OO_n\equiv {1\over n}{\rm tr}(B^n).}
The correlation functions of interest are
\eqn\corfunctns{(n_1,n_2,\cdots, n_j)\equiv\langle \OO_{n_1}\OO_{n_2}
\cdots \OO_{n_j}\bar\OO_n\rangle;\;\;\;n=\sum_{i=1}^j n_i.}
As usual in DSLST, we will compute these correlation functions
at a point along the Coulomb branch, \bexpv, where the $SU(k)$
gauge symmetry is broken to $U(1)^{k-1}$. Using free field theory,
one finds that the correlators \corfunctns\ are non-zero.
The leading connected diagram in the $1/M_W$
expansion is due to a contraction of a single $B$ out of each $\OO_{n_i}$ 
with a
$B^*$ inside the $\bar \OO_n$. Thus, it goes like $M_W^{2(n-j)}$, and
the dependence
on the locations of the operators, $(x_1,x_2,\cdots, x_j;\bar x)$, is
simple, $(n_1,n_2,\cdots, n_j)\sim \prod_{i=1}^j(x_i-\bar x)^{-4}$
(we will omit it below). Our main interest will be on the dependence
on the $\{n_i\}$; one finds\foot{For $j=1$ one has $(n)=k$.}
\eqn\corni{(n_1,n_2,\cdots, n_j)=k(n-1)(n-2)\cdots (n-j+1)=k{\Gamma(n)\over
\Gamma(n-j+1)}}
where $n=\sum_i n_i$ and $k$ is the number of fivebranes (or the rank of the
gauge group plus one).

The string theory analysis of \S6 shows that the single string vertex
operators with the quantum numbers of $\OO_n$, which were denoted by
$\BB_n$ in \S6, have the property that the analogs of the correlators
\corfunctns, \corni\ vanish for them (to leading order in the $1/M_W$
expansion). Our interpretation of this fact is that the correspondence
between $\OO_n$ and $\BB_n$ is non-trivial. The symmetries
allow a general mixing of the form
\eqn\mixops{
\BB_n=\OO_n+\sum_l{1\over l!}\alpha_{n_1,n_2,\cdots, n_l}
\OO_{n_1}\cdots\OO_{n_l}}
where the integers $n_i$ are summed over, subject to the constraint
that their sum is $n$. The basic idea is that to match the vanishing
of the string theory correlation function in the low energy gauge
theory, one should consider instead of \corfunctns\ the correlation
function $\langle \BB_{n_1}\BB_{n_2}\cdots \BB_{n_j} \bBB_n\rangle$
and fix the coefficients $\alpha_{n_1,n_2,\cdots, n_l}$ such that it
vanishes.  It is clear that there is the same number of free
parameters ($\alpha_{n_1,n_2,\cdots, n_l}$) and equations (due to the
vanishing of the above correlators).  So, we expect to be able to find
a solution.

As we will see, something surprising (from the current perspective) happens,
and the coefficients $\alpha$ that one finds this way actually do not depend
on all the $n_i$, but only on their sum, $n$, and on $l$. Thus, we will
denote the coefficients $\alpha$ by $\alpha_{n;l}$. We will find that they
are given by the simple expression
\eqn\alnl{\alpha_{n;l}=\left({1-n\over k}\right)^{l-1}.}

\noindent
The mixing \mixops\ leads to the following expansion for the correlation
function of interest:
\eqn\vancor{\eqalign{0=
\langle \BB_{n_1}\BB_{n_2}\cdots
\BB_{n_j}\bBB_n\rangle
=&(n_1,n_2,\cdots, n_j)+\alpha_{n;2}\sum (n,\cdots,n)(n,\cdots,n)\cr
&+\alpha_{n;3}\sum (n,\cdots,n)(n,\cdots,n)(n,\cdots,n)+\cdots,\cr}}
where in each term one sums over all different orderings of the
$n_i$. We next give the few lowest of these equations, and use them to
determine $\alpha_{n;l}$ for small $l$.

For $j=2$, we have
\eqn\jtwo{(n_1,n_2)+\alpha_{n;2}(n_1)(n_2)=0}
Using \corni\ to evaluate the first term, and the fact that $(n_1)=(n_2)=k$,
we find that $\alpha_{n;2}=-(n-1)/k$, in agreement with the general
expression \alnl.

For $j=3$, we have
\eqn\jthree{(n_1,n_2,n_3)+
\alpha_{n;2}\left[(n_1,n_2)(n_3)+(n_1,n_3)(n_2)+(n_2,n_3)(n_1)\right]
+\alpha_{n;3}(n_1)(n_2)(n_3)=0}
Using \corni\ and the form of $\alpha_{n;2}$ found previously, one concludes
that
\eqn\threeconst{
k(n-1)(n-2)-k^2{n-1\over k}(2n-3)+k^3\alpha_{n;3}=0}
This leads to $\alpha_{n;3}=(n-1)^2/k^2$, again in agreement with \alnl.

For $j=4$, we have (in hopefully self explanatory notation)
\eqn\jfour{\eqalign{
&(1,2,3,4)+\cr
&\alpha_{n;2}[(1,2)(3,4)+(1,3)(2,4)+(1,4)(2,3)+\cr
&(1,2,3)(4)+(1,2,4)(3)+(1,3,4)(2)+(2,3,4)(1)]+\cr
&\alpha_{n;3}[(1,2)(3)(4)+(1,3)(2)(4)+(1,4)(2)(3)+\cr
&(2,3)(1)(4)+(2,4)(1)(3)+(3,4)(1)(2)]+\cr
&\alpha_{n;4}(1)(2)(3)(4) = 0.\cr
}}
Substituting the known results one finds
\eqn\fourconst{
(n-1)(n-2)(n-3)-(n-1)(3n^2-12n+11)+
(n-1)^2(3n-6)
+k^3\alpha_{n;4}=0}
Thus, $\alpha_{n;4}=-(n-1)^3/k^3$, as in \alnl.

For $j=5$, the full expression is somewhat long, so we only give the
analog of \fourconst\ for this case. It is:
\eqn\fiveconst{\eqalign{
&(n-1)(n-2)(n-3)(n-4)-(n-1)(4n^3-30n^2+70n-50)+\cr
&(n-1)^2(6n^2-30n+35)-(n-1)^3(4n-10)+k^4\alpha_{n;5}=0.\cr
}}
Again, this is in agreement with \alnl.

A direct continuation of the above approach to higher values of
$j$ seems impractical. To get more general results we will use
a property of the calculations described above that seems quite
non-trivial but that we have not been able to prove in general. Apriori,
one might expect the coefficient of each $\alpha_{n;l}$ in equations
like \vancor, \jthree\ -- \fiveconst\ to depend on the individual $n_i$,
but the explicit formulae displayed above {\it always} depend only
on $n=\sum n_i$. This is why the coefficients $\alpha$ in \mixops\
can be taken to depend only on $n$ and $l$.

If we assume this property persists for arbitrarily high $l$ (an assumption
that is perhaps not unlikely since we have seem that 
all terms with $l\le 5$ do satisfy this
property), we can choose particularly convenient values for the $n_i$,
and use them to determine the mixing coefficients $\alpha_{n;l}$. This
is what we do next.

Consider the special case
\eqn\speccase{(n_1,\cdots, n_j)=(n,0,0,\cdots,0)}
This is outside the range of interest for our application, but
for proving the mathematical statement about polynomials
it is as good as any other choice.
In order to study this case we need to know what are
$(0,0,\cdots,0)\equiv (0^l)$, and $(n,0^j)$. Using \corni\ it is
not difficult to see that
\eqn\zerol{\eqalign{
&(0^l)=k(-1)^{l-1}(l-1)!\cr
&(n,0^j)=k{(n-1)!\over (n-j-1)!}\cr
}}
The expansion \vancor\ simplifies significantly in this case.
One can write it as follows:
\eqn\expzero{{1\over j!}(n,0^j)+
\sum_{j_1+j_2=j}{\alpha_{n;2}\over j_1!j_2!}(n,0^{j_1})(0^{j_2})
+\sum_{j_1+j_2+j_3=j}{\alpha_{n;3}\over 2!}{1\over j_1!j_2!j_3!}
(n,0^{j_1})(0^{j_2})(0^{j_3})+\cdots=0.}
Now, define the following function of an auxiliary variable $x$:
\eqn\fnx{f_n(x)=\sum_{j=0}^\infty{1\over j!}(n,0^j) x^j=k(1+x)^{n-1}.}
Note that although the sum over $j$ runs all the way to infinity,
$f_n(x)$ is actually a polynomial, since the coefficients $(n,0^j)$
vanish for $j\ge n$. This is particularly clear in the representation
with $\Gamma$ functions in \corni. In order to prove that the $\alpha_{n;l}$
take the form \alnl\ we would like to show that multiplying \expzero\
by $x^j$ and summing over $j$ gives a vanishing answer (or more precisely
an $x$ independent constant -- see below).

The sum over $j$ actually simplifies \expzero\ significantly, since now
one can sum independently over $j_1$, $j_2$, etc. To be precise, $j_1$
is now summed from $0$ to infinity, while $j_2, j_3$, etc are summed
from $1$ to $\infty$. Performing the sum over the $j_i$ (using \alnl) one finds
the following expression:
\eqn\aftersum{f_n(x)\sum_{l=1}^\infty {(1-n)^{l-1}\over (l-1)!}
\left[\log(1+x)\right]^{l-1}=f_n(x) e^{(1-n)\log(1+x)}=
f_n(x)(1+x)^{1-n}=k.}
We see that we did not get  quite zero, but the only non-zero term
is a constant. It is easy to understand where it comes from. Equation
\expzero\ only holds for $j>0$; for $j=0$ only the first term is there,
and it is equal to $(n)=k$. This is the non-zero term that we found in
\aftersum. We see that under the assumption that the $\alpha$'s
really only depend on $n$ and $l$, they must be given by the expression
\alnl.

\listrefs
\end